\documentclass[11pt]{article}
\newcounter{ourcount}
\usepackage{amsmath,amsfonts,amssymb,graphicx,epsfig,pdflscape,multirow,theorem,gensymb}
\usepackage[matrix,arrow,curve]{xy} 
\numberwithin{equation}{section}
\graphicspath{{./eps/}}
\usepackage[nosort]{cite}
\usepackage[usenames]{color}
\usepackage{pstricks}
\usepackage{pst-plot}
\usepackage[backref=false]{hyperref}
\usepackage[capitalise,noabbrev]{cleveref} 
\hypersetup{
colorlinks=true,
citecolor=red,
linkcolor=darkblue,
urlcolor=darkblue
}
\definecolor{darkblue}{rgb}{0,0,.8}
\definecolor{red}{rgb}{1,0,0}

\theorembodyfont{\itshape} 
\theoremheaderfont{\scshape}
\theoremstyle{plain}

\newtheorem{Proposition}{Proposition}[section]

\numberwithin{equation}{section}

\newcommand{\nc}{\newcommand}

\def\arxiv#1#2{\href{http://arxiv.org/abs/#1}{\textsf{arXiv:#1 #2}}}
\nc{\ir}{\mathrm{i}}
\nc{\eE}{\mathsf{e}}
\nc{\bib}{\bibitem}
\nc{\be}{\begin{equation}}
\nc{\ee}{\end{equation}}
\nc{\chit}{\raisebox{0.25ex}{$\chi$}}
\nc{\Dbh}{\mbox{\boldmath $\hat D$}}
\nc{\Dh}{\mbox{$\hat D$}}
\nc{\Dbb}{\mbox{\boldmath $\bar D$}}
\nc{\Dbm}{\mbox{\boldmath $\mathcal D$}}
\nc{\Dbt}{\mbox{\boldmath $\tilde{D}$}}
\nc{\Tbt}{\mbox{\boldmath $\tilde{T}$}}
\nc{\Atwoone}{A_2^{(1)}}

\nc{\db}{\mbox{\boldmath $d$}}
\nc{\Ab}{\mbox{\boldmath $A$}}
\nc{\Abt}{\mbox{\boldmath $\tilde{A}$}}
\nc{\Bb}{\mbox{\boldmath $B$}}
\nc{\Bbt}{\mbox{\boldmath $\tilde{B}$}}
\nc{\Cb}{\mbox{\boldmath $C$}}
\nc{\Db}{\mbox{\boldmath $D$}}
\nc{\eb}{\mbox{\boldmath $e$}}
\nc{\Fb}{\mbox{\boldmath $F$}}
\nc{\Fbt}{\mbox{\boldmath $\tilde{F}$}}
\nc{\fb}{\mbox{\boldmath $f$}}
\nc{\fbt}{\mbox{\boldmath $\tilde{f}$}}
\nc{\Gb}{\mbox{\boldmath $G$}}
\nc{\Hb}{\mbox{\boldmath $H$}}
\nc{\Ib}{\mbox{\boldmath $I$}}
\nc{\Jb}{\mbox{\boldmath $J$}}
\nc{\Kb}{\mbox{\boldmath $K$}}
\nc{\Lb}{\mbox{\boldmath $L$}}
\nc{\Mb}{\mbox{\boldmath $M$}}
\nc{\Pb}{\mbox{\boldmath $P$}}
\nc{\Qb}{\mbox{\boldmath $Q$}}
\nc{\Rb}{\mbox{\boldmath $R$}}
\nc{\Tb}{\mbox{\boldmath $T$}}
\nc{\Tbb}{\mbox{\boldmath $\bar T$}}
\nc{\Tbm}{\mbox{\boldmath $\mathcal T$}}
\nc{\tb}{\mbox{\boldmath $t$}}
\nc{\Ub}{\mbox{\boldmath $U$}}
\nc{\Vb}{\mbox{\boldmath $V$}}
\nc{\Wb}{\mbox{\boldmath $W$}}
\nc{\Xb}{\mbox{\boldmath $x$}}
\nc{\Yb}{\mbox{\boldmath $y$}}
\nc{\Zb}{\mbox{\boldmath $z$}}
\nc{\Lambdab}{\boldsymbol{\Lambda}}

\nc{\stan}{\mathsf{W}}
\newrgbcolor{darkgreen}{0., 0.733333, 0.0621042}
\definecolor{lightblue}{rgb}{.7,.7,1}
\definecolor{lightestblue}{rgb}{.95,.95,1}
\definecolor{lightlightblue}{rgb}{.85,.85,1}
\definecolor{midblue}{rgb}{.7,.7,1}

\nc{\elegant}{1.5pt}
\nc{\moyen}{1.0pt}
\nc{\mince}{0.5pt}

\def\loopa{
\psframe[linewidth=.25pt](0,0)(1,1)
}
\def\loopb{
\psframe[linewidth=.25pt](0,0)(1,1)
\psarc[linewidth=1.5pt,linecolor=blue](0,1){.5}{-90}{0}
}
\def\loopc{
\psframe[linewidth=.25pt](0,0)(1,1)
\psarc[linewidth=1.5pt,linecolor=blue](1,0){.5}{90}{180}
}
\def\loopd{
\psframe[linewidth=.25pt](0,0)(1,1)
\psarc[linewidth=1.5pt,linecolor=blue](0,0){.5}{0}{90}
}
\def\loope{
\psframe[linewidth=.25pt](0,0)(1,1)
\psarc[linewidth=1.5pt,linecolor=blue](1,1){.5}{180}{270}
}
\def\loopf{
\psframe[linewidth=.25pt](0,0)(1,1)
\psline[linewidth=1.5pt,linecolor=blue](0.5,0)(0.5,1)
}
\def\loopg{
\psframe[linewidth=.25pt](0,0)(1,1)
\psline[linewidth=1.5pt,linecolor=blue](0,0.5)(1,0.5)
}
\def\looph{
\psframe[linewidth=.25pt](0,0)(1,1)
\psarc[linewidth=1.5pt,linecolor=blue](1,0){.5}{90}{180}
\psarc[linewidth=1.5pt,linecolor=blue](0,1){.5}{-90}{0}
}
\def\loopi{
\psframe[linewidth=.25pt](0,0)(1,1)
\psarc[linewidth=1.5pt,linecolor=blue](0,0){.5}{0}{90}
\psarc[linewidth=1.5pt,linecolor=blue](1,1){.5}{180}{270}
}
\def\loopid{
\psframe[linewidth=.25pt](0,0)(1,1)
\psarc[linewidth=1.5pt,linecolor=blue,linestyle=dashed,dash=2pt 2pt](1,0){.5}{90}{180}
\psarc[linewidth=1.5pt,linecolor=blue,linestyle=dashed,dash=2pt 2pt](0,1){.5}{-90}{0}
}

\def\tba#1{{\mbox{\boldmath $t$}}^{#1}}
\def\tbb#1{{\mbox{\boldmath $\tilde t$}}^{#1}}
\def\Xba{\mbox{\boldmath $x$}}
\def\Xbb{\mbox{\boldmath $\tilde x$}}

\def\facegrid#1#2{
\psframe[fillstyle=solid,fillcolor=lightlightblue,linewidth=0pt]#1#2
\psgrid[gridlabels=0pt,subgriddiv=1]#1#2}

\def\wobblylinev#1{
\psparametricplot[linecolor=blue,linewidth=1.2pt,plotpoints=1435]{0}{#1}{0.05 1500 t mul sin mul t}       
}

\def\wobblyarc#1#2#3{
\psparametricplot[linecolor=blue,linewidth=1.2pt,plotpoints=1435]{#2}{#3}{0.05 40 #1 t mul mul sin mul #1 add t cos mul 0.05 40 #1 t mul mul sin mul #1 add t sin mul}
}

\def\wobblyarctwo#1#2#3{
\psparametricplot[linecolor=blue,linewidth=1.2pt,plotpoints=1435]{#2}{#3}{0.07 20 #1 t mul mul sin mul #1 add t cos mul 0.07 20 #1 t mul mul sin mul #1 add t sin mul}
}

\def\young#1#2{
\multiput(0,0)(.1,0){#2}{\psline[linewidth=0.02cm]{-}(0,0)(0,.1)(.1,.1)(.1,0)(0,0)}
\multiput(0,.1)(.1,0){#1}{\psline[linewidth=0.02cm]{-}(0,0)(0,.1)(.1,.1)(.1,0)(0,0)}
}

\def\specialcircle#1{
\pspolygon[fillstyle=solid,fillcolor=black,linewidth=0.5pt](#1,#1)(#1,-#1)(-#1,-#1)(-#1,#1)
}

\def\specialsquare#1{
\pspolygon[fillstyle=solid,fillcolor=white,linewidth=0.5pt](#1,#1)(#1,-#1)(-#1,-#1)(-#1,#1)
}

\renewcommand{\ge}{\geqslant}
\renewcommand{\le}{\leqslant}

\def\qbinom#1#2{
\left[
\begin{matrix}
#1 \\ #2
\end{matrix}
\right]
}

\nc{\proof}{{\scshape Proof.\ }} 				
\nc{\eproof}{{\hfill \rule{0.5em}{0.5em}\medskip}}

\begin{document}

\topmargin -5mm
\oddsidemargin 5mm

\vspace*{-2cm}

\setcounter{page}{1}

\vspace{22mm}
\begin{center}
{\huge {\bf Fusion hierarchies, $\boldsymbol T$-systems\\[0.2cm] 
and $\boldsymbol Y$-systems 
for
the $\boldsymbol{A_2^{(1)}}$ models}}

\vspace{10mm}
{\Large Alexi Morin-Duchesne$^\dagger$, Paul A. Pearce$^\ddagger$, J\o rgen Rasmussen$^\ast$}
\\[.4cm]
{\em {}$^\dagger$Universit\'e catholique de Louvain, Institut de Recherche en Math\'ematique et Physique}\\
{\em Chemin du Cyclotron 2, 1348 Louvain-la-Neuve, Belgium}
\\[.4cm]
{\em {}$^\ddagger$School of Mathematics and Statistics, University of Melbourne}\\
{\em Parkville, Victoria 3010, Australia}
\\[.4cm]
{\em {}$^\ast$School of Mathematics and Physics, University of Queensland}\\
{\em St Lucia, Brisbane, Queensland 4072, Australia}
\\[.4cm]
{\tt alexi.morin-duchesne\,@\,uclouvain.be\qquad \tt papearce\,@\,unimelb.edu.au\qquad  \tt j.rasmussen\,@\,uq.edu.au}
\end{center}


\vspace{8mm}
\centerline{{\bf{Abstract}}}
\vskip.4cm
\noindent 
The family of $A^{(1)}_2$ models on the square lattice includes a dilute loop model, a $15$-vertex model and, at roots of unity, a family of RSOS models. 
The fused transfer matrices of the general 
loop and vertex models are shown to satisfy $s\ell(3)$-type fusion hierarchies. We use these to derive explicit $T$- and $Y$-systems of functional equations. At roots of unity, we further derive closure identities for the functional relations and show that the universal $Y$-system closes finitely. The  $A^{(1)}_2$  RSOS models are shown to satisfy the same functional and closure identities but with finite truncation.

\vspace{.5cm}
\noindent\textbf{Keywords:} Loop models, vertex models, RSOS models, fusion hierarchies, $T$-systems, $Y$-systems\\

\newpage
\tableofcontents

\newpage
\hyphenpenalty=30000

\setcounter{footnote}{0}

\section{Introduction}

Yang-Baxter equations (YBEs) and exactly solvable lattice models~\cite{BaxterBook} lie at the core of many developments in two-dimensional classical and one-dimensional quantum physics.
It is well known~\cite{Bazh85,Jimbo86} that Lie algebras may be used to classify solutions of the YBEs.
The simplest and most studied models are based on $s\ell(2)$ or $A_1^{(1)}$. 
The representations can be of various types, naturally associated with the 
vertex~\cite{Lieb67,Baxter72}, Restricted-Solid-on-Solid (RSOS)~\cite{ABF84,FB85,BR90,WNS92} and loop~\cite{Nienhuis82,BloteNienhuis89,Nienhuis90} models.
In fact, there are
mappings~\cite{NW93,YB95} relating the different types of representations. The loop models at roots of unity~\cite{PRZ2006} have seen a recent resurgence of interest due to their relation to logarithmic Conformal Field Theory (CFT)~\cite{LCFT}.

Moving beyond $s\ell(2)$, there is an extensive literature on models with $s\ell(3)$ symmetry. The following list of works on the various $A_2^{(1)}$ and $A_2^{(2)}$ models is therefore not meant to be exhaustive.
For example, $A_2^{(1)}$ vertex models have been considered in \cite{KR82,BabelonEtAl1982,dV89,AlcarazMartins1990,Resh91,dVGR94,ZinnJustin1998,SW00} and $A_2^{(2)}$ vertex models in \cite{IK81,AMN95}. 
Similarly, $A_2^{(1)}$ RSOS models have been studied in \cite{JMO88,DFZ90,ZP95} and $A_2^{(2)}$ RSOS models in \cite{WNS92,WPSN94,GW95,ZPG95}. Finally, the $A_2^{(1)}$ loop models were studied in \cite{NW93,R14,DEI16} and the $A_2^{(2)}$ loop models in \cite{Nienhuis90,DJS2010,PSAPR2012}. The vertex and loop models admit general
values for the crossing parameter $\lambda$ whereas, for RSOS models, $\lambda$ is restricted to rational multiples of $\pi$. 

In this paper, we focus on the critical $A_2^{(1)}$ models with $q=\eE^{\ir \lambda}$ a root of unity. In the case of vertex and loop models, we are thus considering a 
countable dense set of points on the continuous critical line. The critical $A_2^{(1)}$ loop model is defined in \cite{NW93}. The vertex weights of the associated $U_q(\widehat{sl}(3))$ 15-vertex model are given in \cite{Jimbo86} and the face weights of the associated RSOS models are given in \cite{JMO88}. 
The continuum scaling limit of the $A_2^{(1)}$ models is described~\cite{BazhHK2002} by a $W_3$ 
conformal field theory.

A standard method to obtain the spectra of lattice models and their associated CFTs is first to establish functional equations on the lattice in the form of fusion hierarchies, $T$-systems and $Y$-systems. Fusion hierarchies were first obtained in 1989 by Bazhanov and Reshetikhin~\cite{BR1989} in the context of the $s\ell(2)$ RSOS models. 
The $Y$-system and associated Thermodynamic Bethe Ansatz (TBA) equations for the ground state of $s\ell(2)$ scattering theories were  extensively studied in 1991 by Zamolodchikov~\cite{Zam1991a,Zam1991b}. A systematic derivation of the $T$- and $Y$-systems for all excitations in the context of $s\ell(2)$ RSOS models was obtained in 1992 by Kl\"umper and Pearce~\cite{KP92}. Generalizations of $T$- and $Y$-systems to higher-rank models, including models with $s\ell(3)$ symmetry,
are considered in \cite{KNS94,KunibaNS9310,ZP95,KNS2011}. The $T$- and $Y$-systems for the $s\ell(2)$ logarithmic minimal loop models were obtained in \cite{MDPR14}.

Once the lattice functional equations are obtained, a primary goal is to solve the 
equations in the continuum scaling limit. Following \cite{KP92}, this can be
achieved by converting the universal $Y$-system into 
non-linear integral equations in the form of TBA equations and obtaining the finite-size corrections using techniques involving dilogarithms.
This program has been carried to completion for the Ising model, tricritical Ising model, hard hexagons (or ${\Bbb Z}_3$ parafermions) and the Yang-Lee model~\cite{OPW1996,KP91,KP92,OPW97,BDP15}. More recently, in the context of loop models, the program has also been completed for critical dense polymers~\cite{PR07,PRVcyl2010,PRV1210,MDPR13} and critical bond percolation on the square 
lattice~\cite{MDKP17}.

\begin{figure}
\begin{center}
$
\psset{unit=2.2}
\begin{pspicture}[shift=-0.9](0,-0.5)(4,3.4)
\multiput(0,0)(1,0){5}{\psdot(0,0)}
\multiput(0.5,0.707)(1,0){4}{\psdot(0,0)}
\multiput(1,1.414)(1,0){3}{\psdot(0,0)}
\multiput(1.5,2.121)(1,0){2}{\psdot(0,0)}
\multiput(2,2.828)(1,0){1}{\psdot(0,0)}
\multiput(0,0)(1,0){4}{\psline[arrowscale=1.4,arrowinset=0.2]{->}(0,0)(0.275, 0.3883)\psline{-}(0,0)(0.5,0.707)\psline[arrowscale=1.4,arrowinset=0.2]{>-}(0.725, 0.3883)(1,0)\psline{-}(0.5,0.707)(1,0)\psline[arrowscale=1.4,arrowinset=0.2]{<-}(0.44,0)(1,0)\psline{-}(0,0)(1,0)}
\multiput(0.5,0.707)(1,0){3}{\psline[arrowscale=1.4,arrowinset=0.2]{->}(0,0)(0.275, 0.3883)\psline{-}(0,0)(0.5,0.707)\psline[arrowscale=1.4,arrowinset=0.2]{>-}(0.725, 0.3883)(1,0)\psline{-}(0.5,0.707)(1,0)\psline[arrowscale=1.4,arrowinset=0.2]{<-}(0.44,0)(1,0)\psline{-}(0,0)(1,0)}
\multiput(1,1.414)(1,0){2}{\psline[arrowscale=1.4,arrowinset=0.2]{->}(0,0)(0.275, 0.3883)\psline{-}(0,0)(0.5,0.707)\psline[arrowscale=1.4,arrowinset=0.2]{>-}(0.725, 0.3883)(1,0)\psline{-}(0.5,0.707)(1,0)\psline[arrowscale=1.4,arrowinset=0.2]{<-}(0.44,0)(1,0)\psline{-}(0,0)(1,0)}
\multiput(1.5,2.121)(1,0){1}{\psline[arrowscale=1.4,arrowinset=0.2]{->}(0,0)(0.275, 0.3883)\psline{-}(0,0)(0.5,0.707)\psline[arrowscale=1.4,arrowinset=0.2]{>-}(0.725, 0.3883)(1,0)\psline{-}(0.5,0.707)(1,0)\psline[arrowscale=1.4,arrowinset=0.2]{<-}(0.44,0)(1,0)\psline{-}(0,0)(1,0)}
\multiput(4,0)(-0.5,0.707){5}{\psline{-}(0,0)(0.4,0)}
\multiput(4,0)(-0.5,0.707){5}{\psline{-}(0,0)(0.2, 0.2824)}
\rput(0,-0.13){$\varnothing$}
\rput(0.95,-0.23){\young{1}{1}}
\rput(1.9,-0.23){\young{2}{2}}
\rput(2.85,-0.23){\young{3}{3}}
\rput(3.8,-0.23){\young{4}{4}}
\rput(0.33,0.707){\young{1}{0}}
\rput(0.74,1.414){\young{2}{0}}
\rput(1.18,2.121){\young{3}{0}}
\rput(1.53,2.828){\young{4}{0}}
\rput(1.3,0.7271){\young21}
\rput(2.2,0.7271){\young32}
\rput(3.1,0.7271){\young43}
\rput(1.7,1.434){\young31}
\rput(2.6,1.434){\young42}
\rput(2.1,2.141){\young41}
\multiput(4.25,0.15)(-0.5,0.707){5}{...}
\end{pspicture}
$
\caption{The infinite (dominant integral) $s\ell(3)$ weight lattice.
The heights at the nodes are labelled by 
Young diagrams.}
\label{fig:su3weights}
\end{center}
\end{figure}
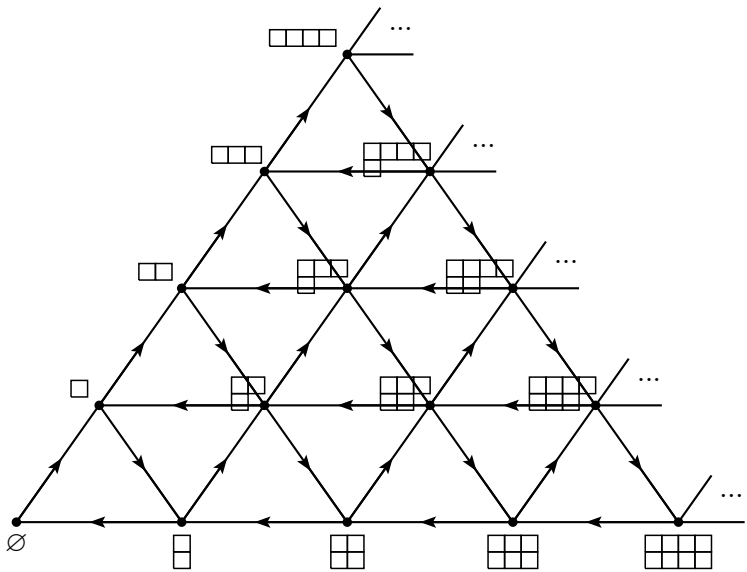

In this paper, we derive the fusion hierarchy and $T$- and $Y$-systems for the $A_2^{(1)}$ loop and vertex models on the cylinder. The construction is based on identities in the planar dilute Temperley-Lieb algebra similar to those given by Kuperberg \cite{Kuperberg96} for the $s\ell(3)$ spider. 
The transfer tangles $\Tb^{m,n}(u)$ of the $A_2^{(1)}$ models 
are labelled by pairs $(m,n)$ of integers associated to nodes of the $su(3)$ weight lattice, see \cref{fig:su3weights}.
For the general $A_2^{(1)}$ loop and vertex models, the resulting $Y$-systems are infinite. 
For generic values of $q$ the $Y$-systems do not close but, for roots of unity, they close finitely and we find the closure relations explicitly.
Notably, the ensuing form of the closed $Y$-system is considerably more complicated than the $D$-type Dynkin diagram structure familiar from the $s\ell(2)$ models \cite{KSS98,MDKP17},
thus revealing a rich underlying $s\ell(3)$ structure. The precise form of these closed equations strongly resembles those of the related so-called complex $su(3)$ Toda theory~\cite{SW00}.
In strong contrast, for the $A_2^{(1)}$ RSOS models, the known $Y$-systems~\cite{ZP95} truncate at a finite level.
The actual {\em solution} of the universal $A_2^{(1)}$
$Y$-systems in the various representations is beyond the scope of the present paper.

The layout of the paper is as follows. In \cref{sec:A21algebras}, we discuss the $A_2^{(1)}$ models in the language of diagrammatic algebras and present
the representations relevant in the loop and vertex models. 
The diagrammatic calculus is developed further in \cref{sec:calculus} to include local relations, 
Wenzl-Jones projectors, fused face operators and braid limits. The fused transfer tangles, fusion hierarchies, $T$-systems and $Y$-systems are presented in \cref{sec:Ftm.fr}, along with the closure relations. 
The details of the proofs are relegated to \cref{app:cov,app:proofs}. 
In \cref{sec:conclusion}, we conclude with some 
remarks and a discussion of related open problems.
In \cref{app:RSOS}, we review the definition of the $A_2^{(1)}$ RSOS model and argue that the functional relations obtained in \cref{sec:Ftm.fr} also hold for this model.

\section{$\boldsymbol{A_2^{(1)}}$ models and diagrammatic algebras}\label{sec:A21algebras}

\subsection[The $A_2^{(1)}$ loop model]{The $\boldsymbol{A_2^{(1)}}$ loop model}

The $A_2^{(1)}$ loop model is a face model on the square lattice, where each face takes on one of seven possible local configurations. The corresponding face weights are encoded in the elementary face operator, defined as a linear combination of seven possible tiles by
\be
\begin{pspicture}[shift=-.40](0,0)(1,1)
\facegrid{(0,0)}{(1,1)}
\psarc[linewidth=0.025]{-}(0,0){0.16}{0}{90}
\rput(.5,.5){$u$}
\end{pspicture}
\ = s_1(-u)
\bigg(\ 
\begin{pspicture}[shift=-.40](0,0)(1,1)
\facegrid{(0,0)}{(1,1)}
\rput[bl](0,0){\loopa}
\end{pspicture}
\ + \ 
\begin{pspicture}[shift=-.40](0,0)(1,1)
\facegrid{(0,0)}{(1,1)}
\rput[bl](0,0){\looph}
\end{pspicture}
\ \bigg)
+ t \ \
\begin{pspicture}[shift=-.40](0,0)(1,1)
\facegrid{(0,0)}{(1,1)}
\rput[bl](0,0){\loopb}
\end{pspicture}
\ + \frac 1t\ \
\begin{pspicture}[shift=-.40](0,0)(1,1)
\facegrid{(0,0)}{(1,1)}
\rput[bl](0,0){\loopc}
\end{pspicture}
\ + s_0(u) \bigg(\ 
\begin{pspicture}[shift=-.40](0,0)(1,1)
\facegrid{(0,0)}{(1,1)}
\rput[bl](0,0){\loopf}
\end{pspicture} 
\ + \ 
\begin{pspicture}[shift=-.40](0,0)(1,1)
\facegrid{(0,0)}{(1,1)}
\rput[bl](0,0){\loopg}
\end{pspicture}
\ + \ 
\begin{pspicture}[shift=-.40](0,0)(1,1)
\facegrid{(0,0)}{(1,1)}
\rput[bl](0,0){\loopi}
\end{pspicture}
\ \bigg)
\label{eq:faceop}
\ee
where
\be
s_k(u) = \frac{\sin(k\lambda+u)}{\sin \lambda}.
\ee
Here, $u$ is the spectral parameter, the crossing parameter $\lambda$ parameterises the fugacity $\beta$ of the contractible loops as
\be
\beta = 2 \cos \lambda = q+q^{-1}, \qquad q = \eE^{\ir \lambda},
\label{betaq}
\ee
and $t$ is a gauge parameter. Since the face operator fails to be crossing symmetric,
\be
\begin{pspicture}[shift=-.40](0,0)(1,1)
\facegrid{(0,0)}{(1,1)}
\psarc[linewidth=0.025]{-}(0,0){0.16}{0}{90}
\rput(.5,.5){$u$}
\end{pspicture}\ \neq \
\begin{pspicture}[shift=-.40](0,0)(1,1)
\facegrid{(0,0)}{(1,1)}
\psarc[linewidth=0.025]{-}(1,0){0.16}{90}{180}
\rput(.5,.5){\small$\lambda-u$}
\end{pspicture}\ \ ,
\ee
it is
customary to 
work with {\em two} elementary face operators. These operators are 
assigned the 
Dynkin
labels $(1,0)$ and $(0,1)$ of the two fundamental $s\ell(3)$ representations:
\be\label{eq:two.face.ops}
\begin{pspicture}[shift=-.40](0,0)(1,1)
\facegrid{(0,0)}{(1,1)}
\psarc[linewidth=0.025]{-}(0,0){0.16}{0}{90}
\rput(.5,.7){\tiny $(1,0)$}
\rput(.5,.40){$u$}
\end{pspicture} 
\ =\
\begin{pspicture}[shift=-.40](0,0)(1,1)
\facegrid{(0,0)}{(1,1)}
\psarc[linewidth=0.025]{-}(0,0){0.16}{0}{90}
\rput(.5,.5){$u$}
\end{pspicture}\ \ , \qquad \qquad
\begin{pspicture}[shift=-.40](0,0)(1,1)
\facegrid{(0,0)}{(1,1)}
\psarc[linewidth=0.025]{-}(0,0){0.16}{0}{90}
\rput(.5,.7){\tiny $(0,1)$}
\rput(.5,.40){$u$}
\end{pspicture} 
\ =\
\begin{pspicture}[shift=-.40](0,0)(1,1)
\facegrid{(0,0)}{(1,1)}
\psarc[linewidth=0.025]{-}(1,0){0.16}{90}{180}
\rput(.5,.5){\small$\lambda-u$}
\end{pspicture}\ \ .
\ee

\begin{figure}
\begin{center}
\psset{unit=0.5}
\begin{pspicture}[shift=-.40](0,0)(12,12)
\facegrid{(0,0)}{(12,12)}
\rput(0,11){\loopd}\rput(1,11){\loopf}\rput(2,11){\loope}\rput(3,11){\loopd}\rput(4,11){\loope}\rput(5,11){\looph}\rput(6,11){\loopi}\rput(7,11){\loopd}\rput(8,11){\loopa}\rput(9,11){\loope}\rput(10,11){\looph}\rput(11,11){\loopi}
\rput(0,10){\looph}\rput(1,10){\loopi}\rput(2,10){\loopg}\rput(3,10){\loopb}\rput(4,10){\loopc}\rput(5,10){\loopi}\rput(6,10){\loopi}\rput(7,10){\loopi}\rput(8,10){\loopg}\rput(9,10){\loopg}\rput(10,10){\looph}\rput(11,10){\loopi}
\rput(0,9){\loopi}\rput(1,9){\loopi}\rput(2,9){\loopg}\rput(3,9){\loopg}\rput(4,9){\loopi}\rput(5,9){\looph}\rput(6,9){\looph}\rput(7,9){\loopi}\rput(8,9){\loopg}\rput(9,9){\loopd}\rput(10,9){\loope}\rput(11,9){\looph}
\rput(0,8){\loopb}\rput(1,8){\loopf}\rput(2,8){\loopa}\rput(3,8){\loopc}\rput(4,8){\loopi}\rput(5,8){\looph}\rput(6,8){\loopb}\rput(7,8){\loopf}\rput(8,8){\loopc}\rput(9,8){\loopi}\rput(10,8){\loopg}\rput(11,8){\loopi}
\rput(0,7){\loopg}\rput(1,7){\loopi}\rput(2,7){\loopd}\rput(3,7){\loopf}\rput(4,7){\loope}\rput(5,7){\loopi}\rput(6,7){\loopg}\rput(7,7){\loopi}\rput(8,7){\loopi}\rput(9,7){\looph}\rput(10,7){\loopd}\rput(11,7){\loope}
\rput(0,6){\loopc}\rput(1,6){\loopi}\rput(2,6){\looph}\rput(3,6){\loopb}\rput(4,6){\loopc}\rput(5,6){\loopi}\rput(6,6){\loopg}\rput(7,6){\loopi}\rput(8,6){\loopb}\rput(9,6){\loopf}\rput(10,6){\loopf}\rput(11,6){\loopa}
\rput(0,5){\loope}\rput(1,5){\loopi}\rput(2,5){\looph}\rput(3,5){\loopd}\rput(4,5){\loope}\rput(5,5){\looph}\rput(6,5){\loopd}\rput(7,5){\loope}\rput(8,5){\loopd}\rput(9,5){\loope}\rput(10,5){\looph}\rput(11,5){\loopd}
\rput(0,4){\loopg}\rput(1,4){\looph}\rput(2,4){\loopb}\rput(3,4){\loopf}\rput(4,4){\loopc}\rput(5,4){\loopb}\rput(6,4){\loopf}\rput(7,4){\loopa}\rput(8,4){\loopf}\rput(9,4){\loopc}\rput(10,4){\looph}\rput(11,4){\loopi}
\rput(0,3){\loopd}\rput(1,3){\loope}\rput(2,3){\loopg}\rput(3,3){\looph}\rput(4,3){\looph}\rput(5,3){\loopd}\rput(6,3){\loope}\rput(7,3){\loopg}\rput(8,3){\loopi}\rput(9,3){\looph}\rput(10,3){\loopi}\rput(11,3){\loopi}
\rput(0,2){\loopf}\rput(1,2){\loopa}\rput(2,2){\loopc}\rput(3,2){\loopi}\rput(4,2){\loopb}\rput(5,2){\loopf}\rput(6,2){\loopc}\rput(7,2){\loopg}\rput(8,2){\looph}\rput(9,2){\looph}\rput(10,2){\loopb}\rput(11,2){\loopf}
\rput(0,1){\loopi}\rput(1,1){\loopd}\rput(2,1){\loope}\rput(3,1){\looph}\rput(4,1){\loopg}\rput(5,1){\loopi}\rput(6,1){\loopi}\rput(7,1){\loopd}\rput(8,1){\loope}\rput(9,1){\looph}\rput(10,1){\loopg}\rput(11,1){\loopi}
\rput(0,0){\loopb}\rput(1,0){\loopf}\rput(2,0){\loopc}\rput(3,0){\loopb}\rput(4,0){\loopc}\rput(5,0){\loopi}\rput(6,0){\looph}\rput(7,0){\loopb}\rput(8,0){\loopa}\rput(9,0){\loopf}\rput(10,0){\loopc}\rput(11,0){\looph}
\end{pspicture}
\caption{A configuration of the loop model on the $12 \times 12$ torus. 
Enumerating the rows from the bottom, the tiles making up the odd rows come from the face operator $(1,0)$, while the ones making up the even rows come from the face operator $(0,1)$.}
\label{fig:loop.config}
\end{center}
\end{figure}
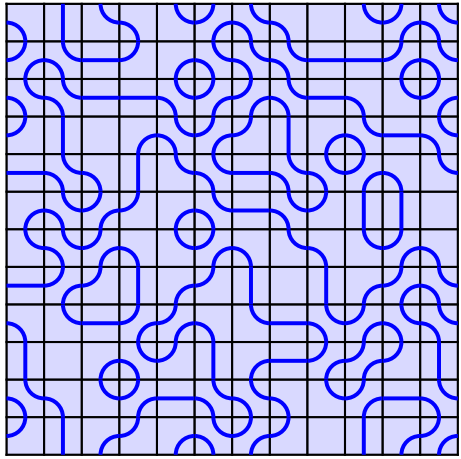

We study the model on the $M\times N$ torus. A configuration of the loop model is a choice of a face configuration
for each of the $MN$ faces. An example is given in \cref{fig:loop.config}. Nodes that are not visited by a loop segment are said to be 
vacant; they are occasionally indicated by small black discs in the diagrams below.
In the statistical model, the vacancies have weight $1$. 
In addition to contractible loops, non-contractible loops may appear and are assigned the loop fugacity $\alpha$.
The weight $W_\sigma$ of a configuration $\sigma$ and the partition function $Z$ are then given by
\be
W_\sigma = \alpha^{n_\alpha}\beta^{n_\beta} \prod_{f} w_f, \qquad Z = \sum_{\sigma} W_\sigma,
\ee
where 
$n_\alpha$ and $n_\beta$ are the number of non-contractible and contractible loops, respectively,
$w_f$ are the local face weights appearing in
\eqref{eq:faceop}, and $\prod_f$ is a product over the $MN$ faces.

Our calculations in later sections distinguish between values of $q$ that are {\it roots of unity} and those that are not. The roots of unity values are parameterised as $q=\eE^{\ir \lambda}$ as in \eqref{betaq}, with
\be
\lambda = \lambda_{p,p'} = \frac{\pi(p'-p)}{p'}, \qquad p,p' \in \mathbb N, \qquad 1\le p < p', \qquad \textrm{gcd}(p,p')=1.
\ee

\subsection{Diagrammatic algebras}

The diagrammatic algebra that we use to describe the $\Atwoone$ models is a subalgebra, $\mathsf {pA}_N(\alpha,\beta)$, 
of the periodic dilute Temperley-Lieb algebra $\mathsf{pdTL}_N(\alpha,\beta)$,
as outlined in the following.

First,
$\mathsf{pdTL}_N(\alpha,\beta)$ is an algebra of connectivity diagrams drawn inside a rectangle with periodic boundary conditions in the horizontal direction. The rectangle has $N$ 
nodes on the top 
edge and $N$ 
nodes on the bottom edge. 
Every node is either connected to another node by a loop segment or left vacant, in such a way that the loop segments are 
non-intersecting.
Because of the periodicity of the rectangle, the loop segments can travel via the back of the cylinder. Here are two examples of connectivities in $\mathsf{pdTL}_6(\alpha,\beta)$:
\be\label{eq:conn.ex}
\begin{pspicture}[shift=-0.3](-0.0,0)(2.4,0.8)
\pspolygon[fillstyle=solid,fillcolor=lightlightblue,linecolor=black,linewidth=0pt](0,0)(0,0.8)(2.4,0.8)(2.4,0)(0,0)
\psarc[linecolor=blue,linewidth=\elegant]{-}(1.6,0){0.2}{0}{180}
\psbezier[linecolor=blue,linewidth=\elegant]{-}(0.6,0.8)(0.6,0.4)(1.4,0.4)(1.4,0.8)
\psarc[linecolor=blue,linewidth=\elegant]{-}(0,0.8){0.2}{-90}{0}
\psarc[linecolor=blue,linewidth=\elegant]{-}(2.4,0.8){0.2}{180}{-90}
\pscircle[fillstyle=solid,fillcolor=black](1.0,0.8){0.04}
\pscircle[fillstyle=solid,fillcolor=black](0.2,0){0.04}
\pscircle[fillstyle=solid,fillcolor=black](0.6,0){0.04}
\pscircle[fillstyle=solid,fillcolor=black](2.2,0){0.04}
\psbezier[linecolor=blue,linewidth=\elegant]{-}(1.8,0.8)(1.8,0.1)(3.4,0.5)(3.4,0)
\rput(-2.4,0){\psbezier[linecolor=blue,linewidth=\elegant]{-}(1.8,0.8)(1.8,0.1)(3.4,0.5)(3.4,0)}
\psframe[fillstyle=solid,linecolor=white,linewidth=0pt](-1.2,0)(0,0.8)
\psframe[fillstyle=solid,linecolor=white,linewidth=0pt](2.4,0)(3.6,0.8)
\end{pspicture}\ \ ,\qquad\qquad
\begin{pspicture}[shift=-0.3](-0.0,0)(2.4,0.8)
\pspolygon[fillstyle=solid,fillcolor=lightlightblue,linecolor=black,linewidth=0pt](0,0)(0,0.8)(2.4,0.8)(2.4,0)(0,0)
\pscircle[fillstyle=solid,fillcolor=black](0.2,0.8){0.04}
\pscircle[fillstyle=solid,fillcolor=black](1.0,0){0.04}
\psarc[linecolor=blue,linewidth=\elegant]{-}(2.0,0.8){0.2}{180}{0}
\psbezier[linecolor=blue,linewidth=\elegant]{-}(0.6,0.8)(0.6,0.4)(-0.2,0.4)(-0.2,0)\rput(2.4,0){\psbezier[linecolor=blue,linewidth=\elegant]{-}(0.6,0.8)(0.6,0.4)(-0.2,0.4)(-0.2,0)}
\psbezier[linecolor=blue,linewidth=\elegant]{-}(1.0,0.8)(1.0,0.4)(0.2,0.4)(0.2,0)
\psbezier[linecolor=blue,linewidth=\elegant]{-}(0.6,0)(0.6,0.4)(1.4,0.4)(1.4,0.0)
\psbezier[linecolor=blue,linewidth=\elegant]{-}(1.4,0.8)(1.4,0.4)(1.8,0.4)(1.8,0)
\psframe[fillstyle=solid,linecolor=white,linewidth=0pt](-0.4,0)(0,0.8)
\psframe[fillstyle=solid,linecolor=white,linewidth=0pt](2.4,0)(3.6,0.8)
\end{pspicture}\ \ .
\ee
The product $c_1c_2$ of two connectivity diagrams in $\mathsf{pdTL}_N(\alpha,\beta)$ is defined by juxtaposition: 
One places $c_2$ atop $c_1$, reads the connectivity diagram between the top and bottom edges of the ensuing rectangle,
and replaces each contractible or non-contractible loop by a scalar factor of the corresponding fugacity, $\beta$ and $\alpha$, respectively.
Moreover, if a loop segment connects to a vacant site, $c_1c_2$ is set to zero. Here are three examples to illustrate:
\begin{subequations}
\begin{alignat}{2}
&\begin{pspicture}[shift=-0.7](-0.0,0)(2.4,1.6)
\pspolygon[fillstyle=solid,fillcolor=lightlightblue,linecolor=black,linewidth=0pt](0,0)(0,0.8)(2.4,0.8)(2.4,0)(0,0)
\psarc[linecolor=blue,linewidth=\elegant]{-}(1.6,0){0.2}{0}{180}
\psbezier[linecolor=blue,linewidth=\elegant]{-}(0.6,0.8)(0.6,0.4)(1.4,0.4)(1.4,0.8)
\psarc[linecolor=blue,linewidth=\elegant]{-}(0,0.8){0.2}{-90}{0}
\psarc[linecolor=blue,linewidth=\elegant]{-}(2.4,0.8){0.2}{180}{-90}
\pscircle[fillstyle=solid,fillcolor=black](1.0,0.8){0.04}
\pscircle[fillstyle=solid,fillcolor=black](0.2,0){0.04}
\pscircle[fillstyle=solid,fillcolor=black](0.6,0){0.04}
\pscircle[fillstyle=solid,fillcolor=black](2.2,0){0.04}
\psbezier[linecolor=blue,linewidth=\elegant]{-}(1.8,0.8)(1.8,0.1)(3.4,0.5)(3.4,0)
\rput(-2.4,0){\psbezier[linecolor=blue,linewidth=\elegant]{-}(1.8,0.8)(1.8,0.1)(3.4,0.5)(3.4,0)}
\rput(0,0.8){
\pspolygon[fillstyle=solid,fillcolor=lightlightblue,linecolor=black,linewidth=0pt](0,0)(0,0.8)(2.4,0.8)(2.4,0)(0,0)
\pscircle[fillstyle=solid,fillcolor=black](0.2,0.8){0.04}
\pscircle[fillstyle=solid,fillcolor=black](1.0,0){0.04}
\psarc[linecolor=blue,linewidth=\elegant]{-}(2.0,0.8){0.2}{180}{0}
\psbezier[linecolor=blue,linewidth=\elegant]{-}(0.6,0.8)(0.6,0.4)(-0.2,0.4)(-0.2,0)\rput(2.4,0){\psbezier[linecolor=blue,linewidth=\elegant]{-}(0.6,0.8)(0.6,0.4)(-0.2,0.4)(-0.2,0)}
\psbezier[linecolor=blue,linewidth=\elegant]{-}(1.0,0.8)(1.0,0.4)(0.2,0.4)(0.2,0)
\psbezier[linecolor=blue,linewidth=\elegant]{-}(0.6,0)(0.6,0.4)(1.4,0.4)(1.4,0.0)
\psbezier[linecolor=blue,linewidth=\elegant]{-}(1.4,0.8)(1.4,0.4)(1.8,0.4)(1.8,0)
}
\psframe[fillstyle=solid,linecolor=white,linewidth=0pt](-1.2,0)(0,1.6)
\psframe[fillstyle=solid,linecolor=white,linewidth=0pt](2.4,0)(3.6,1.6)
\end{pspicture} \ \ = \beta \ \ 
\begin{pspicture}[shift=-0.3](-0.0,0)(2.4,0.8)
\pspolygon[fillstyle=solid,fillcolor=lightlightblue,linecolor=black,linewidth=0pt](0,0)(0,0.8)(2.4,0.8)(2.4,0)(0,0)
\pscircle[fillstyle=solid,fillcolor=black](0.2,0.8){0.04}
\pscircle[fillstyle=solid,fillcolor=black](0.2,0){0.04}
\pscircle[fillstyle=solid,fillcolor=black](0.6,0){0.04}
\pscircle[fillstyle=solid,fillcolor=black](2.2,0){0.04}
\psarc[linecolor=blue,linewidth=\elegant]{-}(0.8,0.8){0.2}{180}{0}
\psarc[linecolor=blue,linewidth=\elegant]{-}(2.0,0.8){0.2}{180}{0}
\psarc[linecolor=blue,linewidth=\elegant]{-}(1.6,0){0.2}{0}{180}
\psbezier[linecolor=blue,linewidth=\elegant](1.4,0.8)(1.4,0.48)(1.6,0.42)(2.4,0.32)
\psbezier[linecolor=blue,linewidth=\elegant](1.0,0)(1.0,0.22)(0.7,0.24)(0,0.32)
\end{pspicture}\ \ ,
\\[0.3cm]
&\begin{pspicture}[shift=-0.7](-0.0,0)(2.4,1.6)
\pspolygon[fillstyle=solid,fillcolor=lightlightblue,linecolor=black,linewidth=0pt](0,0)(0,0.8)(2.4,0.8)(2.4,0)(0,0)
\psarc[linecolor=blue,linewidth=\elegant]{-}(2.0,0.8){0.2}{180}{0}
\psarc[linecolor=blue,linewidth=\elegant]{-}(0.4,0.8){0.2}{180}{0}
\psarc[linecolor=blue,linewidth=\elegant]{-}(1.6,0){0.2}{0}{180}
\psbezier[linecolor=blue,linewidth=\elegant]{-}(0.2,0)(0.2,0.4)(1.0,0.4)(1.0,0.8)
\psbezier[linecolor=blue,linewidth=\elegant]{-}(0.6,0)(0.6,0.5)(2.2,0.5)(2.2,0)
\pscircle[fillstyle=solid,fillcolor=black](1.0,0){0.04}
\pscircle[fillstyle=solid,fillcolor=black](1.4,0.8){0.04}
\rput(0,0.8){
\pspolygon[fillstyle=solid,fillcolor=lightlightblue,linecolor=black,linewidth=0pt](0,0)(0,0.8)(2.4,0.8)(2.4,0)(0,0)
\psarc[linecolor=blue,linewidth=\elegant]{-}(1.2,0.8){0.2}{180}{0}
\pscircle[fillstyle=solid,fillcolor=black](1.4,0){0.04}
\pscircle[fillstyle=solid,fillcolor=black](1.8,0){0.04}
\pscircle[fillstyle=solid,fillcolor=black](0.2,0.8){0.04}
\pscircle[fillstyle=solid,fillcolor=black](0.6,0.8){0.04}
\psbezier[linecolor=blue,linewidth=\elegant]{-}(0.2,0.0)(0.2,0.4)(-0.2,0.4)(-0.2,0.8)\rput(2.4,0){\psbezier[linecolor=blue,linewidth=\elegant]{-}(0.2,0.0)(0.2,0.4)(-0.2,0.4)(-0.2,0.8)}
\psbezier[linecolor=blue,linewidth=\elegant]{-}(1.0,0)(1.0,0.4)(2.2,0.4)(2.2,0)
\psbezier[linecolor=blue,linewidth=\elegant]{-}(0.6,0)(0.6,0.4)(1.8,0.4)(1.8,0.8)
}
\psframe[fillstyle=solid,linecolor=white,linewidth=0pt](-1.2,0)(0,1.6)
\psframe[fillstyle=solid,linecolor=white,linewidth=0pt](2.4,0)(3.6,1.6)
\end{pspicture} \ \ = 0\  ,
\\[0.3cm]
&\begin{pspicture}[shift=-0.7](-0.0,0)(2.4,1.6)
\pspolygon[fillstyle=solid,fillcolor=lightlightblue,linecolor=black,linewidth=0pt](0,0)(0,0.8)(2.4,0.8)(2.4,0)(0,0)
\pscircle[fillstyle=solid,fillcolor=black](1.4,0){0.04}
\pscircle[fillstyle=solid,fillcolor=black](2.2,0){0.04}
\psarc[linecolor=blue,linewidth=\elegant]{-}(0.8,0){0.2}{0}{180}
\psarc[linecolor=blue,linewidth=\elegant]{-}(1.6,0.8){0.2}{180}{0}
\psbezier[linecolor=blue,linewidth=\elegant](0.2,0)(0.2,0.5)(1.8,0.5)(1.8,0)
\psbezier[linecolor=blue,linewidth=\elegant](0.6,0.8)(0.6,0.4)(-0.2,0.4)(-0.2,0.8)\rput(2.4,0){\psbezier[linecolor=blue,linewidth=\elegant](0.6,0.8)(0.6,0.4)(-0.2,0.4)(-0.2,0.8)}
\psframe[fillstyle=solid,linecolor=white,linewidth=0pt](-0.23,0)(0,1.6)
\psframe[fillstyle=solid,linecolor=white,linewidth=0pt](2.4,0)(4.0,1.6)
\rput(0,0.8){
\pspolygon[fillstyle=solid,fillcolor=lightlightblue,linecolor=black,linewidth=0pt](0,0)(0,0.8)(2.4,0.8)(2.4,0)(0,0)
\pscircle[fillstyle=solid,fillcolor=black](0.2,0){0.04}
\pscircle[fillstyle=solid,fillcolor=black](1.0,0){0.04}
\pscircle[fillstyle=solid,fillcolor=black](1.4,0.8){0.04}
\pscircle[fillstyle=solid,fillcolor=black](1.8,0.8){0.04}
\psarc[linecolor=blue,linewidth=\elegant]{-}(2.0,0){0.2}{0}{180}
\psarc[linecolor=blue,linewidth=\elegant]{-}(0.4,0.8){0.2}{180}{0}
\psbezier[linecolor=blue,linewidth=\elegant](1.0,0.8)(1.0,0.35)(-0.2,0.35)(-0.2,0.8)\rput(2.4,0){\psbezier[linecolor=blue,linewidth=\elegant](1.0,0.8)(1.0,0.35)(-0.2,0.35)(-0.2,0.8)}
\psbezier[linecolor=blue,linewidth=\elegant](0.6,0)(0.6,0.4)(1.4,0.4)(1.4,0)
\psframe[fillstyle=solid,linecolor=white,linewidth=0pt](-0.23,0)(0,1.6)
\psframe[fillstyle=solid,linecolor=white,linewidth=0pt](2.4,0)(4.0,1.6)
}
\psframe[fillstyle=solid,linecolor=white,linewidth=0pt](-1.2,0)(0,1.6)
\psframe[fillstyle=solid,linecolor=white,linewidth=0pt](2.4,0)(3.6,1.6)
\end{pspicture} \ \ = \alpha \ \ 
\begin{pspicture}[shift=-0.3](-0.0,0)(2.4,0.8)
\pspolygon[fillstyle=solid,fillcolor=lightlightblue,linecolor=black,linewidth=0pt](0,0)(0,0.8)(2.4,0.8)(2.4,0)(0,0)
\pscircle[fillstyle=solid,fillcolor=black](1.4,0.8){0.04}
\pscircle[fillstyle=solid,fillcolor=black](1.8,0.8){0.04}
\pscircle[fillstyle=solid,fillcolor=black](1.4,0){0.04}
\pscircle[fillstyle=solid,fillcolor=black](2.2,0){0.04}
\psarc[linecolor=blue,linewidth=\elegant]{-}(0.4,0.8){0.2}{180}{0}
\psarc[linecolor=blue,linewidth=\elegant]{-}(0.8,0){0.2}{0}{180}
\psbezier[linecolor=blue,linewidth=\elegant](0.2,0)(0.2,0.5)(1.8,0.5)(1.8,0)
\psbezier[linecolor=blue,linewidth=\elegant](1.0,0.8)(1.0,0.35)(-0.2,0.35)(-0.2,0.8)\rput(2.4,0){\psbezier[linecolor=blue,linewidth=\elegant](1.0,0.8)(1.0,0.35)(-0.2,0.35)(-0.2,0.8)}
\psframe[fillstyle=solid,linecolor=white,linewidth=0pt](-0.23,0)(0,1.6)
\psframe[fillstyle=solid,linecolor=white,linewidth=0pt](2.4,0)(4.0,1.6)
\end{pspicture}\ \ .
\end{alignat}
\end{subequations}

As can be seen in \cref{fig:loop.config}, the number of vacancies on horizontal cuts of the torus is conserved in the $\Atwoone$ loop model. Likewise, the face operator \eqref{eq:faceop}, seen as a connectivity diagram acting from NE to SW, preserves the number of vacancies. We denote by $\mathsf{pdTL}_{N,v}(\alpha,\beta)$ the subalgebra of $\mathsf{pdTL}_{N}(\alpha,\beta)$ obtained by restricting to connectivity diagrams where both the top and bottom edges of the rectangle have exactly $v$ vacancies. 
The algebra $\mathsf{pdTL}_{N,0}(\alpha,\beta)$ then corresponds to the usual periodic Temperley-Lieb algebra. The algebra $\mathsf {pA}_N(\alpha,\beta)$
is defined as the direct sum of these subalgebras:
\be
\mathsf {pA}_N(\alpha,\beta) = \bigoplus_{v=0}^N \mathsf{pdTL}_{N,v}(\alpha,\beta).
\ee
As an example, in \eqref{eq:conn.ex}, only the second connectivity is an element of $\mathsf {pA}_6(\alpha,\beta)$. 

While this paper focuses on lattice models with periodic boundary conditions, 
diagrammatic algebras similar to the ones 
in this section can also be defined for the $A_2^{(1)}$ lattice models defined on the geometry of the strip. In this case, one considers the (ordinary) dilute Temperley-Lieb algebra $\mathsf{dTL}_N(\beta)$. This algebra is generated by connectivity diagrams on the rectangle where loop segments connect nodes of the rectangle pairwise, but without the possibility of travelling via the back of the cylinder. The subalgebras $\mathsf{dTL}_{N,v}(\beta)$ are obtained by restricting to connectivities with exactly $v$ preserved vacancies. The algebra relevant for the $A_2^{(1)}$ lattice models on the strip, $\mathsf {A}_N(\beta)$, is then defined as
\be
\mathsf {A}_N(\beta) = \bigoplus_{v=0}^N \mathsf{dTL}_{N,v}(\beta).
\ee
It is a subalgebra of $\mathsf {pA}_N(\alpha,\beta)$. The algebra $\mathsf {A}_N(\beta)$ is useful to us in the context of periodic boundary conditions. Indeed, in proving that a given set of matrices $\rho(a)$ for $a \in \mathsf {pA}_N(\alpha,\beta)$ realise a representation of $\mathsf {pA}_N(\alpha,\beta)$, a first step is to check that $\rho$ is 
a representation of $\mathsf {A}_N(\beta)$.

\subsection{Standard modules}

\noindent The standard modules $\stan_{N,d,v}$ are defined in terms of planar link states. 
Such a link state is a diagram drawn above a horizontal line segment with $N$ 
nodes, where every node is linked to another node by a loop segment, left vacant, or attached to a defect, in such a way that the
loop segments are non-intersecting. Here, a defect is a vertical loop segment that ties a node to the point at infinity 
above and cannot be overarched. 
A given link state $w$ is characterised by the numbers $d$, $v$ and $a$ of defects, vacancies and arcs it contains.
By construction, these numbers are related by
\be
N = d+v+2a.
\ee
The module $\stan_{N,d,v}$ is defined on the vector space generated by link states with $N$ nodes, $d$ defects and $v$ vacancies. 
Link-state bases for the standard modules for $N=3$ are given by
\be
\begin{array}{ll}
\stan_{3,3,0}:\qquad 
\begin{pspicture}[shift=0](0.0,0)(1.2,0.5)
\psline[linewidth=\mince](0,0)(1.2,0)
\psline[linecolor=blue,linewidth=\elegant]{-}(0.2,0)(0.2,0.4)
\psline[linecolor=blue,linewidth=\elegant]{-}(0.6,0)(0.6,0.4)
\psline[linecolor=blue,linewidth=\elegant]{-}(1.0,0)(1.0,0.4)
\end{pspicture}
&%
\stan_{3,1,0}:\qquad 
\begin{pspicture}[shift=0](0.0,0)(1.2,0.5)
\psline[linewidth=\mince](0,0)(1.2,0)
\psarc[linecolor=blue,linewidth=\elegant]{-}(0.8,0){0.2}{0}{180}
\psline[linecolor=blue,linewidth=\elegant]{-}(0.2,0)(0.2,0.4)
\end{pspicture}
\quad
\begin{pspicture}[shift=0](0.0,0)(1.2,0.5)
\psline[linewidth=\mince](0,0)(1.2,0)
\psarc[linecolor=blue,linewidth=\elegant]{-}(0.4,0){0.2}{0}{180}
\psline[linecolor=blue,linewidth=\elegant]{-}(1.0,0)(1.0,0.4)
\end{pspicture}
\quad
\begin{pspicture}[shift=0](0.0,0)(1.2,0.5)
\psline[linewidth=\mince](0,0)(1.2,0)
\psarc[linecolor=blue,linewidth=\elegant]{-}(0,0){0.2}{0}{90}
\psarc[linecolor=blue,linewidth=\elegant]{-}(1.2,0){0.2}{90}{180}
\psline[linecolor=blue,linewidth=\elegant]{-}(0.6,0)(0.6,0.4)
\end{pspicture}
\\[0.1cm]%
\stan_{3,2,1}:\qquad 
\begin{pspicture}[shift=0](0.0,0)(1.2,0.5)
\psline[linewidth=\mince](0,0)(1.2,0)
\psline[linecolor=blue,linewidth=\elegant]{-}(0.2,0)(0.2,0.4)
\psline[linecolor=blue,linewidth=\elegant]{-}(0.6,0)(0.6,0.4)
\pscircle[fillstyle=solid,fillcolor=black](1.0,0){0.04}
\end{pspicture}
\quad
\begin{pspicture}[shift=0](0.0,0)(1.2,0.5)
\psline[linewidth=\mince](0,0)(1.2,0)
\psline[linecolor=blue,linewidth=\elegant]{-}(0.2,0)(0.2,0.4)
\psline[linecolor=blue,linewidth=\elegant]{-}(1.0,0)(1.0,0.4)
\pscircle[fillstyle=solid,fillcolor=black](0.6,0){0.04}
\end{pspicture}
\quad
\begin{pspicture}[shift=0](0.0,0)(1.2,0.5)
\psline[linewidth=\mince](0,0)(1.2,0)
\psline[linecolor=blue,linewidth=\elegant]{-}(0.6,0)(0.6,0.4)
\psline[linecolor=blue,linewidth=\elegant]{-}(1.0,0)(1.0,0.4)
\pscircle[fillstyle=solid,fillcolor=black](0.2,0){0.04}
\end{pspicture}
&%
\stan_{3,0,3}:\qquad 
\begin{pspicture}[shift=0](0.0,0)(1.2,0.5)
\psline[linewidth=\mince](0,0)(1.2,0)
\pscircle[fillstyle=solid,fillcolor=black](0.2,0){0.04}
\pscircle[fillstyle=solid,fillcolor=black](0.6,0){0.04}
\pscircle[fillstyle=solid,fillcolor=black](1.0,0){0.04}
\end{pspicture}
\\[0.1cm]%
\stan_{3,1,2}:\qquad 
\begin{pspicture}[shift=0](0.0,0)(1.2,0.5)
\psline[linewidth=\mince](0,0)(1.2,0)
\psline[linecolor=blue,linewidth=\elegant]{-}(0.2,0)(0.2,0.4)
\pscircle[fillstyle=solid,fillcolor=black](0.6,0){0.04}
\pscircle[fillstyle=solid,fillcolor=black](1.0,0){0.04}
\end{pspicture}
\quad
\begin{pspicture}[shift=0](0.0,0)(1.2,0.5)
\psline[linewidth=\mince](0,0)(1.2,0)
\psline[linecolor=blue,linewidth=\elegant]{-}(0.6,0)(0.6,0.4)
\pscircle[fillstyle=solid,fillcolor=black](0.2,0){0.04}
\pscircle[fillstyle=solid,fillcolor=black](1.0,0){0.04}
\end{pspicture}
\quad
\begin{pspicture}[shift=0](0.0,0)(1.2,0.5)
\psline[linewidth=\mince](0,0)(1.2,0)
\psline[linecolor=blue,linewidth=\elegant]{-}(1.0,0)(1.0,0.4)
\pscircle[fillstyle=solid,fillcolor=black](0.2,0){0.04}
\pscircle[fillstyle=solid,fillcolor=black](0.6,0){0.04}
\end{pspicture}
\qquad&%
\stan_{3,0,1}:\qquad
\begin{pspicture}[shift=0](0.0,0)(1.2,0.5)
\psline[linewidth=\mince](0,0)(1.2,0)
\pscircle[fillstyle=solid,fillcolor=black](0.2,0){0.04}
\psarc[linecolor=blue,linewidth=\elegant]{-}(0.8,0){0.2}{0}{180}
\end{pspicture}
\quad
\begin{pspicture}[shift=0](0.0,0)(1.2,0.5)
\psline[linewidth=\mince](0,0)(1.2,0)
\pscircle[fillstyle=solid,fillcolor=black](1.0,0){0.04}
\psarc[linecolor=blue,linewidth=\elegant]{-}(0.4,0){0.2}{0}{180}
\end{pspicture}
\quad
\begin{pspicture}[shift=0](0.0,0)(1.2,0.5)
\psline[linewidth=\mince](0,0)(1.2,0)
\pscircle[fillstyle=solid,fillcolor=black](0.6,0){0.04}
\psbezier[linecolor=blue,linewidth=\elegant](0.2,0)(0.2,0.4)(1.0,0.4)(1.0,0)
\end{pspicture}\\[0.1cm]%
&%
\phantom{\stan_{3,0,1}:\hspace{0.1cm} }\qquad
\begin{pspicture}[shift=0](0.0,0)(1.2,0.5)
\psline[linewidth=\mince](0,0)(1.2,0)
\pscircle[fillstyle=solid,fillcolor=black](0.2,0){0.04}
\psbezier[linecolor=blue,linewidth=\elegant](1.0,0)(1.0,0.4)(1.8,0.4)(1.8,0)\rput(-1.2,0){\psbezier[linecolor=blue,linewidth=\elegant](1.0,0)(1.0,0.4)(1.8,0.4)(1.8,0)}
\psframe[fillstyle=solid,linecolor=white,linewidth=0pt](1.2,0)(2.5,0.4)
\psframe[fillstyle=solid,linecolor=white,linewidth=0pt](0,0)(-0.5,0.4)
\end{pspicture}
\quad
\begin{pspicture}[shift=0](0.0,0)(1.2,0.5)
\psline[linewidth=\mince](0,0)(1.2,0)
\pscircle[fillstyle=solid,fillcolor=black](1.0,0){0.04}
\psbezier[linecolor=blue,linewidth=\elegant](0.6,0)(0.6,0.4)(1.4,0.4)(1.4,0)\psbezier[linecolor=blue,linewidth=\elegant](-0.1,0.30)(0.1,0.23)(0.2,0.16)(0.2,0)
\psframe[fillstyle=solid,linecolor=white,linewidth=0pt](1.2,0)(2.5,0.4)
\psframe[fillstyle=solid,linecolor=white,linewidth=0pt](0,0)(-0.2,0.4)
\end{pspicture}
\quad
\begin{pspicture}[shift=0](0.0,0)(1.2,0.5)
\psline[linewidth=\mince](0,0)(1.2,0)
\pscircle[fillstyle=solid,fillcolor=black](0.6,0){0.04}
\psarc[linecolor=blue,linewidth=\elegant]{-}(0,0){0.2}{0}{90}
\psarc[linecolor=blue,linewidth=\elegant]{-}(1.2,0){0.2}{90}{180}
\end{pspicture}
\ \ .
\end{array}
\ee
The standard modules have dimension
\be
\dim \stan_{N,d,v} = \binom{N}{v}\binom{N-v}{a}.
\ee 

The action of $\mathsf {pA}_N(\alpha,\beta)$
on $\stan_{N,d,v}$ is defined as follows. For a connectivity $c$ and a link state $w$, we compute $c w$ by placing $w$ above $c$. The result of this multiplication is either zero or the scalar multiple of a link state. If 
a loop segment connects to a vacancy, $c w$ is set to zero. Otherwise, one reads off
the new link state $w'$ from the connectivity of the nodes on the bottom edge of the diagram $cw$. If there are
less than $d$ defects, $c w$ is set to zero. If there are
exactly $d$ defects, then $c w$ equals $w'$ up to a multiplicative factor. First, factors of $\beta$ and $\alpha$ are respectively included for the contractible and non-contractible loops appearing in the diagram. Second, if $d>0$, a factor of $\omega$, the {\it twist parameter}, is included for each defect that passes through the back of the cylinder towards the left, whereas a factor of $\omega^{-1}$ is included for each defect that passes through in the other direction. 
If no factors of $\alpha, \beta$ or $\omega$ are to be included, then the overall factor is just $1$. Here are two examples to illustrate:
\be
\begin{pspicture}[shift=-0.3](-0.0,0)(2.4,1.2)
\pspolygon[fillstyle=solid,fillcolor=lightlightblue,linecolor=black,linewidth=0pt](0,0)(0,0.8)(2.4,0.8)(2.4,0)(0,0)
\pscircle[fillstyle=solid,fillcolor=black](1.8,0){0.04}
\psarc[linecolor=blue,linewidth=\elegant]{-}(0.4,0){0.2}{0}{180}
\psarc[linecolor=blue,linewidth=\elegant]{-}(1.6,0.8){0.2}{180}{0}
\psbezier[linecolor=blue,linewidth=\elegant]{-}(2.2,0.8)(2.2,0.4)(1.0,0.4)(1.0,0)
\psbezier[linecolor=blue,linewidth=\elegant]{-}(0.2,0.8)(0.2,0.4)(-1.0,0.4)(-1.0,0)\rput(2.4,0){\psbezier[linecolor=blue,linewidth=\elegant]{-}(0.2,0.8)(0.2,0.4)(-1.0,0.4)(-1.0,0)}
\psbezier[linecolor=blue,linewidth=\elegant]{-}(1.0,0.8)(1.0,0.4)(-0.2,0.4)(-0.2,0)\rput(2.4,0){\psbezier[linecolor=blue,linewidth=\elegant]{-}(1.0,0.8)(1.0,0.4)(-0.2,0.4)(-0.2,0)}
\psframe[fillstyle=solid,linecolor=white,linewidth=0pt](-1.4,0)(0,0.8)
\psframe[fillstyle=solid,linecolor=white,linewidth=0pt](2.4,0)(3.6,0.8)
\rput(0,0.8){
\psarc[linecolor=blue,linewidth=\elegant]{-}(1.2,0){0.2}{0}{180}
\pscircle[fillstyle=solid,fillcolor=black](0.6,0){0.04}
\psline[linecolor=blue,linewidth=\elegant]{-}(0.2,0)(0.2,0.4)
\psline[linecolor=blue,linewidth=\elegant]{-}(1.8,0)(1.8,0.4)
\psline[linecolor=blue,linewidth=\elegant]{-}(2.2,0)(2.2,0.4)
}
\end{pspicture}\ \ = \omega^2 \ \ 
\begin{pspicture}[shift=0](0.0,0)(2.4,0.5)
\psline[linewidth=\mince](0,0)(2.4,0)
\pscircle[fillstyle=solid,fillcolor=black](1.8,0){0.04}
\psarc[linecolor=blue,linewidth=\elegant]{-}(0.4,0){0.2}{0}{180}
\psline[linecolor=blue,linewidth=\elegant]{-}(1.0,0)(1.0,0.4)
\psline[linecolor=blue,linewidth=\elegant]{-}(1.4,0)(1.4,0.4)
\psline[linecolor=blue,linewidth=\elegant]{-}(2.2,0)(2.2,0.4)
\end{pspicture}\ \ ,
\qquad\quad
\begin{pspicture}[shift=-0.3](-0.0,0)(2.4,1.2)
\pspolygon[fillstyle=solid,fillcolor=lightlightblue,linecolor=black,linewidth=0pt](0,0)(0,0.8)(2.4,0.8)(2.4,0)(0,0)
\psarc[linecolor=blue,linewidth=\elegant]{-}(0.8,0.8){0.2}{180}{0}
\psbezier[linecolor=blue,linewidth=\elegant]{-}(0.2,0.8)(0.2,0.25)(1.8,0.25)(1.8,0.8)
\pscircle[fillstyle=solid,fillcolor=black](0.2,0){0.04}
\pscircle[fillstyle=solid,fillcolor=black](1.0,0){0.04}
\psarc[linecolor=blue,linewidth=\elegant]{-}(1.6,0){0.2}{0}{180}
\psbezier[linecolor=blue,linewidth=\elegant]{-}(2.2,0)(2.2,0.5)(3.0,0.5)(3.0,0)\rput(-2.4,0){\psbezier[linecolor=blue,linewidth=\elegant]{-}(2.2,0)(2.2,0.5)(3.0,0.5)(3.0,0)}
\psframe[fillstyle=solid,linecolor=white,linewidth=0pt](-0.7,0)(0,0.8)
\psframe[fillstyle=solid,linecolor=white,linewidth=0pt](2.4,0)(3.6,0.8)
\rput(0,0.8){
\pscircle[fillstyle=solid,fillcolor=black](1.4,0){0.04}
\pscircle[fillstyle=solid,fillcolor=black](2.2,0){0.04}
\psarc[linecolor=blue,linewidth=\elegant]{-}(0.8,0){0.2}{0}{180}
\psbezier[linecolor=blue,linewidth=\elegant]{-}(0.2,0)(0.2,0.5)(-0.6,0.5)(-0.6,0)\rput(2.4,0){\psbezier[linecolor=blue,linewidth=\elegant]{-}(0.2,0)(0.2,0.5)(-0.6,0.5)(-0.6,0)}
\psframe[fillstyle=solid,linecolor=white,linewidth=0pt](-1.4,0)(0,0.4)
\psframe[fillstyle=solid,linecolor=white,linewidth=0pt](2.4,0)(3.6,0.4)
}
\end{pspicture}\ \ = \alpha \beta \ \ 
\begin{pspicture}[shift=0](0.0,0)(2.4,0.5)
\psline[linewidth=\mince](0,0)(2.4,0)
\pscircle[fillstyle=solid,fillcolor=black](0.2,0){0.04}
\pscircle[fillstyle=solid,fillcolor=black](1.0,0){0.04}
\psarc[linecolor=blue,linewidth=\elegant]{-}(1.6,0){0.2}{0}{180}
\psbezier[linecolor=blue,linewidth=\elegant]{-}(2.2,0)(2.2,0.5)(3.0,0.5)(3.0,0)\rput(-2.4,0){\psbezier[linecolor=blue,linewidth=\elegant]{-}(2.2,0)(2.2,0.5)(3.0,0.5)(3.0,0)}
\psframe[fillstyle=solid,linecolor=white,linewidth=0pt](-0.25,0)(0,0.4)
\psframe[fillstyle=solid,linecolor=white,linewidth=0pt](2.4,0)(3.2,0.4)
\end{pspicture}\ \ .
\ee
This action defines the 
{\em standard representations} of $\mathsf {pA}_N(\alpha,\beta)$.

\subsection[The $A_2^{(1)}$ vertex model]{The $\boldsymbol{A_2^{(1)}}$ vertex model}

The vertex modules over $\mathsf {pA}_N(\alpha,\beta)$
are defined on the vector space $(\mathbb C^3)^{\otimes N}$. We use the 
standard notation
\be
| {\uparrow} \rangle = \begin{pmatrix}1\\0\\0\end{pmatrix}, \qquad | {0} \rangle = \begin{pmatrix}0\\1\\0\end{pmatrix}, \qquad| {\downarrow} \rangle = \begin{pmatrix}0\\0\\1\end{pmatrix},
\ee
for the canonical basis of $\mathbb C^3$. One obtains a representation of the (non-periodic) dilute Temperley-Lieb algebra 
$\mathsf{dTL_N(\beta)}$
on 
$(\mathbb C^3)^{\otimes N}$, and therefore of $\mathsf {A}_N(\beta)$,
by imposing the following local rules:
\begin{subequations}
\begin{alignat}{2}
&
\begin{pspicture}[shift=-0.5](0,-0.2)(1.0,1.0)
\pspolygon[fillstyle=solid,fillcolor=lightlightblue,linecolor=black,linewidth=0pt](0,0)(0,0.8)(1.0,0.8)(1.0,0)(0,0)
\psbezier[linecolor=blue,linewidth=\elegant]{-}(0.2,0.8)(0.2,0.4)(0.8,0.4)(0.8,0.8)
\rput(0.2,1.0){$_i$}
\rput(0.8,1.0){$_j$}
\end{pspicture}\ \ \longrightarrow q^{1/2} \langle {\uparrow_i\downarrow_j}| + q^{-1/2} \langle {\downarrow_i\uparrow_j}| \ ,
\qquad \quad
\begin{pspicture}[shift=-0.5](0,-0.2)(1.0,1.0)
\pspolygon[fillstyle=solid,fillcolor=lightlightblue,linecolor=black,linewidth=0pt](0,0)(0,0.8)(1.0,0.8)(1.0,0)(0,0)
\psbezier[linecolor=blue,linewidth=\elegant]{-}(0.2,0)(0.2,0.4)(0.8,0.4)(0.8,0.8)
\rput(0.2,-0.2){$_i$}
\rput(0.8,1.0){$_j$}
\end{pspicture}\ \ \longrightarrow \ \ | {\uparrow_i}\rangle\langle{\uparrow_j}| + | {\downarrow_i}\rangle\langle{\downarrow_j}| \ ,
\\[0.1cm]
&
\begin{pspicture}[shift=-0.5](0,-0.2)(1.0,1.0)
\pspolygon[fillstyle=solid,fillcolor=lightlightblue,linecolor=black,linewidth=0pt](0,0)(0,0.8)(1.0,0.8)(1.0,0)(0,0)
\psbezier[linecolor=blue,linewidth=\elegant]{-}(0.2,0)(0.2,0.4)(0.8,0.4)(0.8,0)
\rput(0.2,-0.2){$_i$}
\rput(0.8,-0.2){$_j$}
\end{pspicture}\ \ \longrightarrow q^{1/2} | {\uparrow_i\downarrow_j}\rangle + q^{-1/2} | {\downarrow_i\uparrow_j}\rangle \ ,
\qquad \quad
\begin{pspicture}[shift=-0.5](0,-0.2)(1.0,1.0)
\pspolygon[fillstyle=solid,fillcolor=lightlightblue,linecolor=black,linewidth=0pt](0,0)(0,0.8)(1.0,0.8)(1.0,0)(0,0)
\pscircle[fillstyle=solid,fillcolor=black](0.5,0){0.04}
\rput(0.5,-0.2){$_i$}
\end{pspicture}\ \ \longrightarrow\ \  | {0_i}\rangle \ ,
\qquad \quad
\begin{pspicture}[shift=-0.5](0,-0.2)(1.0,1.0)
\pspolygon[fillstyle=solid,fillcolor=lightlightblue,linecolor=black,linewidth=0pt](0,0)(0,0.8)(1.0,0.8)(1.0,0)(0,0)
\pscircle[fillstyle=solid,fillcolor=black](0.5,0.8){0.04}
\rput(0.5,1.0){$_i$}
\end{pspicture}\ \ \longrightarrow\ \  \langle {0_i}| \ ,
\end{alignat}
\end{subequations}
where the labels $i$ and $j$, as in $|{\uparrow_i}{\downarrow_j}\rangle$, indicate the $i$-th and $j$-th copy in $(\mathbb C^3)^{\otimes N}$. 
For $N=2$, applying this map to each of the seven diagrams in \eqref{eq:faceop}, one obtains the following form for 
$\psset{unit=0.3}
\check R(u) = \, \begin{pspicture}[shift=-0.8](0,0)(2,2)
\pspolygon[fillstyle=solid,fillcolor=lightlightblue](1,0)(0,1)(1,2)(2,1)
\rput(1,1){\scriptsize$u$}
\psarc[linewidth=0.025]{-}(1,0){0.28}{45}{135}
\end{pspicture}\ 
\big|_{(\mathbb C^3)^{\otimes 2}}$: 
\be
\check R(u) = \left(
\begin{array}{ccccccccc}
 s_1(-u) & 0 & 0 & 0 & 0 & 0 & 0 & 0 & 0 \\
 0 & t & 0 & s_0(u) & 0 & 0 & 0 & 0 & 0 \\
 0 & 0 & \eE^{\ir u}  & 0 & 0 & 0 & s_0(u) & 0 & 0 \\
 0 & s_0(u) & 0 & t^{-1} & 0 & 0 & 0 & 0 & 0 \\
 0 & 0 & 0 & 0 & s_1(-u)  & 0 & 0 & 0 & 0 \\
 0 & 0 & 0 & 0 & 0 & t^{-1} & 0 & s_0(u) & 0 \\
 0 & 0 & s_0(u) & 0 & 0 & 0 & \eE^{-\ir u} & 0 & 0 \\
 0 & 0 & 0 & 0 & 0 & s_0(u) & 0 & t & 0 \\
 0 & 0 & 0 & 0 & 0 & 0 & 0 & 0 &  s_1(-u)  \\
\end{array}
\right).
\ee
This is the $\check R(u)$ matrix of the $U_q(\widehat{sl}(3))$-invariant $15$-vertex model. It indeed satisfies the Yang-Baxter equation:
\be
\check R_{12}(u)\check R_{23}(u+v)\check R_{12}(v) = \check R_{23}(v) \check R_{12}(u+v) \check R_{23}(u).
\ee

The map defined above is extended to $\mathsf{pdTL}_{N,v}(\alpha,\beta)$ (and simultaneously to $\mathsf {pA}_N(\alpha,\beta)$)
by including extra prescriptions for diagrams where some loop segments connect via the back of the cylinder. These involve an extra parameter, the {\it twist angle} $\phi$:
\begin{subequations}
\begin{alignat}{2}
&
\begin{pspicture}[shift=-0.5](0,-0.2)(1.0,1.0)
\pspolygon[fillstyle=solid,fillcolor=lightlightblue,linecolor=black,linewidth=0pt](0,0)(0,0.8)(1.0,0.8)(1.0,0)(0,0)
\psbezier[linecolor=blue,linewidth=\elegant]{-}(0.3,0.8)(0.3,0.4)(-0.3,0.4)(-0.3,0.8)\rput(1,0){\psbezier[linecolor=blue,linewidth=\elegant]{-}(0.3,0.8)(0.3,0.4)(-0.3,0.4)(-0.3,0.8)}
\rput(0.3,1.0){$_i$}
\rput(0.7,1.0){$_j$}
\psframe[fillstyle=solid,linecolor=white,linewidth=0pt](-0.35,0)(0,0.8)
\psframe[fillstyle=solid,linecolor=white,linewidth=0pt](1.0,0)(1.35,0.8)
\end{pspicture}\ \ 
\longrightarrow \eE^{-\ir \phi}q^{1/2} \langle {\downarrow_i\uparrow_j}| + \eE^{\ir \phi}q^{-1/2} \langle {\uparrow_i\downarrow_j}| \ ,
\qquad
&&
\begin{pspicture}[shift=-0.5](0,-0.2)(1.0,1.0)
\pspolygon[fillstyle=solid,fillcolor=lightlightblue,linecolor=black,linewidth=0pt](0,0)(0,0.8)(1.0,0.8)(1.0,0)(0,0)
\psbezier[linecolor=blue,linewidth=\elegant]{-}(0.3,0)(0.3,0.4)(-0.3,0.4)(-0.3,0.8)\rput(1,0){\psbezier[linecolor=blue,linewidth=\elegant]{-}(0.3,0)(0.3,0.4)(-0.3,0.4)(-0.3,0.8)}
\rput(0.3,-0.2){$_i$}
\rput(0.7,1.0){$_j$}
\psframe[fillstyle=solid,linecolor=white,linewidth=0pt](-0.35,0)(0,0.8)
\psframe[fillstyle=solid,linecolor=white,linewidth=0pt](1.0,0)(1.35,0.8)
\end{pspicture}\ \ 
\longrightarrow \eE^{-\ir \phi}| {\uparrow_i}\rangle\langle{\uparrow_j}| + \eE^{\ir \phi}| {\downarrow_i}\rangle\langle{\downarrow_j}| \ ,
\\[0.2cm]
&
\begin{pspicture}[shift=-0.5](0,-0.2)(1.0,1.0)
\pspolygon[fillstyle=solid,fillcolor=lightlightblue,linecolor=black,linewidth=0pt](0,0)(0,0.8)(1.0,0.8)(1.0,0)(0,0)
\psbezier[linecolor=blue,linewidth=\elegant]{-}(0.3,0)(0.3,0.4)(-0.3,0.4)(-0.3,0)\rput(1,0){\psbezier[linecolor=blue,linewidth=\elegant]{-}(0.3,0)(0.3,0.4)(-0.3,0.4)(-0.3,0)}
\rput(0.3,-0.2){$_i$}
\rput(0.7,-0.2){$_j$}
\psframe[fillstyle=solid,linecolor=white,linewidth=0pt](-0.35,0)(0,0.8)
\psframe[fillstyle=solid,linecolor=white,linewidth=0pt](1.0,0)(1.35,0.8)
\end{pspicture}\ \ 
\longrightarrow \eE^{\ir \phi} q^{1/2}| {\downarrow_i\uparrow_j}\rangle + \eE^{-\ir \phi}q^{-1/2} | {\uparrow_i\downarrow_j}\rangle \ ,
\qquad
&&
\begin{pspicture}[shift=-0.5](0,-0.2)(1.0,1.0)
\pspolygon[fillstyle=solid,fillcolor=lightlightblue,linecolor=black,linewidth=0pt](0,0)(0,0.8)(1.0,0.8)(1.0,0)(0,0)
\psbezier[linecolor=blue,linewidth=\elegant]{-}(0.3,0.8)(0.3,0.4)(-0.3,0.4)(-0.3,0)\rput(1,0){\psbezier[linecolor=blue,linewidth=\elegant]{-}(0.3,0.8)(0.3,0.4)(-0.3,0.4)(-0.3,0)}
\rput(0.3,1.0){$_i$}
\rput(0.7,-0.2){$_j$}
\psframe[fillstyle=solid,linecolor=white,linewidth=0pt](-0.35,0)(0,0.8)
\psframe[fillstyle=solid,linecolor=white,linewidth=0pt](1.0,0)(1.35,0.8)
\end{pspicture}\ \ 
\longrightarrow \eE^{\ir \phi}| {\uparrow_j}\rangle\langle{\uparrow_i}| + \eE^{-\ir \phi}| {\downarrow_j}\rangle\langle{\downarrow_i}|  \ .
\end{alignat}
\end{subequations}
The 
fugacity
of the non-contractible loops is then parameterised as $\alpha = 2 \cos \phi$.

\goodbreak
\section{Diagrammatic calculus}\label{sec:calculus}

In this section, we develop the diagrammatic calculus that allows us to derive the functional relations satisfied by the transfer tangles  
presented in \cref{sec:Ftm.fr}. 
From here onwards, we set the gauge parameter to $t=1$.\footnote{Because of \eqref{eq:gauge.invariance}, the calculations below are easily generalised to all values of $t \in \mathbb C^*$. In particular, the functional equations of \cref{sec:Ftm.fr} are identical for all $t \in \mathbb C^*$. The 
case $t = \eE^{-\ir u}$ relevant for the RSOS models, see \eqref{eq:faceop.specialgauge}, is special because the braid operators differ from \eqref{eq:braid.ops}. In this case, the functional relations presented in \cref{sec:Ftm.fr} are unchanged, but the asymptotic behavior of the corresponding functions, given in \eqref{eq:braid.T.eig} and \eqref{eq:fused.braid.eig} for $t=1$, is different.}

\subsection{Local relations}\label{sec:local.relations}

The face operator satisfies a number of local relations. First, at $u=0$, the face operator is proportional to the identity:
\be
 
\ee
where the dashed line is an {\it identity strand}, 
corresponding to the sum
\be
\psset{unit=0.8}
%
\ \ .
\ee
In the absence of crossing symmetry, there are two inequivalent Yang-Baxter equations:
\be
%
\label{eq:YBE}
\ee
and likewise two local inversion identities:
\be
\psset{unit=0.7364cm}
%
.
\label{eq:inversions}
\ee
For $u = \lambda$, the face operator factorises into a product of two triangular face operators: 
\be
%
\ .
\label{tri1}
\ee
We also have the following push-through properties:
\begin{subequations}
\begin{alignat}{2}
&
%
\ \ .
\label{eq:pushthru2}
\end{alignat}
\end{subequations}

\subsection{Wenzl-Jones projectors}\label{sec:projs}
In this subsection, we define a two-parameter family of projectors, $P^{m,n}$, denoting them graphically by
\be
 P^{m,n}= \
 %
 \ ,\qquad 
m, n\in
\mathbb{Z}_\geq,
\ee
where $(m,n)\neq(0,0)$.
These projectors are not invariant under reflections
nor rotations. The marker at the bottom-left corner thus serves to indicate the orientation. 

First, we introduce a second triangular face operator,
\be
\label{eq:local.props}
\psset{unit=0.8}
%
\ .
\ee
Unlike the one in \eqref{tri1}, this triangle
is not invariant under rotations. We have the following two local identities:
\be\label{eq:spider.ids}
\psset{unit=0.8}
%
\ee
where the $q$-numbers are
\be
[k] = \frac{q^k - q^{-k}}{q-q^{-1}},\qquad k\in
\mathbb{Z}_\geq,
\ee
and the {\it wavy} loop segment is defined as
\be\label{eq:wavy}
\psset{unit=0.8}
%
\ \ .
\ee
It satisfies the identity
\be
\label{eq:[3]}
\psset{unit=0.8}
%
 \ = q \beta + q^{-2} = [3]. \ee
We note that the relations \eqref{eq:spider.ids} and \eqref{eq:[3]} are identical to those given by Kuperberg \cite{Kuperberg96} for the $s\ell(3)$ spiders.

The projectors $P^{m,0}$ are defined recursively by
\be
\label{eq:Pmnrec}
%
\ \ ,
}
\ee
In particular, the $P_{2,0}$ projector is
\be
%
\ \ .
\ee
These projectors satisfy 
\be\label{eq:projprops}
%
\ , \qquad m \ge n.
\ee 
For $n\in\mathbb{N}$, the $P^{0,n}$ projector is then defined as
\be
P^{0,n}= \
%
\ ,
\ee
where the marker in the upper-right corner indicates that 
$P^{0,n}$ is obtained from $P^{n,0}$ by rotating the latter by 180 degrees. 

The first mixed projector is $P^{1,1}$, defined as
\be
%
\ \ .
\ee
More generally, following \cite{R14}, the $P^{m,n}$ projectors are defined by
\be
\label{eq:Pmn.gen}
%
\ee
where
\be
\qbinom{m}{k} = \frac{[m]!}{[m-k]![k]!}, \qquad [m]! = \prod_{k=1}^m\ [k].
\ee
Some useful identities are
\begin{subequations}
\begin{alignat}{3}
&\psset{unit=0.8}
%
 \ \ .
\label{eq:proj.id.2}
\end{alignat}
\end{subequations}

\subsection{Gauge symmetries}
\label{Sec:Gauge}

We denote a gauge operator on one site by 
\be
\psset{unit=0.8}
%
\ \ .
\ee
For $g = q^k$, we denote the corresponding gauge operator by $%
$\ . 
With this notation,
the wavy loop segment \eqref{eq:wavy} is given by $q^{-2}\times \ %
$\ . 
The elementary face operator satisfies
\be\label{eq:pi.period}
%
\ \ , \qquad g = -1.
\ee
Moreover, we note that the original $t$-dependent face operator of \eqref{eq:faceop} is obtained by applying 
gauge operators to the $t=1$ specialised operator:
\be\label{eq:gauge.invariance}
%
\ \ ,
\ee
where the gauge parameters in the left and right gauge operators are respectively set to $t$ and $t^{-1}$.

We also have the following equality:
\be
\label{eq:gauge.face}
\psset{unit=0.8}
%
\ee
with $k_1$, $k_2$, $k_3$ and $k_4$ satisfying
\be
k_1 + k_3 = -m, \qquad k_2 + k_4 = m, \qquad k_1 + k_4 = -1, \qquad k_2 + k_3 = 1.
\label{kkm}
\ee
The linear system \eqref{kkm} is under-determined and has a one-parameter family of solutions. The left-hand side
of \eqref{eq:gauge.face} appears in the recursive construction \eqref{eq:Pmnrec} of the projectors $P^{m,0}$. By carefully tuning 
the parameters $k_j$, we can then write these projectors as
\be
\label{eq:proj.triangles}
%
\ \, .
\ee

\subsection{Fused face operators}

Fusion is performed in the vertical direction, yielding fused face operators labelled by a pair of non-negative Dynkin labels $(m,n)$.
Each such pair is associated to a node of the $s\ell(3)$ weight lattice, see \cref{fig:su3weights}:
\be
(m,n) \qquad \longleftrightarrow\qquad
\psset{unit=4}
%
 \ \ .
\ee
For $(1,0)$ and $(0,1)$, the face operators are given in \eqref{eq:two.face.ops}. For general $(m,n)$, the fused face operator is defined as
\be
\psset{unit=1.2}
%
\ee
with 
\be u_k = u+k \lambda.\ee
For each pair $(m,n)$, the normalisation removes simple trigonometric factors and ensures that the weight of each fused face operator is a Laurent polynomial given as a linear combination of $\eE^{\ir u}$, $1$ and $\eE^{-\ir u}$.

\subsection{Braid limits}

There are four elementary braid operators: 
\begin{subequations}\label{eq:braid.ops}
\begin{alignat}{2}
&
%
\ \bigg).
\end{alignat}
\end{subequations}
These are obtained from the $u \to \pm \ir \infty$ limits of the elementary face operators:
\be
%
\ .
\ee
The fused braid operators are likewise defined as
\be
%
\ \ .
\ee

\section{Fused transfer matrices and functional relations}\label{sec:Ftm.fr}

In this section, we define a family of transfer tangles that are elements of the algebra $\mathsf{pA}_N(\alpha,\beta)$. We derive a set of functional relations satisfied by these tangles. Because the calculations are performed in the algebra, the corresponding identities hold in all representations of $\mathsf{pA}_N(\alpha,\beta)$.

\subsection{Transfer tangles}

On the cylinder, the two elementary single-row transfer tangles $\Tb^{1,0}(u)$ and $\Tb^{0,1}(u)$ are defined as
\begin{subequations}
\begin{alignat}{2}
&\Tb^{1,0} (u)= \  \
\begin{pspicture}[shift=-0.4](-0.3,0)(4.3,1.0)
\facegrid{(0,0)}{(4,1)}
\psarc[linewidth=0.025]{-}(0,0){0.16}{0}{90}
\psarc[linewidth=0.025]{-}(1,0){0.16}{0}{90}
\psarc[linewidth=0.025]{-}(3,0){0.16}{0}{90}
\psline[linewidth=1.5pt,linecolor=blue,linestyle=dashed,dash=2pt 2pt]{-}(0,0.5)(-0.3,0.5)
\psline[linewidth=1.5pt,linecolor=blue,linestyle=dashed,dash=2pt 2pt]{-}(4,0.5)(4.3,0.5)
\rput(2.5,0.5){$\ldots$}
\rput(0.5,.5){$u$}
\rput(1.5,.5){$u$}
\rput(3.5,.5){$u$}
\end{pspicture} \qquad \quad
\Tb^{0,1} (u)= \  \
\begin{pspicture}[shift=-0.4](-0.3,0)(4.3,1.0)
\facegrid{(0,0)}{(4,1)}
\psarc[linewidth=0.025]{-}(1,0){0.16}{90}{180}
\psarc[linewidth=0.025]{-}(2,0){0.16}{90}{180}
\psarc[linewidth=0.025]{-}(4,0){0.16}{90}{180}
\psline[linewidth=1.5pt,linecolor=blue,linestyle=dashed,dash=2pt 2pt]{-}(0,0.5)(-0.3,0.5)
\psline[linewidth=1.5pt,linecolor=blue,linestyle=dashed,dash=2pt 2pt]{-}(4,0.5)(4.3,0.5)
\rput(2.5,0.5){$\ldots$}
\rput(0.5,.5){\small$\lambda-u$}
\rput(1.5,.5){\small$\lambda-u$}
\rput(3.5,.5){\small$\lambda-u$}
\end{pspicture} \ \ .
\label{Tu01}
\end{alignat}
\end{subequations}
As discussed in \cref{app:cov}, both $\Tb^{1,0}(u)$ and $\Tb^{0,1}(u)$ are elements of $\mathsf {pA}_N(\alpha,\beta)$.
Using the 
Yang-Baxter equations \eqref{eq:YBE} and the local inversion identities \eqref{eq:inversions}, one can show that these transfer tangles are in the same commuting family:
\be
[\Tb^{1,0}(u),\Tb^{1,0}(v)] = 0, \qquad [\Tb^{1,0}(u),\Tb^{0,1}(v)] = 0, \qquad [\Tb^{0,1}(u),\Tb^{0,1}(v)] = 0.
\ee
From \eqref{eq:pi.period}, the transfer tangles satisfy
\be
\Tb^{1,0}(u+\pi) = (-1)^N\Tb^{1,0}(u), \qquad \Tb^{0,1}(u+\pi) = (-1)^N\Tb^{0,1}(u).
\ee

The braid transfer tangles are defined as
\begin{subequations}
\begin{alignat}{2}
\Tb^{1,0}_{\pm\infty}&=\ \ 
\begin{pspicture}[shift=-.42](-0.3,0)(4.3,1)
\facegrid{(-0,0)}{(4,1)}
\psline[linewidth=1.5pt,linecolor=blue,linestyle=dashed,dash=2pt 2pt]{-}(0,0.5)(-0.3,0.5)
\psline[linewidth=1.5pt,linecolor=blue,linestyle=dashed,dash=2pt 2pt]{-}(4,0.5)(4.3,0.5)
\rput(0,0){\rput(.5,.4){\small $\pm \ir \infty$}\rput(.5,.7){\tiny $(1,0)$}\psarc[linewidth=0.025]{-}(0,0){0.16}{0}{90}}
\rput(1,0){\rput(.5,.4){\small $\pm \ir \infty$}\rput(.5,.7){\tiny $(1,0)$}\psarc[linewidth=0.025]{-}(0,0){0.16}{0}{90}}
\rput(3,0){\rput(.5,.4){\small $\pm \ir \infty$}\rput(.5,.7){\tiny $(1,0)$}\psarc[linewidth=0.025]{-}(0,0){0.16}{0}{90}}
\rput(2.53,.51){$\dots$}
\end{pspicture} \ \ = 
\lim_{u\to\pm\ir\infty}\bigg(\frac{e^{\pm\ir\tfrac{\pi-\lambda}3}}{s_0(u)}\bigg)^{\!N} \Tb^{1,0}(u),
\\[0.15cm]
\Tb^{0,1}_{\pm\infty}&=\ \ 
\begin{pspicture}[shift=-.42](-0.3,0)(4.3,1)
\facegrid{(-0,0)}{(4,1)}
\psline[linewidth=1.5pt,linecolor=blue,linestyle=dashed,dash=2pt 2pt]{-}(0,0.5)(-0.3,0.5)
\psline[linewidth=1.5pt,linecolor=blue,linestyle=dashed,dash=2pt 2pt]{-}(4,0.5)(4.3,0.5)
\rput(0,0){\rput(.5,.4){\small $\pm \ir \infty$}\rput(.5,.7){\tiny $(0,1)$}\psarc[linewidth=0.025]{-}(0,0){0.16}{0}{90}}
\rput(1,0){\rput(.5,.4){\small $\pm \ir \infty$}\rput(.5,.7){\tiny $(0,1)$}\psarc[linewidth=0.025]{-}(0,0){0.16}{0}{90}}
\rput(3,0){\rput(.5,.4){\small $\pm \ir \infty$}\rput(.5,.7){\tiny $(0,1)$}\psarc[linewidth=0.025]{-}(0,0){0.16}{0}{90}}
\rput(2.53,.51){$\dots$}
\end{pspicture} \ \ =
\lim_{u\to\pm\ir\infty}\bigg(\frac{e^{\pm2\ir\tfrac{\pi-\lambda}3}}{s_0(u)}\bigg)^{\!N} \Tb^{0,1}(u).
\end{alignat}
\end{subequations}
On the standard modules $\stan_{N,d,v}$, the braid transfer matrices are proportional to the identity, with eigenvalues
\begin{alignat}{2}\label{eq:braid.T.eig}
T^{1,0}_{\pm\infty}\big|_{\stan_{N,d,v}} = \omega\, \eE^{\pm\ir \theta_1} + \eE^{\pm\ir \theta_2} + \omega^{-1} \eE^{\pm\ir \theta_3},
\qquad
T^{0,1}_\infty\big|_{\stan_{N,d,v}} =  \omega\, \eE^{\mp\ir \theta_3} + \eE^{\mp\ir \theta_2} + \omega^{-1} \eE^{\mp\ir \theta_1},
\end{alignat}
where
\be
\theta_1 = \tfrac{\pi-\lambda}3(-a-2d+v), \qquad \theta_2 = \tfrac{\pi-\lambda}3(2a+d-2v),\qquad \theta_3 = \tfrac{\pi-\lambda}3(-a+d+v) = -\theta_1 - \theta_2.
\ee
With the convention $\alpha = \omega+\omega^{-1}$, this result also holds for $d=0$.

\subsection{Fused transfer tangles and fusion hierarchies}

The fused transfer tangles are defined as
\be
\Tb^{m,n} (u)= \  \
\begin{pspicture}[shift=-0.4](-0.3,0)(4.3,1.0)
\facegrid{(0,0)}{(4,1)}
\psarc[linewidth=0.025]{-}(0,0){0.16}{0}{90}
\psarc[linewidth=0.025]{-}(1,0){0.16}{0}{90}
\psarc[linewidth=0.025]{-}(3,0){0.16}{0}{90}
\psline[linewidth=1.5pt,linecolor=blue,linestyle=dashed,dash=2pt 2pt]{-}(0,0.5)(-0.3,0.5)
\psline[linewidth=1.5pt,linecolor=blue,linestyle=dashed,dash=2pt 2pt]{-}(4,0.5)(4.3,0.5)
\rput(2.5,0.5){$\ldots$}
\rput(0.5,.4){$u$}\rput(.5,.7){\tiny $(m,n)$}
\rput(1.5,.4){$u$}\rput(1.5,.7){\tiny $(m,n)$}
\rput(3.5,.4){$u$}\rput(3.5,.7){\tiny $(m,n)$}
\end{pspicture}\ \ .
\ee
The corresponding fused braid transfer tangles are given by
\be
\Tb^{m,n}_{\pm \infty} = \lim_{u\rightarrow \pm \ir \infty}\bigg(\frac{\eE^{\pm\ir(m+2n)\frac{\pi -\lambda}3}}{s_{m+n-1}(u)}\bigg)^{\!N} \Tb^{m,n}(u).
\label{Tmnb}
\ee
We use the notation and initial conditions
\be
\label{eq:notations}
\Tb^{m,n}_k = \Tb^{m,n} (u+k \lambda), \quad \Tb^{0,0}_k = f_{k-1} \Ib, \quad \Tb^{m,-1}_k = \Tb^{-1,n}_k = 0, \quad f_k = \big(s_k(u)\big)^N, \quad \sigma = (-1)^{N}.
\ee
The fused transfer tangles satisfy a set of functional relations known as the {\it fusion hierarchy}. These relations arise
as consequences of the local relations given in \cref{sec:local.relations} and take the form
\begin{subequations}
\label{eq:FH}
\begin{alignat}{2}
\Tb^{m,0}_0\Tb^{1,0}_m &=f_m \Tb^{m-1,1}_0 + f_{m-1} \Tb^{m+1,0}_0,\label{eq:FHa}\\[0.1cm]
\Tb^{0,1}_0\Tb^{0,n}_1 &= \sigma\, f_{-1} \Tb^{1,n-1}_1 + f_{0} \Tb^{0,n+1}_0\label{eq:FHb},\\[0.1cm]
\Tb^{m,0}_0\Tb^{0,n}_m &= f_{m-1} \Tb^{m,n}_0 + \sigma\, \Tb^{m-1,0}_0\Tb^{0,n-1}_{m+1},\label{eq:FHc}
\end{alignat}
\end{subequations}
as shown  
in \cref{app:proofFH}. Comparing with the rule for the tensor product of
an irreducible $s\ell(3)$ representation with the fundamental $(1,0)$ representation,
\begin{alignat}{2}
(m,n) \otimes (1,0) &= \ \ 
\psset{unit=4}
\begin{pspicture}[shift=-0.18](0,-0.1)(0.7,0.2)
\multiput(0,0)(.1,0){4}{\psline[linewidth=0.02cm]{-}(0,0)(0,.1)(.1,.1)(.1,0)(0,0)}
\multiput(0,.1)(.1,0){7}{\psline[linewidth=0.02cm]{-}(0,0)(0,.1)(.1,.1)(.1,0)(0,0)}
\rput(0.2,-0.08){$\underbrace{\ \hspace{1.3cm} \ }_n$}
\rput(0.55,0.02){$\underbrace{\ \hspace{0.85cm} \ }_m$}
\end{pspicture} \ \ \otimes \ \
\begin{pspicture}[shift=-0.23](0,-0.1)(0.1,0.2)
\multiput(0,.1)(.1,0){1}{\psline[linewidth=0.02cm]{-}(0,0)(0,.1)(.1,.1)(.1,0)(0,0)}
\end{pspicture}
\nonumber\\&= \ \
\psset{unit=4}
\begin{pspicture}[shift=-0.18](0,-0.1)(0.8,0.2)
\multiput(0,0)(.1,0){4}{\psline[linewidth=0.02cm]{-}(0,0)(0,.1)(.1,.1)(.1,0)(0,0)}
\multiput(0,.1)(.1,0){8}{\psline[linewidth=0.02cm]{-}(0,0)(0,.1)(.1,.1)(.1,0)(0,0)}
\rput(0.2,-0.08){$\underbrace{\ \hspace{1.3cm} \ }_n$}
\rput(0.6,0.02){$\underbrace{\ \hspace{1.25cm} \ }_{m+1}$}
\end{pspicture} \ \ \oplus \ \
\begin{pspicture}[shift=-0.18](0,-0.1)(0.7,0.2)
\multiput(0,0)(.1,0){5}{\psline[linewidth=0.02cm]{-}(0,0)(0,.1)(.1,.1)(.1,0)(0,0)}
\multiput(0,.1)(.1,0){7}{\psline[linewidth=0.02cm]{-}(0,0)(0,.1)(.1,.1)(.1,0)(0,0)}
\rput(0.25,-0.08){$\underbrace{\ \hspace{1.65cm} \ }_{n+1}$}
\rput(0.6,0.02){$\underbrace{\ \hspace{0.35cm} \ }_{m-1}$}
\end{pspicture} \ \ \oplus \ \
\begin{pspicture}[shift=-0.18](0,-0.1)(0.7,0.2)
\multiput(0,-0.1)(.1,0){1}{\psline[linewidth=0.02cm]{-}(0,0)(0,.1)(.1,.1)(.1,0)(0,0)}
\multiput(0,0)(.1,0){4}{\psline[linewidth=0.02cm]{-}(0,0)(0,.1)(.1,.1)(.1,0)(0,0)}
\multiput(0,.1)(.1,0){7}{\psline[linewidth=0.02cm]{-}(0,0)(0,.1)(.1,.1)(.1,0)(0,0)}
\multiput(0,-0.1)(0,0.1){3}
{\pspolygon[fillstyle=solid,fillcolor=lightgray](0,0)(0,.1)(.1,.1)(.1,0)(0,0)}
\rput(0.25,-0.08){$\underbrace{\ \hspace{0.85cm} \ }_{n-1}$}
\rput(0.55,0.02){$\underbrace{\ \hspace{0.85cm} \ }_m$}
\end{pspicture}
\nonumber\\[0.2cm]&= (m+1,n) \oplus (m-1,n+1) \oplus (m,n-1),
\end{alignat}
we note that \eqref{eq:FHa} is consistent with this rule in the case where $(m,n) = (m,0)$.

The relations \eqref{eq:FH} allow one to express each $\Tb^{m,n}_0$ as a polynomial in the $\Tb^{1,0}_j$ and $\Tb^{0,1}_k$. 
As in \cite[Equation (1.33)]{ZPG95}, these can be put into determinant form, in turn implying
further functional relations obtained by expanding the determinants in terms of minors:
\begin{subequations}
\label{eq:FH2}
\begin{alignat}{2}
f_{m+n-2}\Tb^{m,n}_0 &= \Tb^{m,n-1}_0 \Tb^{0,1}_{m+n-1} -\sigma\, \Tb^{m,n-2}_0 \Tb^{1,0}_{m+n-1} + f_{m+n-1} \Tb^{m,n-3}_0,\label{eq:FH2a}\\[0.1cm]
f_0\Tb^{m,n}_0 &= \Tb^{1,0}_0 \Tb^{m-1,n}_1 - \Tb^{0,1}_0 \Tb^{m-2,n}_2 + \sigma f_{-1} \Tb^{m-3,n}_3.\label{eq:FH2b}
\end{alignat}
\end{subequations}
With the identifications
\be
\label{eq:negTs}
\Tb^{m,n}_0 = - \Tb^{-m-2,m+n+1}_{m+1} = -\sigma^{n+1}\Tb^{m+n+1,-n-2}_0,
\ee
the above fusion hierarchy relations hold for arbitrary $m, n \in \mathbb Z$.

By considering the braid limits of the hierarchy relations in \eqref{eq:FH}, we obtain similar relations satisfied by
the fused braid transfer tangles \eqref{Tmnb}, namely
\begin{subequations}
\begin{alignat}{2}
\Tb^{m,0}_{\pm \infty}\Tb^{1,0}_{\pm \infty} &=\Tb^{m-1,1}_{\pm \infty} + \Tb^{m+1,0}_{\pm \infty},\\[0.1cm]
\Tb^{0,1}_{\pm \infty}\Tb^{0,n}_{\pm \infty} &= \Tb^{1,n-1}_{\pm \infty} + \Tb^{0,n+1}_{\pm \infty},\\[0.1cm]
\Tb^{m,0}_{\pm \infty}\Tb^{0,n}_{\pm \infty} &= \Tb^{m,n}_{\pm \infty} + \Tb^{m-1,0}_{\pm \infty}\Tb^{0,n-1}_{\pm \infty}.\label{eq:FHbraid}
\end{alignat}
\end{subequations}
The eigenvalues $T^{m,n}_{\pm \infty}$ of the braid transfer matrices on $\stan_{N,d,v}$ are given by $s\ell(3)$ Chebyshev polynomials \cite{DFZ90}. For the rectangular Young diagrams, we have
\be
\label{eq:fused.braid.eig}
T^{m,0}_{\pm\infty}|_{\stan_{N,d,v}} = U_m(\omega\, \eE^{\pm \ir \theta_1}, \eE^{\pm \ir \theta_2}), \qquad T^{0,n}_{\pm\infty}|_{\stan_{N,d,v}} = U_n(\omega^{-1} \eE^{\mp \ir \theta_1}, \eE^{\mp \ir \theta_2}),
\ee
with
\be
U_m(y_1,y_2) = \frac{y_1^{m+2}(y_2-y_3)+y_2^{m+2}(y_3-y_1)+y_3^{m+2}(y_1-y_2)}{(y_1-y_2)(y_1-y_3)(y_2-y_3)} , \qquad y_1 y_2 y_3 = 1.
\ee
The values for $T^{m,n}_{\pm \infty}|_{\stan_{N,d,v}}$, $m,n\neq0$,
are obtained from \eqref{eq:fused.braid.eig} using \eqref{eq:FHbraid}.

\subsection[$T$-system and $Y$-system]{$\boldsymbol T$-system and $\boldsymbol Y$-system}

The $T$-system relations 
follow from the fusion hierarchy relations \eqref{eq:FH} and are quadratic relations in the transfer tangles $\Tb^{m,0}$ and $\Tb^{0,n}$:
\begin{subequations}
\label{eq:Trelations}
\begin{alignat}{2}
\Tb^{m,0}_0\Tb^{m,0}_1&= f_m\Tb^{0,m}_0+\Tb^{m+1,0}_0\Tb^{m-1,0}_1,\label{eq:TrelationsA}\\[0.1cm]
\Tb^{0,n}_0\Tb^{0,n}_1&=\sigma^n f_{-1}\Tb^{n,0}_1+\Tb^{0,n+1}_0\Tb^{0,n-1}_1.\label{eq:TrelationsB}
\end{alignat}
\end{subequations}
Their proof relies on an induction argument and is given in \cref{app:proofTsys}.
The $Y$-system is then derived from the $T$-system. The ``tangles" entering the $Y$-system are
\be
\label{eq:t.functions}
\tba{m}_0 =  \frac{\Tb^{m+1,0}_0\Tb^{m-1,0}_1}{f_m \Tb^{0,m}_0}, \qquad \tbb{n}_0 = \sigma^n \frac{\Tb^{0,n+1}_0\Tb^{0,n-1}_1}{f_{-1}\Tb^{n,0}_1}.
\ee
Strictly speaking, as defined here, these are not tangles because of the presence of inverses. Similarly, the $Y$-system equations we exhibit below are not actually equalities between tangles. We express these equations in the manner which is most useful for working with particular matrix representations to extract spectra. Any such given equation can be turned into an equality of tangles by substituting in the definitions for the $\tba{}$'s, rearranging and removing the inverses by multiplying the left and right sides by any transfer tangles that appear 
as inverses.

In terms of the functions $\tba{m}$ and $\tbb{n}$, the $T$-system relations are expressed as
\begin{subequations}\label{eq:Tsystem}
\begin{alignat}{2}
\Tb^{m,0}_0\Tb^{m,0}_1&= f_m\Tb^{0,m}_0(\Ib + \tba{m}_0),\\[0.1cm]
\Tb^{0,n}_0\Tb^{0,n}_1&=\sigma^n f_{-1}\Tb^{n,0}_1(\Ib + \tbb{n}_0).
\end{alignat}
\end{subequations}
The $Y$-system relations then read
\be
\tba{m}_0 \tba{m}_1 = \frac{(\Ib+\tba{m+1}_0)(\Ib+\tba{m-1}_1)}{\Ib+(\tbb{m}_0)^{-1}},\qquad
\tbb{n}_0 \tbb{n}_1 = \frac{(\Ib+\tbb{n+1}_0)(\Ib+\tbb{n-1}_1)}{\Ib+(\tba{n}_1)^{-1}}.
\ee
Their proof is elementary. For instance, the first relation is proven as follows:
\begin{alignat}{2}
\tba{m}_0 \tba{m}_1 &= \frac{(\Tb^{m+1,0}_0\Tb^{m+1,0}_1)(\Tb^{m-1,0}_1\Tb^{m-1,0}_2)}{f_mf_{m+1}\Tb^{0,m}_0\Tb^{0,m}_1} \nonumber\\&= 
\frac{(f_{m+1}\Tb^{0,m+1}_0+\Tb^{m+2,0}_0\Tb^{m,0}_1)(f_m \Tb^{0,m-1}_1+\Tb^{m,0}_1\Tb^{m-2,0}_2)}{f_mf_{m+1}(\sigma^mf_{-1}\Tb^{m,0}_1+\Tb^{0,m+1}_0\Tb^{0,m-1}_1)} \nonumber\\&
= \frac{\displaystyle\Big(\Ib+\frac{\Tb^{m+2,0}_0\Tb^{m,0}_1}{f_{m+1}\Tb^{0,m+1}_0}\Big)\Big(\Ib+\frac{\Tb^{m,0}_1\Tb^{m-2,0}_2}{f_{m}\Tb^{0,m-1}_1}\Big)}{\displaystyle\Big(\Ib+\frac{\sigma^mf_{-1}\Tb^{m,0}_1}{\Tb^{0,m+1}_0\Tb^{0,m-1}_1}\Big)}
\nonumber
= \frac{(\Ib+\tba{m+1}_0)(\Ib+\tba{m-1}_1)}{\Ib+(\tbb{m}_0)^{-1}}.
\end{alignat}
The proof of the second relation is similar.

\subsection{Closure relations at roots of unity}\label{sec:clo}

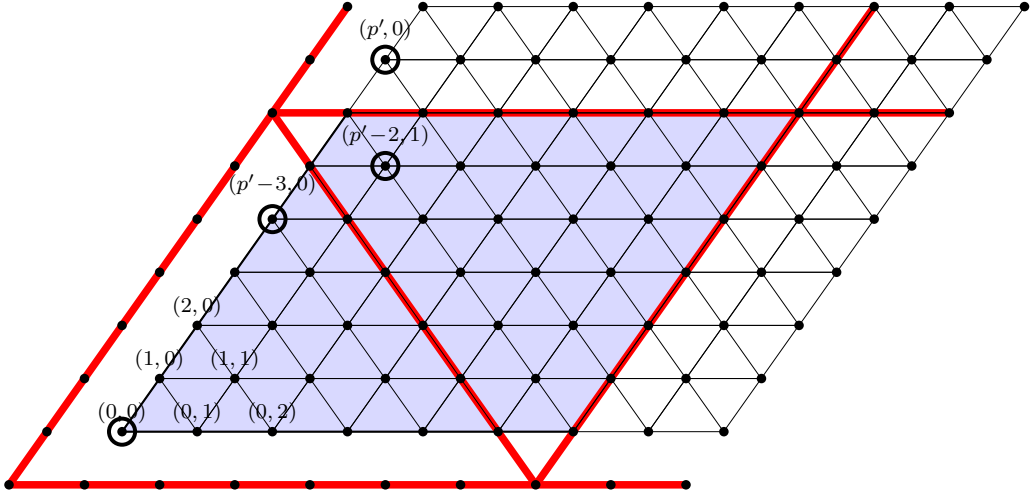
\begin{figure}[h] 
\begin{center}
$
\psset{unit=1}
\begin{pspicture}[shift=-0.9](0,-1.5)(12,7.7)
\pspolygon[fillstyle=solid,fillcolor=lightlightblue](0,0)(3.0,4.242)(9.0,4.242)(6.0,0)
\psline[linecolor=red,linewidth=0.1cm]{-}(2.0,4.242)(11,4.242)
\psline[linecolor=red,linewidth=0.1cm]{-}(2.0,4.242)(5.5,-0.707)
\psline[linecolor=red,linewidth=0.1cm]{-}(-1.5,-0.707)(7.5,-0.707)
\psline[linecolor=red,linewidth=0.1cm]{-}(-1.5,-0.707)(3,5.656)
\psline[linecolor=red,linewidth=0.1cm]{-}(5.5,-0.707)(10,5.656)
\multiput(-1.5,-0.707)(0.5,0.707){10}{\multiput(0,0)(1,0){10}{\psdot(0,0)}}
\psset{linewidth=0.1pt}
\multiput(0,0)(0.5,0.707){8}{\multiput(0,0)(1,0){8}{\psline{-}(0,0)(0.5,0.707)\psline{-}(0.5,0.707)(1,0)\psline{-}(0,0)(1,0)\psline{-}(1,0)(1.5,0.707)\psline{-}(0.5,0.707)(1.5,0.707)}}
\pscircle[linewidth=1.5pt,linecolor=black](3.5,4.949){.2}
\pscircle[linewidth=1.5pt,linecolor=black](3.5,3.535){.2}
\pscircle[linewidth=1.5pt,linecolor=black](2.0,2.828){.2}
\pscircle[linewidth=1.5pt,linecolor=black](0,0){.2}
\rput(0,0.25){\scriptsize$(0,0)$}
\rput(1,0.25){\scriptsize$(0,1)$}
\rput(2,0.25){\scriptsize$(0,2)$}
\rput(0.5,0.957){\scriptsize$(1,0)$}
\rput(1.5,0.957){\scriptsize$(1,1)$}
\rput(1,1.664){\scriptsize$(2,0)$}
\rput(2.0,3.288){\scriptsize$(p'\!-\!3,0)$}
\rput(3.5,5.35){\scriptsize$(p'\!,0)$}
\rput(3.5,3.94){\scriptsize$(p'\!-\!2,1)$}
\end{pspicture}
$
\caption{
The restricted set of $s\ell(3)$ weights for $\lambda = \lambda_{p,p'}$ contains the nodes in the shaded region. The lowest row and leftmost diagonal correspond to transfer tangles $\Tb^{-1,n}$ and $\Tb^{m,-1}$ that are zero tangles. The four terms entering the fusion closure relation \eqref{eq:cloFH1} are circled. The $(0,0)$ term involving $\Jb$ in the fusion closure relation is obtained by a translation into the elementary domain. The other two terms are obtained by reflections across the red ``critical" lines.
}
\label{fig:folding}
\end{center}
\end{figure}

For $\lambda=\lambda_{p,p'}$, we have the following closure relations:
\begin{subequations}
\label{eq:cloFH}
\begin{alignat}{2}
\Tb^{p'\!,0}_0 &= \Tb^{p'-2,1}_1 -\sigma\, \Tb^{p'-3,0}_2 + f_{-1}\Jb,\label{eq:cloFH1}\\[0.1cm]
\Tb^{0,p'}_0 &= \sigma\, \Tb^{1,p'-2}_0 -\Tb^{0,p'-3}_1 + f_{-1}\Kb,
\end{alignat}
\end{subequations}
where the tangles $\Jb$ and $\Kb$ are independent of $u$:
\begin{subequations}
\begin{alignat}{2}\label{eq:JKbraids}
\kappa\, \Jb &= \Tb^{p'\!,0}_\infty + \Tb^{p'-3,0}_\infty - \Tb^{p'-2,1}_\infty, \qquad &&\kappa = \exp\Big( \tfrac{\ir \pi N}3(3p'-2p)\Big),\\[0.15cm]
\tilde\kappa\, \Kb &= \Tb^{0,p'}_\infty + \Tb^{0,p'-3}_\infty - \Tb^{1,p'-2}_\infty, \qquad &&\tilde \kappa = \exp\Big( \tfrac{\ir \pi N}3(3p'-p)\Big).
\end{alignat}
\end{subequations}
The four terms of the first of the closure relations in \eqref{eq:cloFH} are identified among the $s\ell(3)$ weights in 
\cref{fig:folding}.

On the standard modules, $\Jb$ and $\Kb$ act as scalar multiples of the identity, with eigenvalues
\begin{subequations}
\begin{alignat}{2}
\kappa\, \Jb\big|_{\stan_{N,d,v}} &= 
\omega^{p'} \eE^{\frac{\ir \pi p}3(-a-2d+v)} + \eE^{\frac{\ir \pi p}3(2a+d-2v)} + \omega^{-p'}\eE^{\frac{\ir \pi p}3(-a+d+v)},\\[0.15cm]
\tilde \kappa\, \Kb\big|_{\stan_{N,d,v}} &= \omega^{p'} \eE^{\frac{\ir \pi p}3(a-d-v)} + \eE^{\frac{\ir \pi p}3(-2a-d+2v)} + \omega^{-p'}\eE^{\frac{\ir \pi p}3(a+2d-v)}.
\end{alignat}
\end{subequations}
The proof of \eqref{eq:cloFH} is given in \cref{app:proofclosure}. 
There, we also prove the more general closure relations \eqref{eq:genclo} for $\Tb^{p'+j,k}$ and $\Tb^{k,p'+j}$. More generally, for arbitrary $m,n$, $\Tb^{m,n}$ is expressed as a linear combination of tangles in a restricted set, namely the $\Tb^{j,k}$ with $0 \le j,k \le p'-1$, 
see \cref{fig:folding}.
In many cases, this linear combination is obtained after applying the closure relations multiple times. For example, applying \eqref{eq:genclo} three times, we find
\be
\Tb^{p'\!,p'}_0 =\Tb_2^{p'-2,p'-2}- \sigma^{p'}\Tb^{0,0}_{p'} + \sigma\Jb\, \Tb^{1,p'-2}_0 + \Kb\,\Tb^{p'-2,1}_1 + f_{-1} \Jb \Kb.
\ee

\subsection[Closure of the $Y$-system]{Closure of the $\boldsymbol Y$-system}\label{sec:cloY}

\begin{figure}[h] 
\centering
$
\psset{unit=2.2}
\begin{pspicture}[shift=-0.9](0,-2)(6,2)
\psline[linecolor=blue]{-}(0,0.5)(4,0.5)
\psline[linecolor=blue]{-}(0,-0.5)(4,-0.5)
\psline[linecolor=red]{-}(0,0.5)(0,-0.5)\psline[linecolor=red]{-}(1,0.5)(1,-0.5)\psline[linecolor=red]{-}(2,0.5)(2,-0.5)\psline[linecolor=red]{-}(3,0.5)(3,-0.5)\psline[linecolor=red]{-}(4,0.5)(4,-0.5)
\psline{-}(4,-0.5)(4.75,0)\psline{-}(4,0.5)(4.75,0)\psline{-}(4,-0.5)(6,0)\psline{-}(4,0.5)(6,0)
\psline[linecolor=blue]{-}(4,0.5)(4.5,1.5)\psline[linecolor=blue]{-}(4,0.5)(5.5,1.5)\psline[linecolor=blue]{-}(4,0.5)(6.5,1.5)\psline[linecolor=blue]{-}(4,-0.5)(4.5,-1.5)\psline[linecolor=blue]{-}(4,-0.5)(5.5,-1.5)\psline[linecolor=blue]{-}(4,-0.5)(6.5,-1.5)
\psline{-}(4.75,0)(4.5,1.5)\psline{-}(4.75,0)(5.5,1.5)\psline{-}(4.75,0)(6.5,1.5)\psline{-}(4.75,0)(4.5,-1.5)\psline{-}(4.75,0)(5.5,-1.5)\psline{-}(4.75,0)(6.5,-1.5)
\psline{-}(6,0)(4.5,1.5)\psline{-}(6,0)(5.5,1.5)\psline{-}(6,0)(6.5,1.5)\psline{-}(6,0)(4.5,-1.5)\psline{-}(6,0)(5.5,-1.5)\psline{-}(6,0)(6.5,-1.5)
\psline[linecolor=red]{-}(4.5,1.5)(4.5,1.8)\psline[linestyle=dashed,dash=0.9pt 1pt,linecolor=red]{-}(4.5,1.8)(4.5,1.9)
\psline[linecolor=red]{-}(4.5,1.5)(4.65,1.8)\psline[linestyle=dashed,dash=1pt 1pt,linecolor=red]{-}(4.65,1.8)(4.7,1.9)
\psline[linecolor=red]{-}(4.5,1.5)(4.8,1.8)\psline[linestyle=dashed,dash=1pt 1pt,linecolor=red]{-}(4.8,1.8)(4.9,1.9)
\psline[linecolor=red]{-}(5.5,1.5)(5.35,1.8)\psline[linestyle=dashed,dash=1pt 1pt,linecolor=red]{-}(5.35,1.8)(5.3,1.9)
\psline[linecolor=red]{-}(5.5,1.5)(5.5,1.8)\psline[linestyle=dashed,dash=0.9pt 1pt,linecolor=red]{-}(5.5,1.8)(5.5,1.9)
\psline[linecolor=red]{-}(5.5,1.5)(5.65,1.8)\psline[linestyle=dashed,dash=1pt 1pt,linecolor=red]{-}(5.65,1.8)(5.7,1.9)
\psline[linecolor=red]{-}(6.5,1.5)(6.5,1.8)\psline[linestyle=dashed,dash=0.9pt 1pt,linecolor=red]{-}(6.5,1.8)(6.5,1.9)
\psline[linecolor=red]{-}(6.5,1.5)(6.35,1.8)\psline[linestyle=dashed,dash=1pt 1pt,linecolor=red]{-}(6.35,1.8)(6.3,1.9)
\psline[linecolor=red]{-}(6.5,1.5)(6.2,1.8)\psline[linestyle=dashed,dash=1pt 1pt,linecolor=red]{-}(6.2,1.8)(6.1,1.9)
\psline[linecolor=red]{-}(4.5,-1.5)(4.5,-1.8)\psline[linestyle=dashed,dash=0.9pt 1pt,linecolor=red]{-}(4.5,-1.8)(4.5,-1.9)
\psline[linecolor=red]{-}(4.5,-1.5)(4.65,-1.8)\psline[linestyle=dashed,dash=1pt 1pt,linecolor=red]{-}(4.65,-1.8)(4.7,-1.9)
\psline[linecolor=red]{-}(4.5,-1.5)(4.8,-1.8)\psline[linestyle=dashed,dash=1pt 1pt,linecolor=red]{-}(4.8,-1.8)(4.9,-1.9)
\psline[linecolor=red]{-}(5.5,-1.5)(5.35,-1.8)\psline[linestyle=dashed,dash=1pt 1pt,linecolor=red]{-}(5.35,-1.8)(5.3,-1.9)
\psline[linecolor=red]{-}(5.5,-1.5)(5.5,-1.8)\psline[linestyle=dashed,dash=0.9pt 1pt,linecolor=red]{-}(5.5,-1.8)(5.5,-1.9)
\psline[linecolor=red]{-}(5.5,-1.5)(5.65,-1.8)\psline[linestyle=dashed,dash=1pt 1pt,linecolor=red]{-}(5.65,-1.8)(5.7,-1.9)
\psline[linecolor=red]{-}(6.5,-1.5)(6.5,-1.8)\psline[linestyle=dashed,dash=0.9pt 1pt,linecolor=red]{-}(6.5,-1.8)(6.5,-1.9)
\psline[linecolor=red]{-}(6.5,-1.5)(6.35,-1.8)\psline[linestyle=dashed,dash=1pt 1pt,linecolor=red]{-}(6.35,-1.8)(6.3,-1.9)
\psline[linecolor=red]{-}(6.5,-1.5)(6.2,-1.8)\psline[linestyle=dashed,dash=1pt 1pt,linecolor=red]{-}(6.2,-1.8)(6.1,-1.9)
\pscurve[linewidth=0.075cm,linecolor=blue]{-}(6,0)(6.25,0.06)(6.5,0)(6.25,-0.06)(6,0)
\pscurve[linewidth=0.025cm,linecolor=white]{-}(6,0)(6.25,0.06)(6.5,0)(6.25,-0.06)(6,0)
\pscurve[linewidth=0.075cm,linecolor=blue]{-}(4.75,0)(4.5,0.06)(4.25,0)(4.5,-0.06)(4.75,0)
\pscurve[linewidth=0.025cm,linecolor=white]{-}(4.75,0)(4.5,0.06)(4.25,0)(4.5,-0.06)(4.75,0)
\psline[linewidth=0.075cm,linecolor=blue]{-}(4.75,0)(6,0)
\psline[linewidth=0.025cm,linecolor=white]{-}(4.75,0)(6,0)
\multiput(0,-0.5)(1,0){5}{\pscircle[linewidth=1.5pt,linecolor=black,fillstyle=solid,fillcolor=white](0,0){.07}}
\multiput(0,0.5)(1,0){5}{\pscircle[linewidth=1.5pt,linecolor=black,fillstyle=solid,fillcolor=white](0,0){.07}}
\pscircle[linewidth=1.5pt,linecolor=black,fillstyle=solid,fillcolor=white](4.75,0){.07}
\pscircle[linewidth=1.5pt,linecolor=black,fillstyle=solid,fillcolor=white](6,0){.07}
\pscircle[linewidth=1.5pt,linecolor=black,fillstyle=solid,fillcolor=white](4.5,1.5){.07}
\pscircle[linewidth=1.5pt,linecolor=black,fillstyle=solid,fillcolor=white](5.5,1.5){.07}
\pscircle[linewidth=1.5pt,linecolor=black,fillstyle=solid,fillcolor=white](6.5,1.5){.07}
\pscircle[linewidth=1.5pt,linecolor=black,fillstyle=solid,fillcolor=white](4.5,-1.5){.07}
\pscircle[linewidth=1.5pt,linecolor=black,fillstyle=solid,fillcolor=white](5.5,-1.5){.07}
\pscircle[linewidth=1.5pt,linecolor=black,fillstyle=solid,fillcolor=white](6.5,-1.5){.07}
\rput(0,0.675){\scriptsize$\tba{1}$}\rput(1,0.675){\scriptsize$\tba{2}$}\rput(2,0.675){\scriptsize$\cdots$}\rput(2.95,0.675){\scriptsize$\tba{p'-3}$}\rput(3.85,0.675){\scriptsize$\tba{p'-2}$}
\rput(0,-0.675){\scriptsize$\tbb{1}$}\rput(1,-0.675){\scriptsize$\tbb{2}$}\rput(2,-0.675){\scriptsize$\cdots$}\rput(2.95,-0.675){\scriptsize$\tbb{p'-3}$}\rput(3.85,-0.675){\scriptsize$\tbb{p'-2}$}
\rput(4.25,1.5){\scriptsize$\Xba$}\rput(5.25,1.5){\scriptsize$\Xba$}\rput(6.25,1.5){\scriptsize$\Xba$}
\rput(4.25,-1.55){\scriptsize$\Xbb$}\rput(5.25,-1.55){\scriptsize$\Xbb$}\rput(6.25,-1.55){\scriptsize$\Xbb$}
\rput(4.975,0.07){\scriptsize$\Yb$}
\rput(6.12,0.12){\scriptsize$\Zb$}
\end{pspicture}
$
\caption{An artist's impression of the closed $Y$-system. Blue and red lines respectively indicate contributions to the $Y$-system in the numerator and denominator. Black lines indicate contributions that appear in the denominator in one direction but in the numerator in the other. The edges tying $\Yb$ to $\Zb$, $\Yb$ to itself, and $\Zb$ to itself, are doubled. Each $\Xba$ is tied to all three copies of $\Xbb$ via the outer dashed edges.}
\label{fig:DynkinY}
\end{figure}
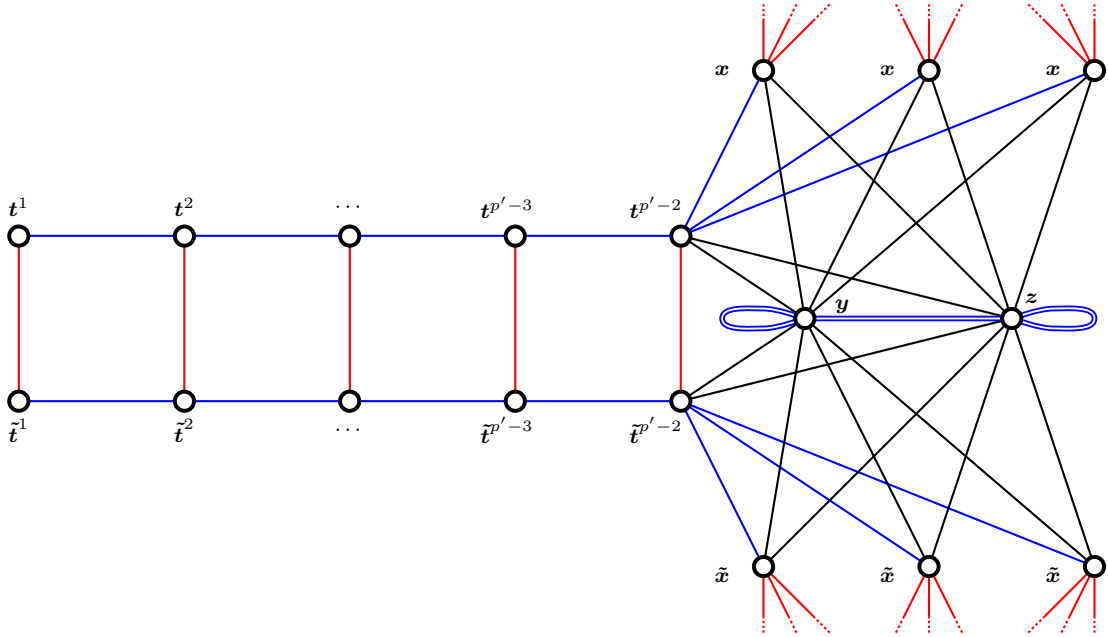

Using the closure relations for the fused transfer tangles for $\lambda=\lambda_{p,p'}$, one finds closure relations for the $Y$-system:
\begin{subequations}
\label{eq:closureY}
\begin{alignat}{2}
\Ib + \tba{p'-1}_0 &= \frac{\displaystyle\bigg(\Ib + \sigma^{p'-p} \Jb \frac{\Tb^{p'-2,0}_1}{\Tb^{0,p'-1}_0}+ \sigma^{p'-p} \Kb \Big(\frac{\Tb^{p'-2,0}_1}{\Tb^{0,p'-1}_0}\Big)^2+ \sigma^{p}\Big(\frac{\Tb^{p'-2,0}_1}{\Tb^{0,p'-1}_0}\Big)^3\bigg)}{\displaystyle\bigg(\Ib - \sigma\frac{\Tb^{p'-2,0}_1\Tb^{0,p'-2}_0}{\Tb^{p'-1,0}_0\Tb^{0,p'-1}_0}\bigg)\bigg(\Ib -\sigma \frac{\Tb^{p'-2,0}_1\Tb^{0,p'-2}_1}{\Tb^{p'-1,0}_1\Tb^{0,p'-1}_0}\bigg)},
\label{eq:cloY1}\\[0.15cm]
\Ib + \tbb{p'-1}_0 &= \frac{\displaystyle\bigg(\Ib + \sigma^{p'+1} \Kb \frac{\Tb^{0,p'-2}_1}{\Tb^{p'-1,0}_1}+ \sigma^{p'} \Jb \Big(\frac{\Tb^{0,p'-2}_1}{\Tb^{p'-1,0}_1}\Big)^2+ \sigma^{p+1}\Big(\frac{\Tb^{0,p'-2}_1}{\Tb^{p'-1,0}_1}\Big)^3\bigg)}{\displaystyle\bigg(\Ib - \sigma\frac{\Tb^{p'-2,0}_2\Tb^{0,p'-2}_1}{\Tb^{p'-1,0}_1\Tb^{0,p'-1}_1}\bigg)\bigg(\Ib -\sigma \frac{\Tb^{p'-2,0}_1\Tb^{0,p'-2}_1}{\Tb^{p'-1,0}_1\Tb^{0,p'-1}_0}\bigg)}.\label{eq:cloY2}
\end{alignat}
\end{subequations}
The proof is given in \cref{app:proofYclosure}.
The numerators in \eqref{eq:cloY1} and \eqref{eq:cloY2} can be factorized by setting
\be
\Jb = \sigma^{p'-p} (\eE^{\ir \Lambdab_1}+\eE^{\ir \Lambdab_2}+\eE^{\ir \Lambdab_3}), \qquad \Kb = \sigma^{p'} (\eE^{-\ir \Lambdab_1}+\eE^{-\ir \Lambdab_2}+\eE^{-\ir \Lambdab_3}).
\ee
On the standard modules, the eigenvalues of $\eE^{\ir \Lambdab_j}$ have simple expressions:
\be
\eE^{\ir \Lambda_1}\big|_{\stan_{N,d,v}} = \omega^{p'} (-1)^{p(N-v-a)}, \qquad \eE^{\ir \Lambda_2}\big|_{\stan_{N,d,v}} = (-1)^{p v}, \qquad \eE^{\ir \Lambda_3}\big|_{\stan_{N,d,v}} = \omega^{-p'}(-1)^{pa}.
\ee
The relations \eqref{eq:closureY} can then be written in factorised form as
\begin{subequations}
\begin{alignat}{2}
\Ib + \tba{p'-1}_0 &= \frac{(\Ib+\eE^{\ir \Lambdab_1}\Xba_0)(\Ib+\eE^{\ir \Lambdab_2}\Xba_0)(\Ib+\eE^{\ir \Lambdab_3}\Xba_0)}{(\Ib+\Yb_0)(\Ib+\Zb_0)},\\[0.15cm]
\Ib + \tbb{p'-1}_0 &= \frac{(\Ib+\eE^{-\ir \Lambdab_1}\Xbb_1)(\Ib+\eE^{-\ir \Lambdab_2}\Xbb_1)(\Ib+\eE^{-\ir \Lambdab_3}\Xbb_1)}{(\Ib+\Yb_1)(\Ib+\Zb_0)},
\end{alignat}
\end{subequations}
where
\be
\Xba_0 = \frac{\Tb^{p'-2,0}_1}{\Tb^{0,p'-1}_0}, \qquad \Xbb_0 = - \frac{\Tb^{0,p'-2}_0}{\Tb^{p'-1,0}_0}, \qquad \Yb_0 = \sigma \Xba_0 \Xbb_0, \qquad \Zb_0 = \sigma \Xba_0 \Xbb_1.
\ee
Using the relations \eqref{eq:Trelations} and \eqref{eq:closureY}, it is straightforward to write down the relations for $\Xba_0\Xba_1$ and $\Xbb_0\Xbb_1$: 
\begin{subequations}
\begin{alignat}{2}
\Xba_0\Xba_1 &=\frac{- \sigma^{p'}(\Ib+ \tba{p'-2}_1)(\Ib+\Yb_1)(\Ib+\Zb_0)}{\big(\Ib+\eE^{-\ir \Lambdab_1}\Xbb_1\big)\big(\Ib+\eE^{-\ir \Lambdab_2}\Xbb_1^{-1}\big)\big(\Ib+\eE^{-\ir \Lambdab_3}\Xbb_1\big)},
\\[.15cm] 
\Xbb_0\Xbb_1 &=\frac{\sigma^{p'-1}(\Ib+ \tbb{p'-2}_0)(\Ib+\Yb_0)(\Ib+\Zb_0)}{\big(\Ib+\eE^{\ir \Lambdab_1}\Xba_0\big)\big(\Ib+\eE^{\ir \Lambdab_2}\Xba_0^{-1}\big)\big(\Ib+\eE^{\ir \Lambdab_3}\Xba_0\big)},
\end{alignat}
\end{subequations}
and likewise for $\Yb_0\Yb_1$ and $\Zb_0\Zb_1$:
\begin{subequations}
\begin{alignat}{2}
\Yb_0\Yb_1 &= \frac{-\sigma (\Ib + \tba{p'-2}_1)(\Ib + \tbb{p'-2}_0)(\Ib+\Yb_0)(\Ib+\Yb_1)(\Ib+\Zb_0)^2}{\big(\Ib+\eE^{\ir \Lambdab_1}\Xba_0\big)\big(\Ib+\eE^{\ir \Lambdab_2}\Xba_0^{-1}\big)\big(\Ib+\eE^{\ir \Lambdab_3}\Xba_0\big)\big(\Ib+\eE^{-\ir \Lambdab_1}\Xbb_1\big)\big(\Ib+\eE^{-\ir \Lambdab_2}\Xbb_1^{-1}\big)\big(\Ib+\eE^{-\ir \Lambdab_3}\Xbb_1\big)},
\\[.15cm] 
\Zb_0\Zb_1 &= \frac{-\sigma (\Ib + \tba{p'-2}_1)(\Ib + \tbb{p'-2}_1)(\Ib+\Yb_1)^2(\Ib+\Zb_0)(\Ib+\Zb_1)}{\big(\Ib+\eE^{\ir \Lambdab_1}\Xba_1\big)\big(\Ib+\eE^{\ir \Lambdab_2}\Xba_1^{-1}\big)\big(\Ib+\eE^{\ir \Lambdab_3}\Xba_1\big)\big(\Ib+\eE^{-\ir \Lambdab_1}\Xbb_1\big)\big(\Ib+\eE^{-\ir \Lambdab_2}\Xbb_1^{-1}\big)\big(\Ib+\eE^{-\ir \Lambdab_3}\Xbb_1\big)}.
\end{alignat}
\end{subequations}
The resulting closed $Y$-system is illustrated in \cref{fig:DynkinY}. It is similar but not identical to the one given by 
Saleur and Wehefritz-Kaufmann \cite{SW00} for the complex $su(3)$ Toda theory.

\section{Conclusion}\label{sec:conclusion}

By now, the CFTs associated with $s\ell(2)$ or $A_1^{(1)}$ models, and the conformal spectra in their various representations, are quite well understood. 
In contrast,
the $A_2^{(1)}$ theories, with their underlying $s\ell(3)$ structures, pose more challenges and their spectra and integrability properties are not so well understood. In this paper, we have derived functional equations in the form of fusion hierarchies, $T$-systems and $Y$-systems for the periodic $A_2^{(1)}$ transfer 
tangles.
For $q=e^{\ir\lambda}$ a root of unity, we have obtained explicit closure relations for these systems.
These functional relations hold in any representation of the algebra $\mathsf{pA}_N(\alpha,\beta)$.

For the $A_2^{(1)}$ vertex and loop models, the root of unity cases represent a countable dense set of points on the continuous critical line. The structure of the closed $Y$-systems reveals a rich $s\ell(3)$ structure which is significantly more complicated than the known $D$-type Dynkin diagram structure of the $s\ell(2)$ models. Interestingly, the precise structure of our equations strongly resembles the structure of the related complex $su(3)$ Toda theory~\cite{SW00}. 
In \cref{app:RSOS}, we argue that the fusion hierarchy relations, $T$-system and $Y$-system equations obtained in \cref{sec:Ftm.fr} also apply for the $A_2^{(1)}$ RSOS model.
In this case, the closure of the fusion hierarchy is simpler and takes the form of truncation relations, which are nevertheless 
compatible with the more general closure relation \eqref{eq:cloFH}.

Our derivations of the functional relations were performed using diagrammatic calculus and apply directly to tangles in the algebra $\mathsf{pA}_N(\alpha,\beta)$. The identities that we used are in fact inspired by similar identities in Kuperberg's $s\ell(3)$ spiders. In particular, the relations \eqref{eq:spider.ids} are the 
analogues 
of the following spider relations:
\be
\begin{pspicture}[shift=-1.1](0.5,-1.2)(1.5,1.2)
\psline[linewidth=0.75pt,arrowsize=0.125cm]{->}(1,0.5)(1,1)
\psline[linewidth=0.75pt,arrowsize=0.125cm]{-}(1,0.5)(1,1.25)
\psline[linewidth=0.75pt,arrowsize=0.125cm]{->}(1,-1.25)(1,-0.75)
\psline[linewidth=0.75pt,arrowsize=0.125cm]{-}(1,-1.25)(1,-0.5)
\psbezier[linewidth=0.75pt,arrowsize=0.125cm]{-}(1,-0.5)(1.5,-0.5)(1.5,0.5)(1,0.5)
\psbezier[linewidth=0.75pt,arrowsize=0.125cm]{-}(1,-0.5)(0.5,-0.5)(0.5,0.5)(1,0.5)
\psline[linewidth=0.75pt,arrowsize=0.125cm]{<-}(1.362,-0.15)(1.362,0.15)
\psline[linewidth=0.75pt,arrowsize=0.125cm]{<-}(0.638,-0.15)(0.638,0.15)
\end{pspicture} \ = [2] \ \
\begin{pspicture}[shift=-0.65](0.2,0)(0.75,1.3)
\psline[linewidth=0.75pt,arrowsize=0.125cm]{->}(0.375,0)(0.375,0.95)
\psline[linewidth=0.75pt]{-}(0.375,0)(0.375,1.5)
\end{pspicture} ,\qquad\qquad
\begin{pspicture}[shift=-.90](0,-1)(2,1)
\psline[linewidth=0.75pt,arrowsize=0.125cm]{<-}(0.427,-0.151)(0.427,-0.15)
\psbezier[linewidth=0.75pt]{-}(0.5,-0.5)(0.4,-0.5)(0.4,0.5)(0.5,0.5)
\psline[linewidth=0.75pt,arrowsize=0.125cm]{<-}(1.572,0.151)(1.572,0.15)
\psline[linewidth=0.75pt,arrowsize=0.125cm]{<-}(1.151,0.572)(1.15,0.572)
\psbezier[linewidth=0.75pt]{-}(1.5,-0.5)(1.6,-0.5)(1.6,0.5)(1.5,0.5)
\psbezier[linewidth=0.75pt]{-}(0.5,-0.5)(0.5,-0.6)(1.5,-0.6)(1.5,-0.5)
\psline[linewidth=0.75pt,arrowsize=0.125cm]{<-}(0.849,-0.572)(0.85,-0.572)
\psbezier[linewidth=0.75pt]{-}(0.5,0.5)(0.5,0.6)(1.5,0.6)(1.5,0.5)
\psline[linewidth=0.75pt,arrowsize=0.125cm]{-}(0.5,0.5)(0,1)
\psline[linewidth=0.75pt,arrowsize=0.125cm]{->}(0.5,0.5)(0.15,0.85)
\psline[linewidth=0.75pt,arrowsize=0.125cm]{-}(0.5,-0.5)(0,-1)
\psline[linewidth=0.75pt,arrowsize=0.125cm]{->}(0,-1)(0.4,-0.6)
\psline[linewidth=0.75pt,arrowsize=0.125cm]{->}(2,1)(1.6,0.6)
\psline[linewidth=0.75pt,arrowsize=0.125cm]{-}(1.5,0.5)(2,1)
\psline[linewidth=0.75pt,arrowsize=0.125cm]{-}(1.5,-0.5)(2,-1)
\psline[linewidth=0.75pt,arrowsize=0.125cm]{->}(1.5,-0.5)(1.85,-0.85)
\end{pspicture} \ \ = \ \ 
\begin{pspicture}[shift=-.90](0.5,-1)(1.5,1)
\psbezier[linewidth=0.75pt]{-}(0.5,-0.5)(0.7,-0.5)(0.7,0.5)(0.5,0.5)
\psline[linewidth=0.75pt,arrowsize=0.125cm]{->}(0.65,0.15)(0.65,0.16)
\psline[linewidth=0.75pt,arrowsize=0.125cm]{->}(1.35,-0.15)(1.35,-0.16)
\psbezier[linewidth=0.75pt]{-}(1.5,-0.5)(1.3,-0.5)(1.3,0.5)(1.5,0.5)
\end{pspicture} \ \ + \ \ 
\begin{pspicture}[shift=-.90](0.5,-1)(1.5,1)
\psbezier[linewidth=0.75pt]{-}(0.5,0.5)(0.5,0.3)(1.5,0.3)(1.5,0.5)
\psline[linewidth=0.75pt,arrowsize=0.125cm]{->}(0.85,0.35)(0.849,0.35)
\psbezier[linewidth=0.75pt]{-}(0.5,-0.5)(0.5,-0.3)(1.5,-0.3)(1.5,-0.5)
\psline[linewidth=0.75pt,arrowsize=0.125cm]{->}(1.15,-0.35)(1.151,-0.35)
\end{pspicture}\ \ .
\ee
The connection between the representation theory of $\mathsf{pA}_N(\alpha,\beta)$ and the $s\ell(3)$ spider algebra still remains to be understood.

Following the derivation of the functional relations, a natural next step is to convert the universal $Y$-systems into TBA equations. In principle, these integral equations can be solved in the continuum scaling limit for the finite-size corrections and conformal data, 
including the central charges, conformal weights and finitized characters. This looks formidable in the general case but should be manageable at least in some prototypical examples. We plan to pursue this analysis in the near future.

The $Y$-systems are universal~\cite{CMP2001} so they must apply to all boundary conditions and all topologies. It is therefore of interest to obtain the same $Y$-systems for double-row 
transfer matrices on the strip with various integrable boundary conditions. These boundary conditions are expected to be conjugate to the scaling operators in the theory and will depend on the representation of the $A_2^{(1)}$ model under consideration.

It is expected that there will be similarities but also some subtle differences in the $s\ell(3)$ structures of the $A_2^{(1)}$ and $A_2^{(2)}$ models, in particular at roots of unity. 
The $A_2^{(2)}$ vertex model is the Izergin-Korepin 19-vertex model~\cite{IK81}. But perhaps most importantly, the $A_2^{(2)}$ RSOS models include the Ising model in a magnetic field~\cite{WNS92,WPSN94} and the logarithmic $A_2^{(2)}$ loop models include site percolation on the triangular lattice~\cite{NienhuisGiens}. It would therefore be of much interest to extend the current derivation of the $T$- and $Y$-systems and their closure relations to the $A_2^{(2)}$ models.

\goodbreak

\section*{Acknowledgments} 

AMD is supported by the FNRS fellowship CR28075116. 
JR is supported by the Australian Research Council under the Discovery Project scheme, project number DP160101376.
AMD and PAP acknowledge the hospitality of the University of Queensland where part of this work was done.
AMD thanks Azat Gainutdinov, David Ridout, Yvan Saint-Aubin and Paul Zinn-Justin for useful discussions.

\goodbreak
\appendix

\section{Conservation of vacancies}\label{app:cov}

In this appendix, we show that the transfer tangle $\Tb^{1,0}(u)$ is an element of $\mathsf {pA}_N(\alpha,\beta)$.
The proof for $\Tb^{0,1}(u)$ uses the same ideas. To start, it is obvious from the definition \eqref{Tu01} that $\Tb^{1,0}(u)$ is an element of $\mathsf{pdTL}_N(\alpha,\beta)$. To prove that $\Tb^{1,0}(u) \in \mathsf {pA}_N(\alpha,\beta)$,
we show that it preserves the number of vacancies. It is easy to see that this is true for $N=1$. For $N>1$, our proof is 
by induction on $N$.

Let us make the following remark: 
An element $c\in \mathsf{pdTL}_N(\alpha,\beta)$ is vacancy-preserving if and only if the element $\tilde c\in \mathsf{pdTL}_{N+k}(\alpha,\beta)$, obtained by adding $2k$ vacant sites to $c$ 
in the $k$ rightmost positions of the top and bottom edges,
is vacancy-preserving. In terms of diagrams, this means that
\be 
\begin{pspicture}[shift=-0.5](0,-0.2)(1.0,1.0)
\pspolygon[fillstyle=solid,fillcolor=lightlightblue,linecolor=black,linewidth=0pt](0,0)(0,0.8)(1.0,0.8)(1.0,0)(0,0)
\rput(0.5,0.4){$c$}
\end{pspicture}
\ \ \in \mathsf {pA}_N(\alpha,\beta)
\qquad \longleftrightarrow\qquad
\begin{pspicture}[shift=-0.5](0,-0.2)(1.8,1.0)
\pspolygon[fillstyle=solid,fillcolor=lightlightblue,linecolor=black,linewidth=0pt](0,0)(0,0.8)(1.8,0.8)(1.8,0)(0,0)
\psline[linewidth=0.5pt,linestyle=dashed,dash=2pt 2pt]{-}(1.0,0)(1.0,0.8)
\pscircle[fillstyle=solid,fillcolor=black](1.1,0){0.04}\pscircle[fillstyle=solid,fillcolor=black](1.1,0.8){0.04}
\pscircle[fillstyle=solid,fillcolor=black](1.3,0){0.04}\pscircle[fillstyle=solid,fillcolor=black](1.3,0.8){0.04}
\pscircle[fillstyle=solid,fillcolor=black](1.5,0){0.04}\pscircle[fillstyle=solid,fillcolor=black](1.5,0.8){0.04}
\pscircle[fillstyle=solid,fillcolor=black](1.7,0){0.04}\pscircle[fillstyle=solid,fillcolor=black](1.7,0.8){0.04}
\psline[linewidth=1.5pt,linecolor=blue](1.0,0.2)(1.8,0.2)
\psline[linewidth=1.5pt,linecolor=blue](1.0,0.6)(1.8,0.6)
\rput(1.4,0.32){.}
\rput(1.4,0.4){.}
\rput(1.4,0.48){.}
\rput(0.5,0.4){$c$}
\end{pspicture}
\ \
 \in \mathsf {pA}_{N+k}(\alpha,\beta),
\ee
where the number of horizontal loop segments in the rightmost diagram is equal to
the number of loop segments in $c$ that travel via the back of the cylinder.

For $N>1$, we expand the leftmost face operator in $\Tb^{1,0}(u)$ and obtain seven terms:
\begin{alignat}{2}
&\Tb^{1,0} (u)= s_1(-u) \bigg(
\psset{unit=0.7}
\begin{pspicture}[shift=-0.4](-0.3,0)(4.3,1.0)
\facegrid{(0,0)}{(4,1)}
\psarc[linewidth=0.025]{-}(1,0){0.16}{0}{90}
\psarc[linewidth=0.025]{-}(3,0){0.16}{0}{90}
\psline[linewidth=1.5pt,linecolor=blue,linestyle=dashed,dash=2pt 2pt]{-}(0,0.5)(-0.3,0.5)
\psline[linewidth=1.5pt,linecolor=blue,linestyle=dashed,dash=2pt 2pt]{-}(4,0.5)(4.3,0.5)
\pscircle[fillstyle=solid,fillcolor=black](0,0.5){0.06}
\pscircle[fillstyle=solid,fillcolor=black](0.5,0){0.06}
\pscircle[fillstyle=solid,fillcolor=black](1,0.5){0.06}
\pscircle[fillstyle=solid,fillcolor=black](0.5,1){0.06}
\rput(2.5,0.5){$\ldots$}
\rput(1.5,.5){$u$}
\rput(3.5,.5){$u$}
\end{pspicture}\ + \
\begin{pspicture}[shift=-0.4](-0.3,0)(4.3,1.0)
\facegrid{(0,0)}{(4,1)}
\psarc[linewidth=0.025]{-}(1,0){0.16}{0}{90}
\psarc[linewidth=0.025]{-}(3,0){0.16}{0}{90}
\psline[linewidth=1.5pt,linecolor=blue,linestyle=dashed,dash=2pt 2pt]{-}(0,0.5)(-0.3,0.5)
\psline[linewidth=1.5pt,linecolor=blue,linestyle=dashed,dash=2pt 2pt]{-}(4,0.5)(4.3,0.5)
\rput(2.5,0.5){$\ldots$}
\rput(0,0){\looph}
\rput(1.5,.5){$u$}
\rput(3.5,.5){$u$}
\end{pspicture}
\bigg)\ + \
\begin{pspicture}[shift=-0.4](-0.3,0)(4.3,1.0)
\facegrid{(0,0)}{(4,1)}
\psarc[linewidth=0.025]{-}(1,0){0.16}{0}{90}
\psarc[linewidth=0.025]{-}(3,0){0.16}{0}{90}
\psline[linewidth=1.5pt,linecolor=blue,linestyle=dashed,dash=2pt 2pt]{-}(0,0.5)(-0.3,0.5)
\psline[linewidth=1.5pt,linecolor=blue,linestyle=dashed,dash=2pt 2pt]{-}(4,0.5)(4.3,0.5)
\rput(0,0){\loopb}
\pscircle[fillstyle=solid,fillcolor=black](1,0.5){0.06}
\pscircle[fillstyle=solid,fillcolor=black](0.5,0){0.06}
\rput(2.5,0.5){$\ldots$}
\rput(1.5,.5){$u$}
\rput(3.5,.5){$u$}
\end{pspicture}
\\[0.2cm]&\psset{unit=0.7}
+ \
\begin{pspicture}[shift=-0.4](-0.3,0)(4.3,1.0)
\facegrid{(0,0)}{(4,1)}
\psarc[linewidth=0.025]{-}(1,0){0.16}{0}{90}
\psarc[linewidth=0.025]{-}(3,0){0.16}{0}{90}
\psline[linewidth=1.5pt,linecolor=blue,linestyle=dashed,dash=2pt 2pt]{-}(0,0.5)(-0.3,0.5)
\psline[linewidth=1.5pt,linecolor=blue,linestyle=dashed,dash=2pt 2pt]{-}(4,0.5)(4.3,0.5)
\rput(0,0){\loopc}
\pscircle[fillstyle=solid,fillcolor=black](0,0.5){0.06}
\pscircle[fillstyle=solid,fillcolor=black](0.5,1){0.06}
\rput(2.5,0.5){$\ldots$}
\rput(1.5,.5){$u$}
\rput(3.5,.5){$u$}
\end{pspicture}\  + s_0(u)
\bigg( 
\begin{pspicture}[shift=-0.4](-0.3,0)(4.3,1.0)
\facegrid{(0,0)}{(4,1)}
\psarc[linewidth=0.025]{-}(1,0){0.16}{0}{90}
\psarc[linewidth=0.025]{-}(3,0){0.16}{0}{90}
\psline[linewidth=1.5pt,linecolor=blue,linestyle=dashed,dash=2pt 2pt]{-}(0,0.5)(-0.3,0.5)
\psline[linewidth=1.5pt,linecolor=blue,linestyle=dashed,dash=2pt 2pt]{-}(4,0.5)(4.3,0.5)
\rput(0,0){\loopf}
\pscircle[fillstyle=solid,fillcolor=black](0,0.5){0.06}
\pscircle[fillstyle=solid,fillcolor=black](1,0.5){0.06}
\rput(2.5,0.5){$\ldots$}
\rput(1.5,.5){$u$}
\rput(3.5,.5){$u$}
\end{pspicture}\ +\ 
\begin{pspicture}[shift=-0.4](-0.3,0)(4.3,1.0)
\facegrid{(0,0)}{(4,1)}
\psarc[linewidth=0.025]{-}(1,0){0.16}{0}{90}
\psarc[linewidth=0.025]{-}(3,0){0.16}{0}{90}
\psline[linewidth=1.5pt,linecolor=blue,linestyle=dashed,dash=2pt 2pt]{-}(0,0.5)(-0.3,0.5)
\psline[linewidth=1.5pt,linecolor=blue,linestyle=dashed,dash=2pt 2pt]{-}(4,0.5)(4.3,0.5)
\pscircle[fillstyle=solid,fillcolor=black](0.5,0){0.06}
\pscircle[fillstyle=solid,fillcolor=black](0.5,1){0.06}
\rput(2.5,0.5){$\ldots$}
\rput(0,0){\loopg}
\rput(1.5,.5){$u$}
\rput(3.5,.5){$u$}
\end{pspicture}\ +\ 
\begin{pspicture}[shift=-0.4](-0.3,0)(4.3,1.0)
\facegrid{(0,0)}{(4,1)}
\psarc[linewidth=0.025]{-}(1,0){0.16}{0}{90}
\psarc[linewidth=0.025]{-}(3,0){0.16}{0}{90}
\psline[linewidth=1.5pt,linecolor=blue,linestyle=dashed,dash=2pt 2pt]{-}(0,0.5)(-0.3,0.5)
\psline[linewidth=1.5pt,linecolor=blue,linestyle=dashed,dash=2pt 2pt]{-}(4,0.5)(4.3,0.5)
\rput(2.5,0.5){$\ldots$}
\rput(0,0){\loopi}
\rput(1.5,.5){$u$}
\rput(3.5,.5){$u$}
\end{pspicture}
\bigg).\nonumber
\end{alignat}
We denote the corresponding diagrams by $b_1, \dots, b_7$ and proceed to show that each one is vacancy-preserving. The diagrams $b_1, \dots, b_4$ can be deformed as follows:
\be\label{eq:b1234}
b_1 = \ 
\begin{pspicture}[shift=-1.9](0,-1)(2,3)
\multiput(0,0)(0.5,0){4}{\psline[linewidth=1.5pt,linecolor=blue,linestyle=dashed,dash=2pt 2pt](0.25,0)(0.25,2)}
\multiput(0,0)(0.5,0.5){3}{\pspolygon[fillstyle=solid,fillcolor=lightlightblue,linecolor=black,linewidth=1pt](0,0.5)(0.5,0)(1,0.5)(0.5,1)\psarc[linewidth=0.015]{-}(0.5,0){0.12}{45}{135}\rput(0.5,0.5){$u$}}
\pspolygon[fillstyle=solid,fillcolor=lightlightblue,linecolor=black,linewidth=1pt](0,2)(0,2.8)(2,2.8)(2,2)
\pspolygon[fillstyle=solid,fillcolor=lightlightblue,linecolor=black,linewidth=1pt](0,0)(0,-0.8)(2,-0.8)(2,0)
\multiput(0,0)(0.5,0){3}{\psbezier[linewidth=1.5pt,linecolor=blue,linestyle=dashed,dash=2pt 2pt](0.25,2)(0.25,2.4)(0.75,2.4)(0.75,2.8)
\psline[linewidth=1.5pt,linecolor=blue,linestyle=dashed,dash=2pt 2pt](0.75,0)(0.75,-0.8)}
\pscircle[fillstyle=solid,fillcolor=black](0.25,-.8){0.05}
\pscircle[fillstyle=solid,fillcolor=black](0.25,0){0.05}
\pscircle[fillstyle=solid,fillcolor=black](1.75,2.0){0.05}
\pscircle[fillstyle=solid,fillcolor=black](0.25,2.8){0.05}
\end{pspicture} \qquad
b_2 = \ 
\begin{pspicture}[shift=-1.9](0,-1)(2,3)
\multiput(0,0)(0.5,0){4}{\psline[linewidth=1.5pt,linecolor=blue,linestyle=dashed,dash=2pt 2pt](0.25,0)(0.25,2)}
\multiput(0,0)(0.5,0.5){3}{\pspolygon[fillstyle=solid,fillcolor=lightlightblue,linecolor=black,linewidth=1pt](0,0.5)(0.5,0)(1,0.5)(0.5,1)\psarc[linewidth=0.015]{-}(0.5,0){0.12}{45}{135}\rput(0.5,0.5){$u$}}
\pspolygon[fillstyle=solid,fillcolor=lightlightblue,linecolor=black,linewidth=1pt](0,2)(0,2.8)(2,2.8)(2,2)
\pspolygon[fillstyle=solid,fillcolor=lightlightblue,linecolor=black,linewidth=1pt](0,0)(0,-0.8)(2,-0.8)(2,0)
\multiput(0,0)(0.5,0){3}{\psbezier[linewidth=1.5pt,linecolor=blue,linestyle=dashed,dash=2pt 2pt](0.25,2)(0.25,2.4)(0.75,2.4)(0.75,2.8)
\psline[linewidth=1.5pt,linecolor=blue,linestyle=dashed,dash=2pt 2pt](0.75,0)(0.75,-0.8)}
\psline[linewidth=1.5pt,linecolor=blue](0.25,0)(0.25,-0.8)
\psbezier[linewidth=1.5pt,linecolor=blue](-0.25,2)(-0.25,2.4)(0.25,2.4)(0.25,2.8)
\psbezier[linewidth=1.5pt,linecolor=blue](1.75,2)(1.75,2.4)(2.25,2.4)(2.25,2.8)
\psframe[fillstyle=solid,linecolor=white,linewidth=0pt](-1,2)(-0.02,2.8)\psframe[fillstyle=solid,linecolor=white,linewidth=0pt](2.02,2)(3,2.8)
\end{pspicture} \qquad
b_3 = \ 
\begin{pspicture}[shift=-1.9](0,-1)(2,3)
\multiput(0,0)(0.5,0){4}{\psline[linewidth=1.5pt,linecolor=blue,linestyle=dashed,dash=2pt 2pt](0.25,0)(0.25,2)}
\multiput(0,0)(0.5,0.5){3}{\pspolygon[fillstyle=solid,fillcolor=lightlightblue,linecolor=black,linewidth=1pt](0,0.5)(0.5,0)(1,0.5)(0.5,1)\psarc[linewidth=0.015]{-}(0.5,0){0.12}{45}{135}\rput(0.5,0.5){$u$}}
\pspolygon[fillstyle=solid,fillcolor=lightlightblue,linecolor=black,linewidth=1pt](0,2)(0,2.8)(2,2.8)(2,2)
\pspolygon[fillstyle=solid,fillcolor=lightlightblue,linecolor=black,linewidth=1pt](0,0)(0,-0.8)(2,-0.8)(2,0)
\multiput(0,0)(0.5,0){3}{\psbezier[linewidth=1.5pt,linecolor=blue,linestyle=dashed,dash=2pt 2pt](0.25,2)(0.25,2.4)(0.75,2.4)(0.75,2.8)
\psline[linewidth=1.5pt,linecolor=blue,linestyle=dashed,dash=2pt 2pt](0.75,0)(0.75,-0.8)}
\pscircle[fillstyle=solid,fillcolor=black](0.25,-.8){0.05}
\pscircle[fillstyle=solid,fillcolor=black](0.25,0){0.05}
\psbezier[linewidth=1.5pt,linecolor=blue](-0.25,2)(-0.25,2.4)(0.25,2.4)(0.25,2.8)
\psbezier[linewidth=1.5pt,linecolor=blue](1.75,2)(1.75,2.4)(2.25,2.4)(2.25,2.8)
\psframe[fillstyle=solid,linecolor=white,linewidth=0pt](-1,2)(-0.02,2.8)\psframe[fillstyle=solid,linecolor=white,linewidth=0pt](2.02,2)(3,2.8)
\end{pspicture} \qquad
b_4 = \ 
\begin{pspicture}[shift=-1.9](0,-1)(2,3)
\multiput(0,0)(0.5,0){4}{\psline[linewidth=1.5pt,linecolor=blue,linestyle=dashed,dash=2pt 2pt](0.25,0)(0.25,2)}
\multiput(0,0)(0.5,0.5){3}{\pspolygon[fillstyle=solid,fillcolor=lightlightblue,linecolor=black,linewidth=1pt](0,0.5)(0.5,0)(1,0.5)(0.5,1)\psarc[linewidth=0.015]{-}(0.5,0){0.12}{45}{135}\rput(0.5,0.5){$u$}}
\pspolygon[fillstyle=solid,fillcolor=lightlightblue,linecolor=black,linewidth=1pt](0,2)(0,2.8)(2,2.8)(2,2)
\pspolygon[fillstyle=solid,fillcolor=lightlightblue,linecolor=black,linewidth=1pt](0,0)(0,-0.8)(2,-0.8)(2,0)
\multiput(0,0)(0.5,0){3}{\psbezier[linewidth=1.5pt,linecolor=blue,linestyle=dashed,dash=2pt 2pt](0.25,2)(0.25,2.4)(0.75,2.4)(0.75,2.8)
\psline[linewidth=1.5pt,linecolor=blue,linestyle=dashed,dash=2pt 2pt](0.75,0)(0.75,-0.8)}
\psline[linewidth=1.5pt,linecolor=blue](0.25,0)(0.25,-0.8)
\pscircle[fillstyle=solid,fillcolor=black](1.75,2.0){0.05}
\pscircle[fillstyle=solid,fillcolor=black](0.25,2.8){0.05}
\end{pspicture}\ \ . 
\ee
Recalling that $\psset{unit=0.3}\begin{pspicture}[shift=-0.8](0,0)(2,2)
\pspolygon[fillstyle=solid,fillcolor=lightlightblue](1,0)(0,1)(1,2)(2,1)
\rput(1,1){\scriptsize$u$}
\psarc[linewidth=0.025]{-}(1,0){0.28}{45}{135}
\end{pspicture}\ $is vacancy-preserving, we see that each diagram in \eqref{eq:b1234} is the product of three elements in $\mathsf{pA}_{N}(\alpha,\beta)$, thus implying that $b_1, b_2, b_3, b_4 \in \mathsf {pA}_{N}(\alpha, \beta)$.
For $b_5$, we use the 
remark
above and embed the diagram in $\mathsf{pdTL}_{N+1}(\alpha,\beta)$:
\be
\psset{unit=0.7cm}
\tilde b_5 = \ \ 
\begin{pspicture}[shift=-0.4](0,0)(5,1.0)
\facegrid{(0,0)}{(5,1)}
\psarc[linewidth=0.025]{-}(1,0){0.16}{0}{90}
\psarc[linewidth=0.025]{-}(3,0){0.16}{0}{90}
\rput(0,0){\loopf}
\pscircle[fillstyle=solid,fillcolor=black](0,0.5){0.06}
\pscircle[fillstyle=solid,fillcolor=black](1,0.5){0.06}
\pscircle[fillstyle=solid,fillcolor=black](4,0.5){0.06}
\pscircle[fillstyle=solid,fillcolor=black](4.5,0){0.06}
\pscircle[fillstyle=solid,fillcolor=black](4.5,1){0.06}
\pscircle[fillstyle=solid,fillcolor=black](5,0.5){0.06}
\rput(2.5,0.5){$\ldots$}
\rput(1.5,.5){$u$}
\rput(3.5,.5){$u$}
\end{pspicture} \ \ = \psset{unit=1cm}\ \ 
\begin{pspicture}[shift=-1.9](-0.5,-1)(2,3)
\multiput(0,0)(0.5,0){4}{\psline[linewidth=1.5pt,linecolor=blue,linestyle=dashed,dash=2pt 2pt](0.25,0)(0.25,2)}
\multiput(0,0)(0.5,0.5){3}{\pspolygon[fillstyle=solid,fillcolor=lightlightblue,linecolor=black,linewidth=1pt](0,0.5)(0.5,0)(1,0.5)(0.5,1)\psarc[linewidth=0.015]{-}(0.5,0){0.12}{45}{135}\rput(0.5,0.5){$u$}}
\pspolygon[fillstyle=solid,fillcolor=lightlightblue,linecolor=black,linewidth=1pt](-0.5,2)(-0.5,2.8)(2,2.8)(2,2)
\pspolygon[fillstyle=solid,fillcolor=lightlightblue,linecolor=black,linewidth=1pt](-0.5,0)(-0.5,-0.8)(2,-0.8)(2,0)
\psline[linewidth=1.5pt,linecolor=blue](-0.25,-0.8)(-0.25,2.8)
\multiput(0,0)(0.5,0){3}{\psbezier[linewidth=1.5pt,linecolor=blue,linestyle=dashed,dash=2pt 2pt](0.25,-0.8)(0.25,-0.4)(0.75,-0.4)(0.75,0)}
\multiput(0,0)(0.5,0){3}{\psline[linewidth=1.5pt,linecolor=blue,linestyle=dashed,dash=2pt 2pt](0.25,2)(0.25,2.8)}
\pscircle[fillstyle=solid,fillcolor=black](0.25,0){0.05}
\pscircle[fillstyle=solid,fillcolor=black](1.75,-0.8){0.05}
\pscircle[fillstyle=solid,fillcolor=black](1.75,2.8){0.05}
\pscircle[fillstyle=solid,fillcolor=black](1.75,2.0){0.05}
\end{pspicture}\ \ .
\ee
We readily see that $\tilde b_5$ is an element of $\mathsf {pA}_{N+1}(\alpha, \beta)$,
from which we infer that $b_5\in \mathsf {pA}_{N}(\alpha, \beta)$
as well. The term $b_6$ can be expressed as
\be
\psset{unit=0.7cm}
b_6 = -b_1 + \ 
\begin{pspicture}[shift=-0.4](-0.3,0)(4.3,1.0)
\facegrid{(0,0)}{(4,1)}
\psarc[linewidth=0.025]{-}(1,0){0.16}{0}{90}
\psarc[linewidth=0.025]{-}(3,0){0.16}{0}{90}
\psline[linewidth=1.5pt,linecolor=blue,linestyle=dashed,dash=2pt 2pt]{-}(1,0.5)(-0.3,0.5)
\psline[linewidth=1.5pt,linecolor=blue,linestyle=dashed,dash=2pt 2pt]{-}(4,0.5)(4.3,0.5)
\pscircle[fillstyle=solid,fillcolor=black](0.5,0){0.06}
\pscircle[fillstyle=solid,fillcolor=black](0.5,1){0.06}
\rput(2.5,0.5){$\ldots$}
\rput(1.5,.5){$u$}
\rput(3.5,.5){$u$}
\end{pspicture}\ .
\label{b6}
\ee
We have already seen that $b_1\in\mathsf {pA}_{N}(\alpha, \beta)$,
while the second term in \eqref{b6} is nothing but the transfer tangle $\Tb^{1,0}(u)$ on $N-1$ nodes. By the induction hypothesis, this also preserves the number of vacancies, so $b_6 \in \mathsf {pA}_{N}(\alpha, \beta)$. 
For $b_7$, we embed the diagram in $\mathsf {pA}_{N+2}(\alpha, \beta)$
and find
\be
\psset{unit=0.7cm}
\tilde b_7 = \ \ 
\begin{pspicture}[shift=-0.4](0,0)(6,1.0)
\facegrid{(0,0)}{(6,1)}
\psarc[linewidth=0.025]{-}(1,0){0.16}{0}{90}
\psarc[linewidth=0.025]{-}(3,0){0.16}{0}{90}
\rput(0,0){\loopi}
\pscircle[fillstyle=solid,fillcolor=black](4.5,0){0.06}
\pscircle[fillstyle=solid,fillcolor=black](4.5,1){0.06}
\pscircle[fillstyle=solid,fillcolor=black](5.5,0){0.06}
\pscircle[fillstyle=solid,fillcolor=black](5.5,1){0.06}
\psline[linewidth=1.5pt,linecolor=blue](4,0.5)(6,0.5)
\rput(2.5,0.5){$\ldots$}
\rput(1.5,.5){$u$}
\rput(3.5,.5){$u$}
\end{pspicture} \ \ = \psset{unit=1cm}\ \ 
\begin{pspicture}[shift=-1.9](-0.5,-1)(2.5,3)
\multiput(0,0)(0.5,0){4}{\psline[linewidth=1.5pt,linecolor=blue,linestyle=dashed,dash=2pt 2pt](0.25,0)(0.25,2)}
\multiput(0,0)(0.5,0.5){3}{\pspolygon[fillstyle=solid,fillcolor=lightlightblue,linecolor=black,linewidth=1pt](0,0.5)(0.5,0)(1,0.5)(0.5,1)\psarc[linewidth=0.015]{-}(0.5,0){0.12}{45}{135}\rput(0.5,0.5){$u$}}
\pspolygon[fillstyle=solid,fillcolor=lightlightblue,linecolor=black,linewidth=1pt](-0.5,2)(-0.5,2.8)(2.5,2.8)(2.5,2)
\pspolygon[fillstyle=solid,fillcolor=lightlightblue,linecolor=black,linewidth=1pt](-0.5,0)(-0.5,-0.8)(2.5,-0.8)(2.5,0)
\psline[linewidth=1.5pt,linecolor=blue](-0.25,0)(-0.25,2.8)
\psline[linewidth=1.5pt,linecolor=blue](2.25,0)(2.25,2)
\psarc[linewidth=1.5pt,linecolor=blue](0,0){.25}{180}{0}
\psarc[linewidth=1.5pt,linecolor=blue](2,2){.25}{0}{180}
\multiput(0,0)(0.5,0){3}{\psbezier[linewidth=1.5pt,linecolor=blue,linestyle=dashed,dash=2pt 2pt](0.25,-0.8)(0.25,-0.4)(0.75,-0.4)(0.75,0)}
\multiput(0,0)(0.5,0){3}{\psline[linewidth=1.5pt,linecolor=blue,linestyle=dashed,dash=2pt 2pt](0.25,2)(0.25,2.8)}
\pscircle[fillstyle=solid,fillcolor=black](1.75,-0.8){0.05}
\pscircle[fillstyle=solid,fillcolor=black](2.25,-0.8){0.05}
\pscircle[fillstyle=solid,fillcolor=black](1.75,2.8){0.05}
\pscircle[fillstyle=solid,fillcolor=black](2.25,2.8){0.05}
\psbezier[linewidth=1.5pt,linecolor=blue](-0.25,-0.8)(-0.25,-0.4)(-0.75,-0.4)(-0.75,0)
\psbezier[linewidth=1.5pt,linecolor=blue](2.75,-0.8)(2.75,-0.4)(2.25,-0.4)(2.25,0)
\psframe[fillstyle=solid,linecolor=white,linewidth=0pt](-1,0)(-0.52,-0.8)\psframe[fillstyle=solid,linecolor=white,linewidth=0pt](2.52,0)(3,-0.8)
\end{pspicture}\ \ .
\ee
We have thus written $\tilde b_7$ as a product of three tangles. The top one decreases the number of vacancies by two, the middle one is vacancy-preserving, and the bottom one increases the number of vacancies by 
two. Overall, $\tilde b_7$ is therefore vacancy-preserving,
implying that $b_7 \in \mathsf {pA}_N(\alpha, \beta)$.
This concludes the proof that $\Tb^{1,0}(u) \in \mathsf {pA}_N(\alpha, \beta)$.

\section{Proofs of functional relations}\label{app:proofs}

\subsection{Fusion hierarchy relations}\label{app:proofFH}

In this subsection, we prove the fusion hierarchy relations \eqref{eq:FH} in the planar algebra. For convenience, we draw the diagrams smaller than in the rest of the paper and remove some of the information that can be easily 
deduced. One property of the fused face operators that is used repeatedly is
\be
\psset{unit=0.5}
\ \ .
\ee
It follows from the recursive definition \eqref{eq:Pmnrec} of the projectors and the local relations \eqref{eq:pushthru2} 
and \eqref{eq:projprops}. To prove \eqref{eq:FHa}, we again use the recursive definition of the projectors:
\be
\psset{unit=0.5}
\prod_{k=0}^{m-1} \big(s_k(u)\big)^N\Tb^{m+1,0}_0 = \ \ 
%
\ \ .
\ee
The first term is recognised as $\prod_{k=0}^{m-2} \big(s_k(u)\big)^N\Tb^{m,0}_0\Tb^{1,0}_m$. For the second term, we apply the second relation in \eqref{eq:pushthru2} $N$ consecutive times and find
\be\label{eq:propagate.N.times}
\psset{unit=0.5}
%
 \ \ = \frac{[m+1]}{[m]}\big(s_m(u)\big)^N \prod_{k=0}^{m-2} \big(s_k(u)\big)^N \Tb^{m-1,1}_0.
\ee
The second relation in \eqref{eq:proj.id.2} was used at the last step. Simplifying the common trigonometric functions yields \eqref{eq:FHa}.

Similarly, to prove 
\eqref{eq:FHb}, we use a reflected version of \eqref{eq:Pmnrec}:
\be
\label{eq:Pmnrec2}
%
\ \ .
\ee
This relation is easily seen to hold from the form \eqref{eq:proj.triangles} of the projectors. We then find
\be
\psset{unit=0.5}
\prod_{k=0}^{n-1} \big(s_k(u)\big)^N\Tb^{0,n+1}_0 = \ \ 
%
\ \ .
\ee
The first term is $\prod_{k=1}^{n-2} \big(s_k(u)\big)^N\Tb^{0,1}_0\Tb^{0,n}_1$. For the second term, we apply the same idea as for \eqref{eq:propagate.N.times}:
\be
\psset{unit=0.5} 
%
\  = \frac{[m+1]}{[m]}\sigma\big(s_{-1}(u)\big)^N \prod_{k=1}^{n-2} \big(s_k(u)\big)^N \Tb^{1,n-1}_1.
\ee
The relation \eqref{eq:FHb} follows by simplifying 
the common factors.

To prove \eqref{eq:FHc}, we use the explicit form \eqref{eq:Pmn.gen} of the projector $P^{m,n}$ and write
\be
\label{eq:Tmn0}
\psset{unit=0.5}
\prod_{\ell=0}^{m+n-2} \big(s_\ell(u)\big)^N\Tb^{m,n}_0 
= \sum_{k=0}^{\textrm{min}(m,n)} (-1)^k\frac{\qbinom{m}{k}\qbinom{n}{k}}{\qbinom{m+n+1}{k}} \,\boldsymbol\tau_k
\ee
where
\be
\boldsymbol\tau_k = \ \
\psset{unit=0.75}
%
 \ \ .
\ee
The integer $k$ counts the number of arcs. For $k=0$, we have 
\be
\boldsymbol\tau_0 = \prod_{\ell=0}^{m-2} \big(s_\ell(u)\big)^N\prod_{\ell=m}^{m+n-2} \big(s_\ell(u)\big)^N\Tb^{m,0}_0\Tb^{0,n}_m.
\ee 
For $k \ge 1$, the smallest arc on the right-hand side is pushed through using the second relation in \eqref{eq:pushthru}:

\be
\label{eq:tauk}
\boldsymbol\tau_k = \big(s_1(u_{m-1})s_1(-u_{m-1})\big)^N\ \
\psset{unit=0.5}
%
\ \ \qquad k \ge 1.
\ee
This diagram can be simplified further. To this end, we note that \eqref{eq:Pmnrec} and \eqref{eq:Pmnrec2} imply that
\be
\label{eq:twoPs}
%
\ \ .
\ee
If there are $k$ wavy arcs with $k>1$, we then use \eqref{eq:twoPs} iteratively to obtain
\begin{alignat}{2}
%
\ \ .
\end{alignat}
Rotating this by $90$ degrees and inserting it in \eqref{eq:tauk}, we find
\be
\boldsymbol \tau_k = \sigma \big(s_m(u)s_{m-2}(u)\big)^N\bigg(\frac{[m+n+2-k][k]}{[m][n]}\Wb_{k-1} +\frac{[m-k][n-k]}{[m][n]} \Wb_k\bigg), \quad 1 \le k \le \textrm{min}(m,n),
\ee
where 
\be
\label{eq:Wk}
\psset{unit=0.65}
\Wb_k = \ \
%
\ee
with $k$ again counting the arcs. Inserting this into the sum in \eqref{eq:Tmn0}, we find that most of the terms cancel pairwise:
\be
\sum_{k=0}^{\textrm{min}(m,n)} (-1)^k\frac{\qbinom{m}{k}\qbinom{n}{k}}{\qbinom{m+n+1}{k}} \boldsymbol\tau_k = \boldsymbol \tau_0-\sigma \big(s_m(u)s_{m-2}(u)\big)^N\Wb_0.
\ee
The
tangle $\Wb_0$ is proportional to $\Tb^{m-1,0}_k\Tb^{0,n-1}_{m+1}$. It follows that
\begin{alignat}{2}
\prod_{\ell=0}^{m+n-2} \big(s_\ell(u)\big)^N\Tb^{m,n}_0 &= \prod_{\ell=0}^{m-2} \big(s_\ell(u)\big)^N\prod_{\ell=m}^{m+n-2} \big(s_\ell(u)\big)^N\Tb^{m,0}_k\Tb^{0,n}_m\nonumber\\[0.1cm]
&- \sigma \big(s_m(u)s_{m-2}(u)\big)^N \prod_{\ell=0}^{m-3} \big(s_\ell(u)\big)^N\prod_{\ell=m+1}^{m+n-2} \big(s_\ell(u)\big)^N\Tb^{m-1,0}_k\Tb^{0,n-1}_{m+1},
\end{alignat}
which simplifies to \eqref{eq:FHc} after removal of common factors.

\subsection[$T$-system relations]{$\boldsymbol T$-system relations}\label{app:proofTsys}

The goal of this subsection is to prove the relations \eqref{eq:Trelations}. In fact, they merely
correspond to the $k=0$ specialisations of the relations in the following proposition.
\begin{Proposition}
For $m,k \in \mathbb Z$, we have
\begin{subequations}
\label{eq:moreTT}
\begin{alignat}{2}
\Tb^{m,0}_0 \Tb^{m-k,0}_{k+1} &= f_m \Tb^{k,m-k}_0 + \Tb^{m+1,0}_0 \Tb^{m-1-k,0}_{k+1},\label{eq:moreTTA}\\[0.1cm]
\Tb^{0,n}_0 \Tb^{0,n+k}_{1} &= \sigma^n f_{-1} \Tb^{n,k}_1 + \Tb^{0,n+1+k}_0 \Tb^{0,n-1}_{1}.\label{eq:moreTTB}
\end{alignat}
\end{subequations}
\end{Proposition}
\proof
We demonstrate
\eqref{eq:moreTTA}; the proof of \eqref{eq:moreTTB} follows similar arguments. We first note that by virtue of \eqref{eq:notations} and
\eqref{eq:negTs}, \eqref{eq:moreTTA} holds trivially for $m=-1,-2,-3$ and $k \in \mathbb Z$. Let us fix  
$m \ge 0$. The proof of \eqref{eq:moreTTA} for this $m$ is 
inductive, and requires that the relation
holds for $m-1, m-2$ and $m-3$. We have
\begin{alignat}{2}
&f_0\Tb^{m,0}_0 \Tb^{m-k,0}_{k+1}  \overset{\textrm{\tiny\eqref{eq:FH2b}}}{=} (\Tb^{1,0}_0 \Tb^{m-1,0}_1 - \Tb^{0,1}_0 \Tb^{m-2,0}_2 + \sigma f_{-1} \Tb^{m-3,0}_3)\Tb^{m-k,0}_{k+1}
\nonumber\\[0.1cm]
&\hspace{0.2cm}= \Tb^{1,0}_0 (\Tb^{m-1,0}_0\Tb^{m-1-(k-1),0}_{(k-1)+1})_1 - \Tb^{0,1}_0(\Tb^{m-2,0}_0 \Tb^{m-2-(k-2),0}_{(k-2)+1})_2 + \sigma f_{-1} (\Tb^{m-3,0}_0)\Tb^{m-3-(k-3),0}_{(k-3)+1})_3
\end{alignat}
where a parenthesis with a subscript indicates that its entire content is shifted accordingly.
At this step, we use the induction hypothesis for $m-1$, $m-2$ and $m-3$:
\begin{alignat}{2}
f_0\Tb^{m,0}_0 \Tb^{m-k,0}_{k+1} &= \Tb^{1,0}_0 (f_m \Tb^{k-1,m-k}_1 + \Tb^{m,0}_1 \Tb^{m-1-k,0}_{k+1}) - \Tb^{0,1}_0(f_m \Tb^{k-2,m-k}_2 + \Tb^{m-1,0}_2 \Tb^{m-1-k,0}_{k+1}) \nonumber\\[0.1cm]
&\hspace{1cm}+ \sigma f_{-1} (f_m \Tb^{k-3,m-k}_3 + \Tb^{m-2,0}_3 \Tb^{m-1-k,0}_{k+1}) \nonumber\\[0.1cm]
& = f_m (\Tb^{1,0}_0\Tb^{k-1,m-k}_1 -\Tb^{0,1}_0 \Tb^{k-2,m-k}_2 +  \sigma f_{-1} \Tb^{k-3,m-k}_3) \nonumber\\[0.1cm]&\hspace{1cm}
+ (\Tb^{1,0}_0\Tb^{m,0}_1- \Tb^{0,1}_0\Tb^{m-1,0}_2+\Tb^{m-2,0}_3)\Tb^{m-1-k,0}_{k+1}
\nonumber\\[0.1cm]&\hspace{-0.25cm} \overset{\textrm{\tiny\eqref{eq:FH2b}}}{=}f_0( \Tb^{k,m-k}_1 + \Tb^{m+1,0}_1\Tb^{m-1-k,0}_{k+1}),
\end{alignat}
thus completing the proof for $m\ge 0$. The proof for $m<-3$ uses a similar inductive argument.
\eproof

\subsection{Closure of the fusion hierarchy}\label{app:proofclosure}

\begin{Proposition}\label{prop:clo}
At $\lambda = \lambda_{p,p'}$, we have the following closure relations:
\begin{subequations}
\begin{alignat}{2}
\Tb^{p'\!,0}_0 &= \Tb^{p'-2,1}_1 -\sigma\, \Tb^{p'-3,0}_2 + f_{-1}\Jb,\label{eq:cloproof1}\\[0.1cm]
\Tb^{0,p'}_0 &= \sigma\, \Tb^{1,p'-2}_0 -\Tb^{0,p'-3}_1 + f_{-1}\Kb,\label{eq:cloproof2}
\end{alignat}
\end{subequations}
where $\Jb$ and $\Kb$ are given by \eqref{eq:JKbraids}.
\end{Proposition}
\proof We consider the transfer tangles with inhomogeneity parameters $\xi^1, \xi^2, \dots, \xi^N$, namely
\be
\Tb^{m,n} (u)= \  \
\begin{pspicture}[shift=-0.4](-0.3,0)(4.3,1.0)
\facegrid{(0,0)}{(4,1)}
\psarc[linewidth=0.025]{-}(0,0){0.16}{0}{90}
\psarc[linewidth=0.025]{-}(1,0){0.16}{0}{90}
\psarc[linewidth=0.025]{-}(3,0){0.16}{0}{90}
\psline[linewidth=1.5pt,linecolor=blue,linestyle=dashed,dash=2pt 2pt]{-}(0,0.5)(-0.3,0.5)
\psline[linewidth=1.5pt,linecolor=blue,linestyle=dashed,dash=2pt 2pt]{-}(4,0.5)(4.3,0.5)
\rput(2.5,0.5){$\ldots$}
\rput(0.5,.4){\scriptsize$u+\xi^1$}\rput(.5,.7){\tiny $(m,n)$}
\rput(1.5,.4){\scriptsize$u+\xi^2$}\rput(1.5,.7){\tiny $(m,n)$}
\rput(3.5,.4){\scriptsize$u+\xi^N$}\rput(3.5,.7){\tiny $(m,n)$}
\end{pspicture}\ .
\ee
These transfer tangles satisfy the fusion hierarchy relations \eqref{eq:FH} and \eqref{eq:FH2} with the function $f_k$ modified to
\be
f_k = \prod_{j=1}^N s_k(u+\xi^j).
\ee
We now specialize the spectral parameter to $u= \lambda - \xi^j$ for some $j$, so that $f_{-1} = 0$. Under this specialisation, from \eqref{eq:FH2b}, we have
\be
f_0 \Tb^{p'\!,0}_0 = \Tb^{1,0}_0 \Tb^{p'-1,0}_1 - \Tb^{0,1}_0 \Tb^{p'-2,0}_2.
\ee
Using the relations
\be
\Tb^{m,n}_{p'+k} = \nu\, \Tb^{m,n}_{p'}, \qquad f_{p'+k} = \nu f_k, \qquad \nu = (-1)^{(p'-p)N},
\ee
we find
\be
 \Tb^{1,0}_0 \Tb^{p'-1,0}_1 = \nu\, \Tb^{p'-1,0}_1  \Tb^{1,0}_{p'} \overset{\textrm{\tiny\eqref{eq:FHa}}}{=} \nu\, (f_{p'}\Tb^{p'-2,1}_1+f_{p'-1}\Tb^{p,0}_1) = f_0 \Tb^{p'-2,1}_1.
\ee
Similarly,
\be
\Tb^{0,1}_0\Tb^{p'-2,0}_2= \nu\, \Tb^{p'-2,0}_2\Tb^{0,1}_{p'}  \overset{\textrm{\tiny\eqref{eq:FHc}}}{=} \nu\,(f_{p'-1}\Tb^{p'-2,1}_2 + \sigma \Tb^{p'-3,0}_2 \Tb^{0,0}_{p'+1}) = \sigma f_0 \Tb^{p'-3,0}_2.
\ee
As a result, we have the equality
\be
\label{eq:spec.clo}
\Tb^{p'\!,0}_0 = \Tb^{p'-2,1}_1 -\sigma\, \Tb^{p'-3,0}_2
\ee
which holds at $u = \lambda - \xi^j$, $j = 1, \dots, N$. From the periodicity property $\Tb^{m,n}(u+\pi) = (-1)^N\Tb^{m,n}(u)$, the equality \eqref{eq:spec.clo} also holds at $u = \lambda - \xi^j + \pi$.

As a function of $u$, $\Tb^{p'\!,0}$ is a Laurent polynomial in $\eE^{\ir u}$ with minimal and maximal powers $\pm N$. To prove \eqref{eq:cloproof1}, we must show that it holds at $2N+1$ points. The previous argument shows that the equality holds at $2N$ points. The last point is at $\ir \infty$. The tangle $\Jb$ is obtained from the braid limit of \eqref{eq:cloproof1}, so \eqref{eq:cloproof1} automatically holds in this limit, completing the proof of \eqref{eq:cloproof1}. The proof of \eqref{eq:cloproof2} uses similar arguments.
\eproof

\begin{Proposition} At $\lambda = \lambda_{p,p'}$, we have the additional closure relations:
\begin{subequations}
\label{eq:extraclo}
\begin{alignat}{2}
\Tb^{p'\!,k}_0 &= \Tb^{p'-2,k+1}_{1}-\sigma^{k+1}\Tb^{p'-k-3,0}_{k+2}+ \Jb\,\Tb^{0,k}_0,\label{eq:extracloa} \\[0.1cm]
\Tb^{k,p'}_0 &= \sigma\,\Tb^{k+1,p'-2}_{0}-\sigma^{k}\Tb^{0,p'-k-3}_{k+1}+ \Kb\,\Tb^{k,0}_0.\label{eq:extraclob}
\end{alignat}
\end{subequations}
\end{Proposition}
\proof
Equation \eqref{eq:extracloa} holds trivially for $k=-1$. By \cref{prop:clo}, it also holds for $k=0$.
For $k=1$, we have
\begin{alignat}{2}
f_{-1}\Tb^{p'\!,1}_0  &\overset{\textrm{\tiny\eqref{eq:FHc}}}{=} \Tb^{p',0}_0\Tb^{0,1}_0-\sigma\,\Tb^{p'-1,0}_0\Tb^{0,0}_1 \overset{\textrm{\tiny\eqref{eq:cloproof1}}}{=} (\Tb^{p'-2,1}_1-\sigma\,\Tb^{p'-3,0}_2 + f_{-1}\Jb)\Tb^{0,1}_0 - \sigma f_0 \Tb^{p'-1,0}_0\nonumber\\
&\overset{\textrm{\tiny\eqref{eq:FH2b}}}{=}(\Tb^{p'-2,1}_1-\sigma\,\Tb^{p'-3,0}_2 + f_{-1}\Jb)\Tb^{0,1}_0 - \sigma(\Tb^{1,0}_0 \Tb^{p'-2,0}_1 - \Tb^{0,1}_0 \Tb^{p'-3,0}_2 + \sigma\, f_{-1}\Tb^{p'-4,0}_3)\nonumber\\
&\overset{\textrm{\tiny\eqref{eq:FH2a}}}{=} f_{-1}(\Tb^{p'-2,2}_1-\Tb^{p'-4,0}_3 + \Jb\, \Tb^{0,1}_0)
\end{alignat}
which is the desired result. The cases $k>1$ follow by induction:
\begin{alignat}{2}
f_{k-2}\Tb^{p'\!,k}_0 &\overset{\textrm{\tiny\eqref{eq:FH2a}}}{=} \Tb^{p'\!,k-1,0}_0\Tb^{0,1}_{k-1} - \sigma\, \Tb^{p'\!,k-2}_0 \Tb^{1,0}_{k-1} + f_{k-1}\Tb^{p'\!,k-3}_0\nonumber\\
&\hspace{-0.035cm}\overset{\textrm{\tiny\eqref{eq:extraclob}}}{=} 
(\Tb^{p'-2,k}_1\Tb^{0,1}_{k-1}-\sigma\,\Tb^{p'-2,k-1}_1\Tb^{1,0}_{k-1} + f_{k-1} \Tb^{p'-2,k-2}_1)\nonumber\\&
\qquad
-\sigma^k(\Tb^{0,1}_{k-1}\Tb^{p'-k-2,0}_{k+1}-\Tb^{1,0}_{k-1}\Tb^{p'-k-1,0}_k+f_{k-1}\Tb^{p'-k,0}_{k-1})
\nonumber\\&
\qquad
+\Jb(\Tb^{0,k-1}_0\Tb^{0,1}_{k-1}-\sigma\,\Tb^{0,k-2}_0\Tb^{1,0}_{k-1}+f_{k-1}\Tb^{0,k-3}_0)\nonumber\\&
\hspace{0.1cm}\overset{\textrm{\tiny\eqref{eq:FH2}}}{=} f_{k-2} (\Tb^{p'-2,k+1}_1 - \sigma^{k+1}\Tb^{p'-k-3,0}_{k+2} + \Jb\, \Tb^{0,k}_0).
\end{alignat}
This
completes the proof of \eqref{eq:extracloa}. The proof of \eqref{eq:extraclob} uses similar arguments. 
\eproof

Using the fusion hierarchy relations \eqref{eq:FH} and \eqref{eq:FH2}, we have also obtained
closure relations in the cases where $m$ and $n$ are greater than $p'$:
\begin{subequations}
\label{eq:genclo}
\begin{alignat}{2}
\Tb^{p'+j,k}_0 &= \Tb^{p'-j-2,j+k+1}_{j+1}-\sigma^{k+1}\Tb^{p'-j-k-3,j}_{j+k+2}+ \Jb\,\Tb^{j,k}_0, \label{eq:gencloa}\\[0.1cm]
\Tb^{k,p'+j}_0 &= \sigma^{j+1}\Tb^{j+k+1,p'-j-2}_{0}-\sigma^{j+k}\Tb^{j,p'-j-k-3}_{k+1}+ \Kb\,\Tb^{k,j}_0.\label{eq:genclob}
\end{alignat}
\end{subequations}
These hold for arbitrary $j,k$, and the proof is again by induction. For $j=-1$, in particular,
these relations are trivial; for instance, \eqref{eq:gencloa} reads $\Tb^{p'-1,k}_0 = \Tb^{p'-1,k}_0$. We believe that the closure relations \eqref{eq:extraclo} are minimal, in the sense that there are no alternate closure relations for $\Tb^{m,n}_0$ with $m,n<p'$.

\subsection[Closure of the $Y$-system]{Closure of the $\boldsymbol Y$-system}\label{app:proofYclosure}

In this subsection, we prove the closure relations for the $Y$-system at roots of unity. We have the following proposition.

\begin{Proposition} For $\lambda = \lambda_{p,p'}$, the fused transfer tangles satisfy the following identities:
\begin{subequations}\label{eq:closureYv2}
\begin{alignat}{2}
&\Big(\Tb^{p'-1,0}_0\Tb^{0,p'-1}_0 - \sigma\, \Tb^{p'-2,0}_1\Tb^{0,p'-2}_0\Big)\Big(\Tb^{p'-1,0}_1\Tb^{0,p'-1}_0 - \sigma\, \Tb^{p'-2,1}_1\Tb^{0,p'-2}_1\Big) =  \label{eq:closureY1}\\
& \qquad f_{p'-1} \Big( (\Tb^{0,p'-1}_0)^3 + \sigma^{p'-p} \Jb (\Tb^{0,p'-1}_0)^2\Tb^{p'-2,0}_1 + \sigma^{p'-p} \Kb\, \Tb^{0,p'-1}_0(\Tb^{p'-2,0}_1)^2+\sigma^p(\Tb^{p'-2,0}_1)^3 \Big),\nonumber\\
&\Big(\Tb^{p'-1,1}_1\Tb^{0,p'-1}_1 - \sigma\, \Tb^{p'-2,0}_2\Tb^{0,p'-2}_1\Big)\Big(\Tb^{p'-1,0}_1\Tb^{0,p'-1}_0 - \sigma\, \Tb^{p'-2,1}_1\Tb^{0,p'-2}_1\Big) =  \label{eq:closureY2}\\
& \qquad \sigma^{p'-1}f_{-1} \Big( (\Tb^{p'-1,0}_1)^3 + \sigma^{p'+1} \Kb (\Tb^{p'-1,0}_1)^2\Tb^{0,p'-2}_1 + \sigma^{p'} \Jb\, \Tb^{p'-1,0}_1(\Tb^{0,p'-2}_1)^2+\sigma^{p+1}(\Tb^{0,p'-2}_1)^3 \Big).\nonumber
\end{alignat}
\end{subequations}
\end{Proposition}
\proof
We demonstrate \eqref{eq:closureY1}; the
proof of \eqref{eq:closureY2} uses similar arguments. First, we
recall that 
\be f_{p'+k} = \sigma^{p'-p} f_{k}, \qquad \Tb^{m,n}_{p'+k} = \sigma^{p'-p}\,\Tb^{m,n}_{k}.\label{eq:perfT}\ee 
We then compute the term $\Tb^{p'-1,0}_0 \Tb^{p'-1,0}_1(\Tb^{0,p'-1}_0)^2$ from the left-hand side of \eqref{eq:closureY1}:
\begin{alignat}{2}
(\Tb^{p'-1,0}_0 \Tb^{p'-1,0}_1)(\Tb^{0,p'-1}_0)^2 &\overset{\textrm{\tiny\eqref{eq:TrelationsA}}}{=} (f_{p'-1}\Tb^{0,p'-1}+ \Tb^{p',0}_0\Tb^{p'-2,0}_1)(\Tb^{0,p'-1}_0)^2\nonumber\\
&\overset{\textrm{\tiny\eqref{eq:perfT}}}{=}f_{p'-1}(\Tb^{0,p'-1}_0)^3 +\sigma^{p'-p} \Tb^{p'-2,0}_1 \Tb^{0,p'-1}_0(\Tb^{p',0}_0\Tb^{0,p'-1}_{p'})\\
&\overset{\textrm{\tiny\eqref{eq:FHc}}}{=}f_{p'-1}(\Tb^{0,p'-1}_0)^3 + \sigma^{p'-p} \Tb^{p'-2,0}_1 \Tb^{0,p'-1}_0(f_{p'-1}\Tb^{p',p'-1}_0+\sigma\, \Tb^{p'-1,0}_0\Tb^{0,p'-2}_{p'+1}). \nonumber
\end{alignat}
The last term is $\sigma\, \Tb^{p'-2,0}_1\Tb^{0,p'-1}_0\Tb^{p'-1,0}_0\Tb^{0,p'-2}_1$ and is recognised as one of the contributions
to the left-hand side of \eqref{eq:closureY1}. For the second term, we use the first closure relation \eqref{eq:extracloa} and obtain
\begin{alignat}{2}
&(\Tb^{p'-1,0}_0 \Tb^{p'-1,0}_1)(\Tb^{0,p'-1}_0)^2 - \sigma\, \Tb^{p'-2,0}_1\Tb^{0,p'-1}_0\Tb^{p'-1,0}_0\Tb^{0,p'-2}_1\nonumber\\
& \qquad = f_{p'-1}(\Tb^{0,p'-1}_0)^3 + \sigma^{p'-p}f_{p'-1} \Tb^{p'-2,0}_1 \Tb^{0,p'-1}_0(\Tb^{p'-2,p'}_1 - \sigma^{p'}\Tb^{-2,0}_{p'+1}+\Jb\, \Tb^{0,p'-1}_0).
\end{alignat}
We know by virtue of \eqref{eq:negTs} that $\Tb^{-2,0}_{p'+1} = 0$. We apply the second closure relation \eqref{eq:extraclob} to $\Tb^{p'-2,p'}_1$ and find 
\begin{alignat}{2}
&(\Tb^{p'-1,0}_0 \Tb^{p'-1,0}_1)(\Tb^{0,p'-1}_0)^2 - \sigma\, \Tb^{p'-2,0}_1\Tb^{0,p'-1}_0\Tb^{p'-1,0}_0\Tb^{0,p'-2}_1 = \sigma^{p'-p+1}f_{p'-1} \Tb^{p'-2,0}_1 \Tb^{0,p'-1}_0\Tb^{p'-1,p'-2}_1 \nonumber\\
& \qquad + f_{p'-1}\Big((\Tb^{0,p'-1}_0)^3+ \sigma^{p'-p} \Jb(\Tb^{0,p'-1}_0)^2 \Tb^{p'-2,0}_1 + \sigma^{p'-p} \Kb\,\Tb^{0,p'-1}_0 (\Tb^{p'-2,0}_1)^2\Big).
\end{alignat}
The last three terms appear on the right-hand side of \eqref{eq:closureY1}. The first term is rewritten as follows:
\begin{alignat}{2}
\sigma^{p'-p+1}f_{p'-1}& \Tb^{p'-2,0}_1 \Tb^{0,p'-1}_0\Tb^{p'-1,p'-2}_1 \overset{\textrm{\tiny\eqref{eq:FHc}}}{=} \sigma^{p'-p+1} \Tb^{p'-2,0}_1 \Tb^{0,p'-1}_0(\Tb^{p'-1,0}_1\Tb^{0,p'-2}_{p'}-\sigma\, \Tb^{p'-2,0}_1 \Tb^{0,p'-3}_{p'+1})\nonumber\\
&= \sigma\, \Tb^{p'-2,0}_1 \Tb^{0,p'-1}_0\Tb^{p'-1,0}_1\Tb^{0,p'-2}_{0} - (\Tb^{p'-2,0}_1)^2(\Tb^{0,p'-1}_0 \Tb^{0,p'-3}_1)\nonumber\\
&\hspace{-0.26cm}\overset{\textrm{\tiny\eqref{eq:TrelationsA}}}{=} \sigma\, \Tb^{p'-2,0}_1 \Tb^{0,p'-1}_0\Tb^{p'-1,0}_1\Tb^{0,p'-2}_{0} - (\Tb^{p'-2,0}_1)^2(\Tb^{0,p'-2}_0 \Tb^{0,p'-2}_1 - \sigma^{p'}f_{-1}\Tb^{p'-2,0}_1).
\end{alignat}
The third term equals $f_{p'-1} \sigma^p (\Tb^{p'-2,0}_1)^3$ and is the last remaining contribution to the right-hand side of \eqref{eq:closureY1}. The first two terms contribute to its left-hand side. Putting all the terms together yields \eqref{eq:closureY1}.
\eproof

It is not hard to see that the relations \eqref{eq:closureYv2} are equivalent to the relations \eqref{eq:closureY} given in \cref{sec:cloY}. For example for \eqref{eq:closureY1}, we factorise $\Tb^{p'-1,0}_0\Tb^{p'-1,0}_1$ from the left-hand side, $f_{p'-1}(\Tb^{0,p'-1}_{0})^3$ from both sides and, recalling that 
\be
\frac{\Tb^{p'-1,0}_0\Tb^{p'-1,0}_1}{f_{p'-1}\Tb^{0,p'-1}_{0}} = (\Ib+\tba{p'-1}_0),\ee
we readily obtain \eqref{eq:cloY1}.

\section{The $\boldsymbol{A_2^{(1)}}$ RSOS models}
\label{app:RSOS}

\subsection[Definition of the $A_2^{(1)}$ RSOS models]{Definition of the $\boldsymbol{A_2^{(1)}}$ RSOS models}

An $\Atwoone$ RSOS model is a member of a family of Interaction-Round-A-Face models 
on the square lattice, 
where the degrees of freedom are ``heights'' attached to the sites.
Each model is defined by a choice of a root of unity $q=\eE^{\ir \lambda}$ corresponding to
$\lambda=\lambda_{p,p'}$ with $p'\ge 5$. In this subsection, we follow the presentation of \cite{DEI16}. The height variables are two-dimensional vectors that live on the following finite part of the $s\ell(3)$ weight lattice:
\be
\mathcal L = \big\{\kappa_1 \boldsymbol{\omega}_1+\kappa_2 \boldsymbol{\omega}_2
\ |\ 
\kappa_1,\kappa_2 \in \mathbb N, \ 
\kappa_1 + \kappa_2 \le p'-3\big\}. 
\ee
Here,
$\boldsymbol{\omega}_1$ and $\boldsymbol{\omega}_2$ are the fundamental $s\ell(3)$ weights satisfying
\be
\boldsymbol{\omega}_1 \cdot \boldsymbol{\omega}_1 = \boldsymbol{\omega}_2 \cdot \boldsymbol{\omega}_2 = \tfrac 23, \qquad \boldsymbol{\omega}_1 \cdot \boldsymbol{\omega}_2 = \tfrac 13,
\ee
where the dot product is the usual Euclidean inner product on $\mathbb R^2$. Thus, $\mathcal L$ is
a finite graph in the form of an equilateral triangle of side length $p'-3$,
as shown in Figure~\ref{fig:su3weights}. The edges of this graph are oriented along the three vectors
\be
\boldsymbol{h}_1 = \boldsymbol{\omega}_1, \qquad \boldsymbol{h}_2 = \boldsymbol{\omega}_2-\boldsymbol{\omega}_1,\qquad \boldsymbol{h}_3 = - \boldsymbol{\omega}_2.
\ee
In \cref{fig:su3weights}, these vectors are oriented as follows:
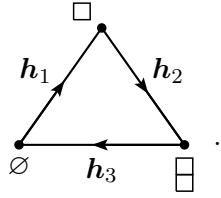
\begin{equation*}
\psset{unit=2.2}
\begin{pspicture}[shift=-0.4](0,-0.4)(1,1)
\psdot(0,0)\psdot(0.5,0.707)\psdot(1,0)
\psline[arrowscale=1.4,arrowinset=0.2]{->}(0,0)(0.275, 0.3883)\psline{-}(0,0)(0.5,0.707)\psline[arrowscale=1.4,arrowinset=0.2]{>-}(0.725, 0.3883)(1,0)\psline{-}(0.5,0.707)(1,0)\psline[arrowscale=1.4,arrowinset=0.2]{<-}(0.44,0)(1,0)\psline{-}(0,0)(1,0)
\rput(0,-0.13){$\varnothing$}
\rput(0.95,-0.23){\young{1}{1}}
\rput(0.33,0.707){\young{1}{0}}
\rput(0.1,0.45){$\boldsymbol{h}_1$}
\rput(0.9,0.45){$\boldsymbol{h}_2$}
\rput(0.5,-0.15){$\boldsymbol{h}_3$}
\end{pspicture}\ \ \ .
\end{equation*}

The local Boltzmann weights of this model are assigned to each face of the 
lattice according to the heights $\boldsymbol a, \boldsymbol b, \boldsymbol c, \boldsymbol d \in \mathcal L$ of the four 
corners of the face,
\be
\begin{pspicture}[shift=-.7](-0.3,-0.3)(1.3,1.3)
\facegrid{(0,0)}{(1,1)}
\psarc[linewidth=0.025]{-}(0,0){0.16}{0}{90}
\rput(.5,.5){$u$}
\rput(-0.2,1.2){$\boldsymbol a$}
\rput(1.2,1.2){$\boldsymbol b$}
\rput(-0.2,-0.2){$\boldsymbol d$}
\rput(1.2,-0.2){$\boldsymbol c$}
\end{pspicture}\ \ .
\ee
These heights are constrained to be neighbours in the oriented $s\ell(3)$ graph, meaning that the states
\be
\boldsymbol{h}_\kappa = \boldsymbol{b}-\boldsymbol{a}, \qquad 
\boldsymbol{h}_\mu = \boldsymbol{d}-\boldsymbol{a}, \qquad 
\boldsymbol{h}_\nu = \boldsymbol{c}-\boldsymbol{d}, \qquad 
\boldsymbol{h}_\sigma = \boldsymbol{c}-\boldsymbol{b}
\ee
satisfy
$\boldsymbol{h}_\kappa, \boldsymbol{h}_\mu, \boldsymbol{h}_\nu, \boldsymbol{h}_\sigma \in \{\boldsymbol{h}_1,\boldsymbol{h}_2,\boldsymbol{h}_3\}.$ The Boltzmann weights take the form
\begin{subequations}
\begin{alignat}{2}
&W\bigg(\begin{matrix}
\boldsymbol a & \boldsymbol a + \boldsymbol{h}_\mu\\
\boldsymbol a + \boldsymbol{h}_\mu& \boldsymbol a + 2 \boldsymbol{h}_\mu 
\end{matrix}\ \bigg|\, u\bigg) = \frac{\sin(\lambda-u)}{\sin \lambda}\\[0.1cm]
&W\bigg(\begin{matrix}
\boldsymbol a & \boldsymbol a + \boldsymbol{h}_\mu\\
\boldsymbol a + \boldsymbol{h}_\mu& \boldsymbol a +  \boldsymbol{h}_\mu +\boldsymbol{h}_\nu 
\end{matrix}\ \bigg|\,u\bigg) = \frac{\sin(\lambda a_{\mu\nu}+u)}{\sin (\lambda a_{\mu\nu})} \qquad && (\mu \neq \nu)\\[0.1cm]
&W\bigg(\begin{matrix}
\boldsymbol a & \boldsymbol a + \boldsymbol{h}_\nu\\
\boldsymbol a + \boldsymbol{h}_\mu& \boldsymbol a +  \boldsymbol{h}_\mu +\boldsymbol{h}_\nu 
\end{matrix}\ \bigg|\,u\bigg) = \frac{\sin u\sin\big(\lambda (a_{\mu\nu}+1)\big)}{\sin \lambda \sin (\lambda a_{\mu\nu})} \qquad && (\mu \neq \nu)
\end{alignat}
\end{subequations}
where $a_{\mu\nu} = (\boldsymbol{a}+\boldsymbol{\rho})\cdot(\boldsymbol{h}_\mu-\boldsymbol{h}_\nu)$
and $\boldsymbol{\rho} = \boldsymbol{\omega}_1 + \boldsymbol{\omega}_2$.
This can be expressed compactly as
\be
W\bigg(\begin{matrix}
\boldsymbol a & \boldsymbol b \\
\boldsymbol d & \boldsymbol c
\end{matrix}\ \bigg|\,u\bigg) = 
\frac{\sin(\lambda - u)}{\sin \lambda}\ \delta_{\boldsymbol b, \boldsymbol d} + 
\frac{\sin u}{\sin \lambda}\ U\bigg(\begin{matrix}
\boldsymbol a & \boldsymbol b \\
\boldsymbol d & \boldsymbol c
\end{matrix}\ \bigg|\,u\bigg) 
\ee
with 
\be
U\bigg(\begin{matrix}
\boldsymbol a & \boldsymbol a + \boldsymbol{h}_\nu\\
\boldsymbol a + \boldsymbol{h}_\mu& \boldsymbol a +  \boldsymbol{h}_\mu +\boldsymbol{h}_\nu 
\end{matrix}\ \bigg|\,u\bigg) = (1-\delta_{\mu\nu}) \ \frac{\sin\big(\lambda (a_{\mu\nu}+1)\big)}{\sin(\lambda a_{\mu\nu})}.
\ee

One can consider $W$ and $U$ as operators acting on words of the form $w = \boldsymbol{a}_0\boldsymbol{a}_1 \dots \boldsymbol{a}_N$ where $\boldsymbol{a}_i\in\mathcal L$ and
$(\boldsymbol{a}_{i+1}-\boldsymbol{a}_{i})\in \{\boldsymbol{h}_1, \boldsymbol{h}_2, \boldsymbol{h}_3\}$. A subscript $j$ on 
$W_j$ or $U_j$ indicates that the operator $W$ or $U$ acts 
non-trivially on the segment $\boldsymbol{a}_{j-1}\boldsymbol{a}_{j}\boldsymbol{a}_{j+1}$ of $w$. Mimicking, for the $s\ell(3)$ case, the construction of Dyck paths for $s\ell(2)$, we depict the word $w$ using a path of length $N$ where each step either goes up, goes down or remains at the same height, corresponding to $\boldsymbol{h}_1$, $\boldsymbol{h}_2$ and $\boldsymbol{h}_3$, 
respectively. 
For instance, the possible words of length three with $\boldsymbol a$ as the first letter are given by
\be\label{eq:pathbasis}
\psset{unit=0.6cm}
\begin{pspicture}[shift=-0.7](0.0,-0.8)(1,0.8)
\rput(-0.35,0){\scriptsize $\boldsymbol a$}
\psline[linewidth=\mince](0,0)(0.4,0.4)(0.8,0.8)
\psdots[dotsize=2.5pt](0,0)(0.4,0.4)(0.8,0.8)
\end{pspicture}\hspace{0.7cm}
\begin{pspicture}[shift=-0.7](0.0,-0.8)(1,0.8)
\rput(-0.35,0){\scriptsize $\boldsymbol a$}
\psline[linewidth=\mince](0,0)(0.4,0.4)(0.93,0.4)
\psdots[dotsize=2.5pt](0,0)(0.4,0.4)(0.93,0.4)
\end{pspicture}\hspace{0.7cm}
\begin{pspicture}[shift=-0.7](0.0,-0.8)(1,0.8)
\rput(-0.35,0){\scriptsize $\boldsymbol a$}
\psline[linewidth=\mince](0,0)(0.4,0.4)(0.8,0)
\psdots[dotsize=2.5pt](0,0)(0.4,0.4)(0.8,0)
\end{pspicture}\hspace{0.7cm}
\begin{pspicture}[shift=-0.7](0.0,-0.8)(1,0.8)
\rput(-0.35,0){\scriptsize $\boldsymbol a$}
\psline[linewidth=\mince](0,0)(0.53,0)(0.93,0.4)
\psdots[dotsize=2.5pt](0,0)(0.53,0)(0.93,0.4)
\end{pspicture}\hspace{0.7cm}
\begin{pspicture}[shift=-0.7](0.0,-0.8)(1,0.8)
\rput(-0.35,0){\scriptsize $\boldsymbol a$}
\psline[linewidth=\mince](0,0)(0.53,0)(1.06,0)
\psdots[dotsize=2.5pt](0,0)(0.53,0)(1.06,0)
\end{pspicture}\hspace{0.7cm}
\begin{pspicture}[shift=-0.7](0.0,-0.8)(1,0.8)
\rput(-0.35,0){\scriptsize $\boldsymbol a$}
\psline[linewidth=\mince](0,0)(0.53,0)(0.93,-0.4)
\psdots[dotsize=2.5pt](0,0)(0.53,0)(0.93,-0.4)
\end{pspicture}\hspace{0.7cm}
\begin{pspicture}[shift=-0.7](0.0,-0.8)(1,0.8)
\rput(-0.35,0){\scriptsize $\boldsymbol a$}
\psline[linewidth=\mince](0,0)(0.4,-0.4)(0.8,0)
\psdots[dotsize=2.5pt](0,0)(0.4,-0.4)(0.8,0)
\end{pspicture}\hspace{0.7cm}
\begin{pspicture}[shift=-0.7](0.0,-0.8)(1,0.8)
\rput(-0.35,0){\scriptsize $\boldsymbol a$}
\psline[linewidth=\mince](0,0)(0.4,-0.4)(0.93,-0.4)
\psdots[dotsize=2.5pt](0,0)(0.4,-0.4)(0.93,-0.4)
\end{pspicture}\hspace{0.7cm}
\begin{pspicture}[shift=-0.7](0.0,-0.8)(1,0.8)
\rput(-0.35,0){\scriptsize $\boldsymbol a$}
\psline[linewidth=\mince](0,0)(0.4,-0.4)(0.8,-0.8)
\psdots[dotsize=2.5pt](0,0)(0.4,-0.4)(0.8,-0.8)
\end{pspicture}\ \ ,
\ee
although some of them may not be allowed for a given $\boldsymbol a$ and level $p'$.

Importantly, if a word $w$ is such that the segment $\boldsymbol{a}_{j-1}\boldsymbol{a}_{j}\boldsymbol{a}_{j+1}$ is allowed, namely it satisfies $\boldsymbol{a}_{j-1},\boldsymbol{a}_{j},\boldsymbol{a}_{j+1} \in \mathcal L$ and $(\boldsymbol{a}_{j}-\boldsymbol{a}_{j-1}),(\boldsymbol{a}_{j+1}-\boldsymbol{a}_j) \in \{\boldsymbol{h}_1,\boldsymbol{h}_2,\boldsymbol{h}_3\}$, then all non-zero contributions in 
$W_j w$ and $U_j w$ 
are along allowed states.
Restricted to words where the letter in position $j-1$ is $\boldsymbol a$, the operator $U_j$ has the following matrix representation:
\be\label{eq:Umat}
U_j = 
\begin{pmatrix}
0 & 0 & 0 & 0 & 0 & 0 & 0 & 0 & 0\\
0 & S_{13}^{\boldsymbol a} & 0 & S_{13}^{\boldsymbol a} & 0 & 0 & 0 & 0 & 0\\
0 & 0 & S_{12}^{\boldsymbol a} & 0 & 0 & 0 & S_{12}^{\boldsymbol a} & 0 & 0\\
0 & S_{31}^{\boldsymbol a} & 0 & S_{31}^{\boldsymbol a} & 0 & 0 & 0 & 0 & 0\\
0 & 0 & 0 & 0 & 0 & 0 & 0 & 0 & 0\\
0 & 0 & 0 & 0 & 0 & S_{32}^{\boldsymbol a} & 0 & S_{32}^{\boldsymbol a} & 0\\
0 & 0 & S_{21}^{\boldsymbol a} & 0 & 0 & 0 & S_{21}^{\boldsymbol a} & 0 & 0\\
0 & 0 & 0 & 0 & 0 & S_{23}^{\boldsymbol a} & 0 & S_{23}^{\boldsymbol a} & 0\\
0 & 0 & 0 & 0 & 0 & 0 & 0 & 0 & 0
\end{pmatrix}, \qquad
S^{\boldsymbol a}_{\mu\nu} = \frac{\sin[\lambda (a_{\mu\nu}+1)]}{\sin[\lambda a_{\mu\nu}]},
\ee
where the ordered basis is \eqref{eq:pathbasis}, and we have supposed that all nine basis states 
are allowed. If some states are disallowed, then the matrix \eqref{eq:Umat} is 
truncated; the corresponding columns and rows are simply not present.
One readily checks that the operators $U_j$ satisfy the $s\ell(3)$ Hecke relations:
\begin{subequations}\label{eq:su3Hecke}
\begin{alignat}{2}
&(U_j)^2 = 2 \cos \lambda \ U_j, \qquad &&U_j U_{j+1}U_j - U_j = U_{j+1}U_j U_{j+1}-U_{j+1},\\[0.1cm]
&U_j U_k = U_k U_j \qquad \big(|j-k|>1\big), \qquad &&(U_{j-1}-U_{j+1}U_jU_{j-1}+U_j)(U_j U_{j+1}U_j-U_j)=0.
\end{alignat}
\end{subequations}

The face operator of the $\Atwoone$ loop model can similarly be written in terms of the Hecke generators if the gauge parameter $t$ is set to the value $t = \eE^{-\ir u}$. Indeed, in this case, from \eqref{eq:faceop} one has
\be\label{eq:faceop.specialgauge}
\psset{unit=0.7364cm}
\begin{pspicture}[shift=-0.95](0,-1.1)(2.0,1.1)
\pspolygon[fillstyle=solid,fillcolor=lightlightblue](0,0)(1,1)(2,0)(1,-1)
\psarc[linewidth=0.025]{-}(1,-1){0.21}{45}{135}
\rput(1,0){$u$}
\end{pspicture} \ \ = \frac{\sin(\lambda - u)}{\sin \lambda} \,\mathbb I + \frac{\sin u}{\sin \lambda} \,\mathbb U,
\ee
where
\begin{subequations}
\begin{alignat}{2}
\mathbb I &= \ \
\psset{unit=0.7364cm}
\begin{pspicture}[shift=-0.95](0,-1.1)(2.0,1.1)
\pspolygon[fillstyle=solid,fillcolor=lightlightblue](0,0)(1,1)(2,0)(1,-1)
\end{pspicture} \ \ + \ \ 
\begin{pspicture}[shift=-0.95](0,-1.1)(2.0,1.1)
\pspolygon[fillstyle=solid,fillcolor=lightlightblue](0,0)(1,1)(2,0)(1,-1)
\psarc[linewidth=1.5pt,linecolor=blue](0,0){.707}{-45}{45}
\end{pspicture} \ \ + \ \ 
\begin{pspicture}[shift=-0.95](0,-1.1)(2.0,1.1)
\pspolygon[fillstyle=solid,fillcolor=lightlightblue](0,0)(1,1)(2,0)(1,-1)
\psarc[linewidth=1.5pt,linecolor=blue](2,0){.707}{135}{-135}
\end{pspicture} \ \ + \ \ 
\begin{pspicture}[shift=-0.95](0,-1.1)(2.0,1.1)
\pspolygon[fillstyle=solid,fillcolor=lightlightblue](0,0)(1,1)(2,0)(1,-1)
\psarc[linewidth=1.5pt,linecolor=blue](0,0){.707}{-45}{45}
\psarc[linewidth=1.5pt,linecolor=blue](2,0){.707}{135}{-135}
\end{pspicture}\ \ ,\\
\mathbb U &= \eE^{\ir \lambda}\ \
\psset{unit=0.7364cm}
\begin{pspicture}[shift=-0.95](0,-1.1)(2.0,1.1)
\pspolygon[fillstyle=solid,fillcolor=lightlightblue](0,0)(1,1)(2,0)(1,-1)
\psarc[linewidth=1.5pt,linecolor=blue](0,0){.707}{-45}{45}
\end{pspicture} \ \ + \eE^{-\ir \lambda} \ \ 
\begin{pspicture}[shift=-0.95](0,-1.1)(2.0,1.1)
\pspolygon[fillstyle=solid,fillcolor=lightlightblue](0,0)(1,1)(2,0)(1,-1)
\psarc[linewidth=1.5pt,linecolor=blue](2,0){.707}{135}{-135}
\end{pspicture} \ \ + \ \ 
\begin{pspicture}[shift=-0.95](0,-1.1)(2.0,1.1)
\pspolygon[fillstyle=solid,fillcolor=lightlightblue](0,0)(1,1)(2,0)(1,-1)
\psline[linewidth=1.5pt,linecolor=blue](0.5,0.5)(1.5,-0.5)
\end{pspicture}\ \ + \ \ 
\begin{pspicture}[shift=-0.95](0,-1.1)(2.0,1.1)
\pspolygon[fillstyle=solid,fillcolor=lightlightblue](0,0)(1,1)(2,0)(1,-1)
\psline[linewidth=1.5pt,linecolor=blue](0.5,-0.5)(1.5,0.5)
\end{pspicture} \ \ + \ \ 
\begin{pspicture}[shift=-0.95](0,-1.1)(2.0,1.1)
\pspolygon[fillstyle=solid,fillcolor=lightlightblue](0,0)(1,1)(2,0)(1,-1)
\psarc[linewidth=1.5pt,linecolor=blue](1,1){.707}{-135}{-45}
\psarc[linewidth=1.5pt,linecolor=blue](1,-1){.707}{45}{135}
\end{pspicture}\ \ .
\end{alignat}
\end{subequations}
One readily verifies that the operators $\mathbb U_j$, where $j$ indicates action at position $j$, satisfy the $s\ell(3)$ Hecke relations \eqref{eq:su3Hecke}.

\subsection{Representations of diagrammatic algebras}

A natural question for this model is the following: Is there a representation of the algebra $\mathsf{pA}_N(\alpha,\beta)$ defined on the RSOS vector space? For the $A_2^{(1)}$ RSOS models defined on the geometry of a strip, we believe that there exist representations of the algebra $\mathsf{A}_N(\beta)$ that underlie these models. For $N=2$, in the ordered basis \eqref{eq:pathbasis}, the elementary tiles are represented by the following matrices:
\begin{alignat}{2}
&\rho\bigg(\psset{unit=0.5cm}
\begin{pspicture}[shift=-0.95](0,-1.1)(2.0,1.1)
\pspolygon[fillstyle=solid,fillcolor=lightlightblue](0,0)(1,1)(2,0)(1,-1)
\end{pspicture}
\bigg) = 
\left(\begin{smallmatrix}
0 & 0 & 0 & 0 & 0 & 0 & 0 & 0 & 0 \\
0 & 0 & 0 & 0 & 0 & 0 & 0 & 0 & 0 \\
0 & 0 & 0 & 0 & 0 & 0 & 0 & 0 & 0 \\
0 & 0 & 0 & 0 & 0 & 0 & 0 & 0 & 0 \\
0 & 0 & 0 & 0 & 1 & 0 & 0 & 0 & 0 \\
0 & 0 & 0 & 0 & 0 & 0 & 0 & 0 & 0 \\
0 & 0 & 0 & 0 & 0 & 0 & 0 & 0 & 0 \\
0 & 0 & 0 & 0 & 0 & 0 & 0 & 0 & 0 \\
0 & 0 & 0 & 0 & 0 & 0 & 0 & 0 & 0 
\end{smallmatrix}\right),
\qquad
&&\rho\bigg(\psset{unit=0.5cm}
\begin{pspicture}[shift=-0.95](0,-1.1)(2.0,1.1)
\pspolygon[fillstyle=solid,fillcolor=lightlightblue](0,0)(1,1)(2,0)(1,-1)
\psarc[linewidth=1.5pt,linecolor=blue](0,0){.707}{-45}{45}
\end{pspicture}
\bigg) =
\left(\begin{smallmatrix}
0 & 0 & 0 & 0 & 0 & 0 & 0 & 0 & 0 \\
0 & 1 & 0 & y_0 & 0 & 0 & 0 & 0 & 0 \\
0 & 0 & 0 & 0 & 0 & 0 & 0 & 0 & 0 \\
0 & 0 & 0 & 0 & 0 & 0 & 0 & 0 & 0 \\
0 & 0 & 0 & 0 & 0 & 0 & 0 & 0 & 0 \\
0 & 0 & 0 & 0 & 0 & 0 & 0 & 0 & 0 \\
0 & 0 & 0 & 0 & 0 & 0 & 0 & 0 & 0 \\
0 & 0 & 0 & 0 & 0 & z_0 & 0 & 1 & 0 \\
0 & 0 & 0 & 0 & 0 & 0 & 0 & 0 & 0 
\end{smallmatrix}\right),
\\[0.2cm]%
&\rho\bigg(\psset{unit=0.5cm}
\begin{pspicture}[shift=-0.95](0,-1.1)(2.0,1.1)
\pspolygon[fillstyle=solid,fillcolor=lightlightblue](0,0)(1,1)(2,0)(1,-1)
\psarc[linewidth=1.5pt,linecolor=blue](2,0){.707}{135}{-135}
\end{pspicture}
\bigg) = 
\left(\begin{smallmatrix}
0 & 0 & 0 & 0 & 0 & 0 & 0 & 0 & 0 \\
0 & 0 & 0 & -y_0 & 0 & 0 & 0 & 0 & 0 \\
0 & 0 & 0 & 0 & 0 & 0 & 0 & 0 & 0 \\
0 & 0 & 0 & 1 & 0 & 0 & 0 & 0 & 0 \\
0 & 0 & 0 & 0 & 0 & 0 & 0 & 0 & 0 \\
0 & 0 & 0 & 0 & 0 & 1 & 0 & 0 & 0 \\
0 & 0 & 0 & 0 & 0 & 0 & 0 & 0 & 0 \\
0 & 0 & 0 & 0 & 0 & -z_0 & 0 & 0 & 0 \\
0 & 0 & 0 & 0 & 0 & 0 & 0 & 0 & 0 
\end{smallmatrix}\right),
\qquad
&&\rho\bigg(\psset{unit=0.5cm}
\begin{pspicture}[shift=-0.95](0,-1.1)(2.0,1.1)
\pspolygon[fillstyle=solid,fillcolor=lightlightblue](0,0)(1,1)(2,0)(1,-1)
\psarc[linewidth=1.5pt,linecolor=blue](0,0){.707}{-45}{45}
\psarc[linewidth=1.5pt,linecolor=blue](2,0){.707}{135}{-135}
\end{pspicture}
\bigg) = 
\left(\begin{smallmatrix}
1 & 0 & 0 & 0 & 0 & 0 & 0 & 0 & 0 \\
0 & 0 & 0 & 0 & 0 & 0 & 0 & 0 & 0 \\
0 & 0 & 1 & 0 & 0 & 0 & 0 & 0 & 0 \\
0 & 0 & 0 & 0 & 0 & 0 & 0 & 0 & 0 \\
0 & 0 & 0 & 0 & 0 & 0 & 0 & 0 & 0 \\
0 & 0 & 0 & 0 & 0 & 0 & 0 & 0 & 0 \\
0 & 0 & 0 & 0 & 0 & 0 & 1 & 0 & 0 \\
0 & 0 & 0 & 0 & 0 & 0 & 0 & 0 & 0 \\
0 & 0 & 0 & 0 & 0 & 0 & 0 & 0 & 1 
\end{smallmatrix}\right),
\\[0.2cm]%
&\rho\bigg(\psset{unit=0.5cm}
\begin{pspicture}[shift=-0.95](0,-1.1)(2.0,1.1)
\pspolygon[fillstyle=solid,fillcolor=lightlightblue](0,0)(1,1)(2,0)(1,-1)
\psline[linewidth=1.5pt,linecolor=blue](0.5,0.5)(1.5,-0.5)
\end{pspicture}
\bigg) = 
\left(\begin{smallmatrix}
0 & 0 & 0 & 0 & 0 & 0 & 0 & 0 & 0 \\
0 & y_1 & 0 & y_2 & 0 & 0 & 0 & 0 & 0 \\
0 & 0 & 0 & 0 & 0 & 0 & 0 & 0 & 0 \\
0 & S_{31}^{\boldsymbol a} & 0 & -y_1 & 0 & 0 & 0 & 0 & 0 \\
0 & 0 & 0 & 0 & 0 & 0 & 0 & 0 & 0 \\
0 & 0 & 0 & 0 & 0 & -z_1 & 0 & S_{32}^{\boldsymbol a} & 0 \\
0 & 0 & 0 & 0 & 0 & 0 & 0 & 0 & 0 \\
0 & 0 & 0 & 0 & 0 & z_2 & 0 & z_1 & 0 \\
0 & 0 & 0 & 0 & 0 & 0 & 0 & 0 & 0 
\end{smallmatrix}\right),
\qquad
&&\rho\bigg(\psset{unit=0.5cm}
\begin{pspicture}[shift=-0.95](0,-1.1)(2.0,1.1)
\pspolygon[fillstyle=solid,fillcolor=lightlightblue](0,0)(1,1)(2,0)(1,-1)
\psline[linewidth=1.5pt,linecolor=blue](0.5,-0.5)(1.5,0.5)
\end{pspicture}
\bigg) = 
\left(\begin{smallmatrix}
0 & 0 & 0 & 0 & 0 & 0 & 0 & 0 & 0 \\
0 & 0 & 0 & (S_{31}^{\boldsymbol a})^{-1} & 0 & 0 & 0 & 0 & 0 \\
0 & 0 & 0 & 0 & 0 & 0 & 0 & 0 & 0 \\
0 & 0 & 0 & 0 & 0 & 0 & 0 & 0 & 0 \\
0 & 0 & 0 & 0 & 0 & 0 & 0 & 0 & 0 \\
0 & 0 & 0 & 0 & 0 & 0 & 0 & 0 & 0 \\
0 & 0 & 0 & 0 & 0 & 0 & 0 & 0 & 0 \\
0 & 0 & 0 & 0 & 0 & (S_{32}^{\boldsymbol a})^{-1} & 0 & 0 & 0 \\
0 & 0 & 0 & 0 & 0 & 0 & 0 & 0 & 0 
\end{smallmatrix}\right),
\\[0.2cm]%
&\rho\bigg(\psset{unit=0.5cm}
\begin{pspicture}[shift=-0.95](0,-1.1)(2.0,1.1)
\pspolygon[fillstyle=solid,fillcolor=lightlightblue](0,0)(1,1)(2,0)(1,-1)
\psarc[linewidth=1.5pt,linecolor=blue](1,1){.707}{-135}{-45}
\psarc[linewidth=1.5pt,linecolor=blue](1,-1){.707}{45}{135}
\end{pspicture}
\bigg) = 
\left(\begin{smallmatrix}
0 & 0 & 0 & 0 & 0 & 0 & 0 & 0 & 0 \\
0 & 0 & 0 & 0 & 0 & 0 & 0 & 0 & 0 \\
0 & 0 & S_{12}^{\boldsymbol a} & 0 & 0 & 0 & S_{12}^{\boldsymbol a} & 0 & 0 \\
0 & 0 & 0 & 0 & 0 & 0 & 0 & 0 & 0 \\
0 & 0 & 0 & 0 & 0 & 0 & 0 & 0 & 0 \\
0 & 0 & 0 & 0 & 0 & 0 & 0 & 0 & 0 \\
0 & 0 & S_{21}^{\boldsymbol a} & 0 & 0 & 0 &  S_{21}^{\boldsymbol a}  & 0 & 0 \\
0 & 0 & 0 & 0 & 0 & 0 & 0 & 0 & 0 \\
0 & 0 & 0 & 0 & 0 & 0 & 0 & 0 & 0 
\end{smallmatrix}\right),
\end{alignat}
where the functions $y_i$ and $z_j$ are given by
\begin{subequations}
\begin{alignat}{4}
y_0 &= 1 - \frac{\eE^{-\ir \lambda}}{S^{\boldsymbol a}_{3,1}}, \qquad &&y_1 = -y_0 S^{\boldsymbol a}_{3,1}, \qquad &&y_2 = y_0 y_1, 
\\[0.1cm]
z_0 &= 1 - \frac{\eE^{-\ir \lambda}}{S^{\boldsymbol a}_{3,2}}, \qquad &&z_1 = -z_0 S^{\boldsymbol a}_{3,2}, \qquad &&z_2 = z_0 z_1. 
\end{alignat}
\end{subequations}

In contrast, for periodic boundary conditions, we do not expect the RSOS models to give representations of $\mathsf{pA}_N(\alpha,\beta)$. Instead, we expect a description of the RSOS models in terms of representations of $\mathsf{pA}_N(\alpha,\beta)$ that mirrors the similar situation for the $A_1^{(1)}$ RSOS models and the periodic Temperley-Lieb algebra. In the $A_1^{(1)}$ case, the matrix representatives of the generators $e_j$ realise a direct sum of representations corresponding to copies of the periodic Temperley-Lieb algebras, each assigned a different value of the fugacity $\alpha$ of the non-contractible loops \cite{BGJSV17}. For fixed $\lambda = \lambda_{p,p'}$, the number of direct summands is finite and the values of $\alpha$ depend on $p$ and $p'$.

\subsection{RSOS functional relations}

Because we do not have a construction of a representation of $\mathsf{pA}_N(\alpha,\beta)$ in the RSOS model, we cannot immediately conclude that the functional relations found in \cref{sec:Ftm.fr} hold for this model. 
Instead, we proceed by comparing
our results with the functional relations given in \cite{ZP95}. In this paper, 
the authors construct, for the $A^{(1)}_{n-1}$ RSOS models, the fused transfer matrices and the functions of the $Y$-system corresponding to rectangular Young diagrams of size $b \times q$. Their transfer matrices are denoted by $\Tb^{(b,q)}(u)$, with $b = 0,1,\dots,n$ and $q \in \mathbb N$. For $n=3$, these 
transfer matrices satisfy the same $T$-systems as those given in \eqref{eq:Tsystem}. The explicit dictionary between the two notations is
as follows:
\be
\Tb^{0,0}(u) = \Tb^{(0,m)}(-u-m\lambda) = \sigma^m\Tb^{(3,m)}(-u), \quad \Tb^{m,0}(u) = \Tb^{(1,m)}(-u), \quad \Tb^{0,m}(u) = \Tb^{(2,m)}(-u).
\ee
The $T$-system allows one to reconstruct each $\Tb^{m,0}(u)$ and $\Tb^{0,n}(u)$ as a polynomial in $\Tb^{1,0}(u)$ and $\Tb^{0,1}(u)$. One can then construct all the other fused transfer matrices from the functional relations \eqref{eq:FH}.

The closure relation obtained in \cite{ZP95} for the RSOS model is simpler than \eqref{eq:cloFH}. It takes the form of truncation relations:
\be
\Tb^{p'-2,0}(u) = \Tb^{0,p'-2}(u) = 0.
\ee
From \eqref{eq:Trelations}, it readily follows that $\Tb^{p'-1,0}(u) = \Tb^{0,p'-1}(u) = 0$ as well. In fact, using \eqref{eq:moreTTA}, we find that
\be
\Tb^{m,n}(u) = 0\qquad \textrm{for}\qquad m+n = p'-2.
\ee
All the matrices corresponding to pairs $(m,n)$ lying on the critical lines (drawn in red in \cref{fig:folding}) are identically zero in the RSOS models.

We now argue that these truncation relations are compatible with the closure relations \eqref{eq:cloFH}. In fact, the RSOS transfer matrices that appear in these closure relations all have a simple dependence on $u$. Indeed, by specializing
\eqref{eq:moreTTA} to $m = p'-1$ and $k=p'-2$, we obtain the relation 
\be
f_{-2}\Tb^{p',0}_0 = -f_{-1} \Tb^{p'-2,1}_0.
\ee 
Because $\Tb^{p',0}_0$ and $\Tb^{p'-2,1}_0$ are 
centred Laurent 
polynomials of degree width $2N$, we deduce that 
\be
\Tb^{p',0}_0 = f_{-1} \Ab, \qquad \Tb^{p'-2,1}_0 = -f_{-2} \Ab,
\ee
where $\Ab$ is independent of $u$. Likewise, using \eqref{eq:moreTTB}, we find $\Tb^{0,p'}_0 = f_{-1}\Abt$ and $\Tb^{1,p'-2}_0 = -\sigma f_{-1}\Abt$, with $\Abt$ independent of $u$. 

We apply a similar idea to determine $\Tb^{p'-3,0}_1$. From \eqref{eq:moreTT}, we 
thus find the relations
\be
\Tb^{0,p'-3}_0\Tb^{0,p'-3}_0 = \sigma^{p'-3} f_{-1} \Tb^{p'-3,0}_1, \qquad \Tb^{p'-3,0}_0\Tb^{p'-3,0}_1= f_{p'-3} \Tb^{0,p'-3}_0.
\ee
Combining these relations yields
\be
 \Tb^{p'-3,0}_0\Tb^{p'-3,0}_1\Tb^{p'-3,0}_2 = \sigma^{p'-3}f_{-3}f_{-2}f_{-1} \Ib, \qquad \Tb^{0,p'-3}_{-1} \Tb^{0,p'-3}_0 \Tb^{0,p'-3}_1 = \sigma^{p'-p}f_{-3}f_{-2}f_{-1} \Ib,
\ee
from which we infer that $\Tb^{p'-3,0}_0 = \sigma^{p'-1} f_{-3} \Bb$ and $\Tb^{0,p'-3}_0 = \sigma^{p'-p} f_{-2} \Bbt$  with $\Bb^3 = \Bbt^3 = \Ib$. The closure relations \eqref{eq:cloFH} then apply to the RSOS model, with 
\be
\Jb = 2 \Ab + \sigma^{p'}\Bb, \qquad \Kb = 2 \Abt + \sigma^{p'-p}\Bbt.
\ee


\end{document}